\documentclass[sigconf, screen, authorversion]{acmart}

\setcopyright{none}
\acmConference[CHI '26]{the 2026 CHI Conference on Human Factors in Computing Systems}{April 13--17, 2026}{Barcelona, Spain}
\acmDOI{10.1145/3772318.3790506}
\acmISBN{979-8-4007-2278-3/2026/04}

\makeatletter
\ifcase\ACM@format@nr
\relax % manuscript
\or % acmsmall
  \def\@authorfont{\large\sffamily}
  \def\@affiliationfont{\small\normalfont}
\or % acmlarge
\or % acmtog
  \def\@authorfont{\LARGE\sffamily}
  \def\@affiliationfont{\large}
\or % sigconf
  \def\@authorfont{\LARGE}
  \def\@affiliationfont{\small}
\or % siggraph
  \def\@authorfont{\normalsize\normalfont}
  \def\@affiliationfont{\normalsize\normalfont}
\or % sigplan
  \def\@authorfont{\Large\normalfont}
  \def\@affiliationfont{\normalsize\normalfont}
\or % sigchi
  \def\@authorfont{\bfseries}
  \def\@affiliationfont{\mdseries}
\or % sigchi-a
  \def\@authorfont{\bfseries}
  \def\@affiliationfont{\mdseries}
\or % acmengage
  \def\@authorfont{\LARGE}
  \def\@affiliationfont{\large}
\or % acmcp
  \def\@authorfont{\large\sffamily}
  \def\@affiliationfont{\small\normalfont}
\fi
\makeatother 

\usepackage{enumitem}
\usepackage{subcaption} 
\usepackage{multirow}
\usepackage{booktabs}
\usepackage{verbatim}
\usepackage{makecell}
\usepackage{stfloats}
\usepackage{siunitx}
\usepackage{tabularx}

\makeatletter
\let\@oldmkabstract\@mkabstract
\def\@mkabstract{\clearpage\@oldmkabstract}
\def\@copyrightyear{\@gobbletwo}
\makeatother

\begin{document}

\title{Bloom: Designing for LLM-Augmented Behavior Change Interactions\vspace{0.2em}}

\author{Matthew J\"{o}rke}
\affiliation{
\department{Computer Science}
\institution{Stanford University}
\city{Stanford}
\state{California}
\country{USA}}
\email{joerke@stanford.edu}

\author{Defne Gen\c{c}}
\authornote{Both authors contributed equally to this research.}
\affiliation{
\department{Computer Science}
\institution{Stanford University}
\city{Stanford}
\state{California}
\country{USA}}
\email{defneg@stanford.edu}

\author{Valentin Teutschbein}
\authornotemark[1]
\affiliation{
\department{Computer Science}
\institution{Hasso Plattner Institute}
\city{Potsdam}
\state{Brandenburg}
\country{Germany}}
\email{valentin.teutschbein@student.hpi.uni-potsdam.de}

\author{Shardul Sapkota}
\affiliation{
\department{Computer Science}
\institution{Stanford University}
\city{Stanford}
\state{California}
\country{USA}}
\email{sapkota@stanford.edu}

\author{Sarah Chung}
\affiliation{
\department{Computer Science}
\institution{Stanford University}
\city{Stanford}
\state{California}
\country{USA}}
\email{chvng@stanford.edu}

\author{Paul Schmiedmayer}
\affiliation{
\department{Mussallem Center for Biodesign}
\institution{Stanford University}
\city{Stanford}
\state{California}
\country{USA}}
\email{schmiedmayer@stanford.edu}

\author{Maria Ines Campero}
\affiliation{
\department{School of Medicine}
\institution{Stanford University}
\city{Stanford}
\state{California}
\country{USA}}
\email{icampero@stanford.edu}

\author{Abby C. King}
\affiliation{
\department{School of Medicine}
\institution{Stanford University}
\city{Stanford}
\state{California}
\country{USA}}
\email{king@stanford.edu}

\author{Emma Brunskill}
\affiliation{
\department{Computer Science}
\institution{Stanford University}
\city{Stanford}
\state{California}
\country{USA}}
\email{ebrun@cs.stanford.edu}

\author{James A. Landay}
\affiliation{
\department{Computer Science}
\institution{Stanford University}
\city{Stanford}
\state{California}
\country{USA}}
\email{landay@stanford.edu}

\renewcommand{\shortauthors}{J{\"o}rke, et al.}

\begin{abstract}
  Large language models (LLMs) offer novel opportunities to support health behavior change, yet existing work has narrowly focused on text-only interactions. Building on decades of HCI research on effective behavior change interactions, we present Bloom, an application for physical activity promotion that integrates an LLM-based health coaching chatbot with existing design strategies and UI elements. As part of Bloom's development, we conducted a redteaming evaluation and contribute a safety benchmark dataset. In a four-week randomized field study (N=54) comparing Bloom to a no-LLM control, we observed important shifts in psychological outcomes: participants in the LLM condition reported stronger beliefs that activity was beneficial, greater enjoyment, and more self-compassion. Both conditions significantly increased physical activity levels, doubling the proportion of participants meeting recommended weekly guidelines, though descriptively, we observed no advantage for the LLM condition in short-term physical activity levels. Instead, our findings suggest that LLMs may be more effective at shifting mindsets that precede longer-term behavior change.
\end{abstract}

\begin{CCSXML}
<ccs2012>
   <concept>
       <concept_id>10003120.10003121.10011748</concept_id>
       <concept_desc>Human-centered computing~Empirical studies in HCI</concept_desc>
       <concept_significance>500</concept_significance>
       </concept>
   <concept>
       <concept_id>10003120.10003121.10003124.10010870</concept_id>
       <concept_desc>Human-centered computing~Natural language interfaces</concept_desc>
       <concept_significance>300</concept_significance>
       </concept>
   <concept>
       <concept_id>10010405.10010444.10010449</concept_id>
       <concept_desc>Applied computing~Health informatics</concept_desc>
       <concept_significance>300</concept_significance>
       </concept>
   <concept>
       <concept_id>10003120.10003138.10011767</concept_id>
       <concept_desc>Human-centered computing~Empirical studies in ubiquitous and mobile computing</concept_desc>
       <concept_significance>300</concept_significance>
       </concept>
   <concept>
       <concept_id>10010147.10010178.10010179</concept_id>
       <concept_desc>Computing methodologies~Natural language processing</concept_desc>
       <concept_significance>100</concept_significance>
       </concept>
 </ccs2012>
\end{CCSXML}

\ccsdesc[500]{Human-centered computing~Empirical studies in HCI}
\ccsdesc[300]{Human-centered computing~Natural language interfaces}
\ccsdesc[300]{Applied computing~Health informatics}
\ccsdesc[300]{Human-centered computing~Empirical studies in ubiquitous and mobile computing}
\ccsdesc[100]{Computing methodologies~Natural language processing}

\keywords{Physical activity, health coaching, behavior change, large language models (LLM), mobile health}

\begin{teaserfigure}
  \centering
  \includegraphics[width=\textwidth]{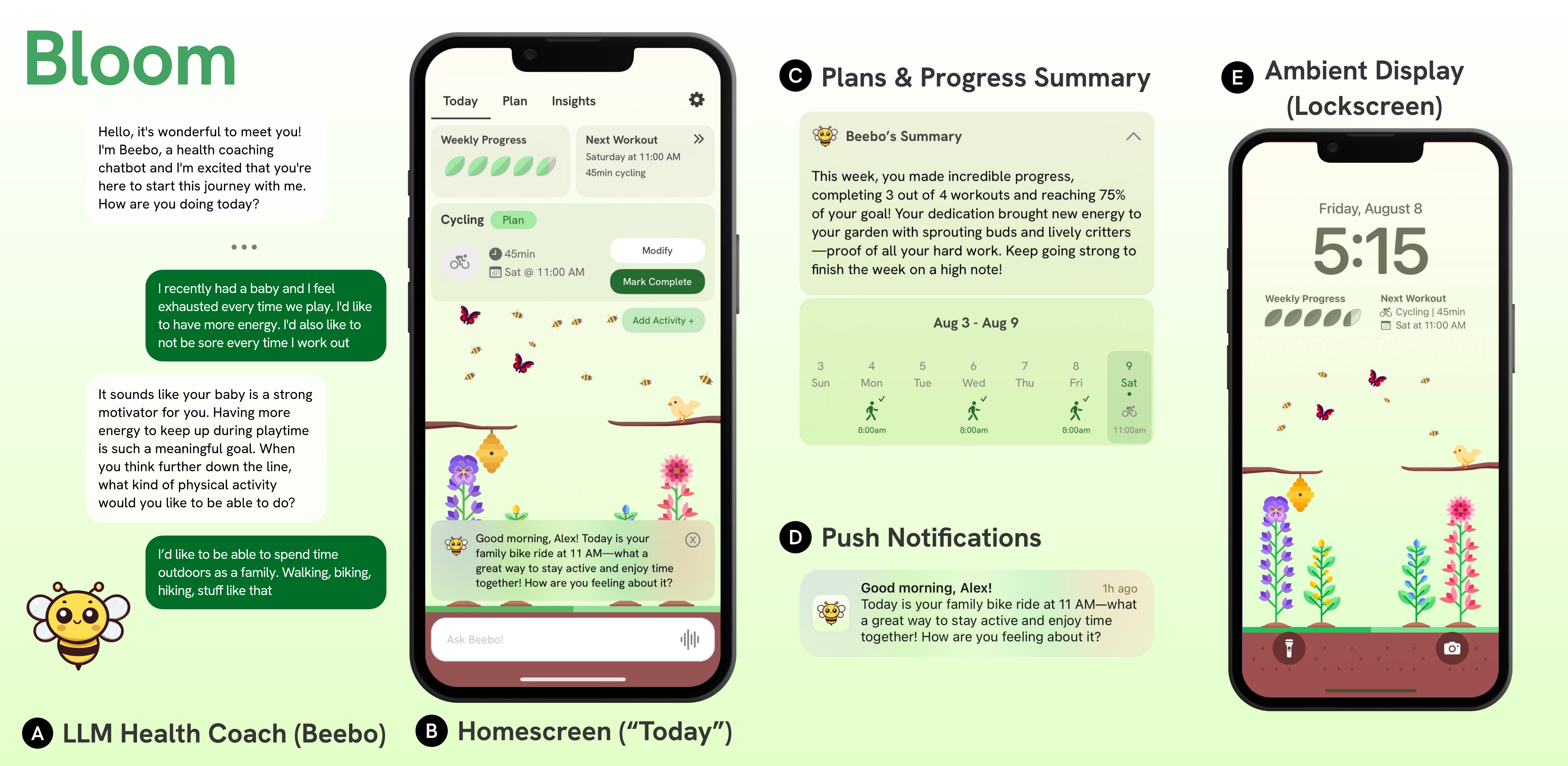}
  \caption{\textbf{Bloom integrates an LLM-based physical activity coaching chatbot (Beebo) with established behavior change interactions.}
    (A) An onboarding conversation with Beebo, an LLM health coaching chatbot.
    (B) The app's home screen, showing weekly progress, upcoming activities, and proactive messages from Beebo.
    (C) LLM-generated activity plans and progress summaries.
    (D) LLM-generated notifications encouraging action or reflection.
    (E) A garden-based ambient display on the user's lockscreen.
  }
  \label{fig:teaser}
  \Description{Overview of the Bloom application. A (left): A conversation between the user and Beebo, the LLM health coach, depicted as a cartoon bee. The user shares that they recently had a baby and expresses a desire for more energy to play. Beebo responds by affirming their motivations and asks what kinds of activities they would like to be able to do in the future. B (center): The homescreen (“Today” screen) of the application, showing a garden ambient display with widgets for weekly progress, upcoming workouts, and a proactive message from Beebo. C (top right): A plan widget, showing a calendar with Monday/Wednesday/Friday walks and a Saturday bike ride, and a progress summary where Bloom congratulates the user for completing three of four workouts. D (bottom right): A push notification reminding the user about their scheduled family bike ride. E (far right): A garden-based ambient display on an iPhone lockscreen, showing a prominent garden wallpaper as well as weekly progress and the next workout widgets.}
\end{teaserfigure}

\maketitle

\section{Introduction}
Regular physical activity (PA) reduces the risk of cardiovascular disease, diabetes, premature mortality, depression, and numerous other important physical and mental health outcomes~\cite{piercy2018physical}, yet one in four adults worldwide and nearly half of the US population fall short of recommended PA guidelines~\cite{bull2020world, cdc2022}. 
Many of the most effective approaches for increasing PA, such as in-person coaching, are resource-intensive and costly to deliver at scale.
This challenge has motivated research in mobile health, which explores how more accessible technologies like smartphones and wearables can be used to deliver scalable, cost-effective, and personalized support~\cite{hicks2023leveraging}. 

Recent advances in large language models (LLMs)~\cite{ouyang2022training, bommasani2022opportunities} have sparked a growing interest in their applications for health behavior change, presenting novel opportunities for improved support. Early research on LLM-based health coaching chatbots suggests that they offer marked improvements in conversational flexibility over prior rule-based systems~\cite{jorke2025gptcoach, wang2025exploring, xu2025goals, heydari2025anatomy}. 
This flexibility could allow LLMs to more accurately implement evidence-based health communication frameworks like motivational interviewing~\cite{miller2023motivational}.
Moreover, LLMs offer an improved capacity to interpret diverse sources of personal context, enabling interventions that are more effectively tailored to an individual~\cite{khasentino2025personal, merrill2026transforming}. 
Unlike prior systems, which primarily rely on quantitative context (e.g., step count) for personalization, LLMs can also leverage \textit{qualitative context}~\cite{jorke2025gptcoach} that is more easily expressed in natural language. 
Qualitative factors such as goals, motivation, values, life circumstances, time constraints, and access to resources are key constructs in behavior change theory and are crucial for effective personalization.

Meanwhile, decades of research in human-computer interaction (HCI) has produced a rich body of interactions for health behavior change applications---from goal setting~\cite{ekhtiar2023goals} and self-tracking~\cite{epstein2020mapping} to ambient feedback~\cite{consolvo2008activity, murnane2023narrative}, and just-in-time nudges~\cite{nahum2018just}---which are both grounded in behavior change theory~\cite{consolvo2009theory, locke2002building, michie2011behaviour} and have demonstrated efficacy~\cite{webb2010using, yang2019comparative}. 
However, existing research on LLMs for behavioral health has focused almost exclusively on text-only interactions, either through chat~\cite{jorke2025gptcoach, wang2025exploring, xu2025goals}, summaries~\cite{merrill2026transforming, khasentino2025personal}, or nudging~\cite{song2025investigating, mantena2025fine}.
This narrow focus overlooks the potential for LLMs to \textit{augment} established interactions, rather than replace them.
Behavior change theory emphasizes that behavior is shaped not just by what is said, but when, how, and through which channels it is delivered~\cite{mohr2014behavioral, michie2011behaviour}. HCI research has shown that multimodal interaction~\cite{turk2014multimodal, oviatt1999ten}, which blends textual, graphical, ambient, or wearable delivered feedback, can help users manage cognitive load~\cite{oviatt2004we}, make behaviors more salient through glanceable or ambient displays~\cite{consolvo2008activity, matthews2007designing, gouveia2016exploring}, and foster affective engagement~\cite{lin2006fish, murnane2020designing}. Taken together, this research suggests that behavior change interactions that augment LLM chat with other interaction modalities may offer greater efficacy. 
 
At the same time, the technology is recent and real-world evaluations of LLMs for behavioral health remain limited. Most studies have relied on static assessments or single-session studies~\cite{mantena2025fine, jorke2025gptcoach, meyer2025llm, srinivas2025substance, heydari2025anatomy}, with few studies reporting on preliminary field deployments~\cite{wang2025exploring, xu2025goals, loerakker2025give}, offering only partial insights into how users engage with LLM-based behavior change systems over time in real-world settings.
Importantly, most prior work assumes the efficacy of LLMs without directly evaluating LLM-based systems relative to established pre-LLM approaches.
As a result, this limits our understanding of the unique design opportunities and requirements for LLM-augmented behavior change systems.

To address this gap, we present \textbf{Bloom}, a mobile application for promoting PA that combines an LLM coaching chatbot with established behavior change interactions, including goal setting, action planning, activity tracking, data visualization, an ambient display, and push notifications (Figure \ref{fig:teaser}). 
Our goal in developing and evaluating Bloom was to surface design insights for \textit{LLM-augmented behavior change interactions} by studying how LLM augmentation shapes user experience and behavior.
Bloom's LLM coach implements conversational strategies from motivational interviewing~\cite{miller2023motivational} and a scientifically validated health coaching program~\cite{king2002stanford}. Conversations with Bloom's chatbot simultaneously serve as standalone interventions and sources of qualitative context for other parts of the system to use for personalization. As part of Bloom's development, we also conducted a redteaming evaluation with four domain experts and created a safety benchmark dataset for LLM coaching with 600 examples to evaluate our system's safety filters.

We evaluate Bloom in a four-week, between-subjects, randomized field study with $N=54$ participants, comparing Bloom to a no-LLM control that removes the LLM coach and all LLM augmentation.
Our comparative study was formative, exploratory, and primarily design-oriented, with the no-LLM control condition allowing us to isolate changes in participant experiences specifically attributable to LLM augmentation. 
A secondary aim was to gather preliminary evidence of Bloom's impact on measurable PA outcomes, noting that our early-stage study was not designed to establish statistically significant treatment-control differences (e.g., as a clinical feasibility or efficacy trial~\cite{czajkowski2015ideas}).
Our evaluation uses a mixed-methods approach, including qualitative coding of semi-structured interviews; pre/post, weekly, and daily survey measures; objective PA data collected from participants' wearables; and application usage logs including chats, plans, and UI interactions.

We observed important shifts in psychological and motivational outcomes, with participants in the LLM condition reporting stronger beliefs that their activity was beneficial to their health. They also noted greater enjoyment of exercise, an expanded appreciation of what ``counts'' as activity, and increased self-compassion when goals were missed.
The LLM condition also produced more varied and personalized plans, with slightly higher completion rates, and participants spent more than five times as much time in the app than control participants.
Both conditions significantly increased their PA relative to a pre-study baseline, doubling the number of participants who met or exceeded the recommended 150 min/week of PA. 
However, across quantitative wearable outcomes, descriptive patterns did not indicate a clear advantage of the LLM condition during the four-week study period.
Instead, our survey and interview findings suggest that LLMs may be particularly valuable for fostering positive mindsets, flexible planning, and sustained engagement---factors that behavior change theory links to longer-term maintenance rather than as immediate drivers of short-term levels of PA. 
We discuss the implications of these findings for designing LLM-augmented interventions, highlighting how the LLM's role as a conversational coach shifted mindsets and motivation more than short-term PA, how qualitative context enabled non-prescriptive interactions that promote agency, and how relational cues increased engagement while introducing design trade-offs around attachment and overreliance.

In summary, this work makes the following contributions:
\begin{enumerate}[leftmargin=*]
    \item The \textbf{Bloom system}, which implements a novel design for integrating an LLM health coaching chatbot with established design strategies and interface elements. Bloom is grounded in evidence-based strategies from behavior change theory and interactions from HCI research. Bloom leverages qualitative context from coaching conversations to personalize a number of \textit{LLM-augmented behavior change interactions}.
    \item A \textbf{safety evaluation} of our LLM-based health coaching agent. We share results, prompts, and a benchmark dataset containing 600 examples to support future research on safety.
    \item Findings from our \textbf{field study} of an LLM-based PA coaching agent, a four-week, between-subjects randomized study with $N=54$ participants comparing Bloom to a no-LLM control. We analyze qualitative coding of pre/post semi-structured interviews, survey responses, wearable PA outcomes, and system interaction logs to assess both behavioral and psychological outcomes.
    \item \textbf{Design implications} drawn from our field study's findings to inform future work on LLM-augmented behavior change systems and LLM health coaching.
\end{enumerate}
\section{Related Work}

In this section, we situate our work within the prior literature on HCI systems for PA behavior change, automated and human health coaching, and LLMs for behavioral health.     

\subsection{HCI Systems for Physical Activity Promotion}

Designing for PA promotion has a long history in HCI~\cite{epstein2020mapping}, with many approaches advocated for in the literature~\cite{consolvo2014designing, klasnja2012healthcare, hicks2023leveraging}. 
One popular line of work draws from the broader field of personal informatics~\cite{li2010stage} and aims to promote behavior change by helping users make sense of their activity patterns through self-monitoring~\cite{korotitsch1999overview, choe2017semi} and reflection~\cite{baumer2015reflective, bentvelzen2022revisiting, cho2022reflection}. Common implementations frequently include visualizations~\cite{aseniero2020activity, choe2015characterizing,kersten2017personal, thudt2015visual} such as step count graphs, often seen in commercial health tracking apps. Other systems employ more qualitative forms of feedback, including ambient displays~\cite{consolvo2008activity, murnane2023narrative, lane2012bewell} and textual feedback~\cite{bentley2013health, consolvo2006design}.
Qualitative feedback can be less overwhelming~\cite{daskalova2017lessons} and promote more positive mindsets~\cite{murnane2020designing} than quantitative feedback.
Nudging represents another class of interventions, which in mobile health are often instantiated as push notification reminders for just-in-time support~\cite{bentley2013power,intille2004ubiquitous, vandelanotte2023increasing, nahum2018just}. Unlike other kinds of reflective or ambient support, just-in-time interventions aim to capture the user's attention and often optimize intervention timing and content to maximize efficacy.

Another prominent body of work focuses on goal setting as a strategy for promoting PA~\cite{ekhtiar2023goals, munson2012exploring}.
Goal setting theory emphasizes that goals are most effective when they are personally meaningful, realistic, and sufficiently challenging~\cite{locke2002building}. 
Users are often asked to set their own PA goals~\cite{gasser2006persuasiveness, lin2006fish}, or are alternatively provided a goal based on their behavior history~\cite{consolvo2006design}. Prior work has also employed conversational agents for collaborative goal setting~\cite{king2020effects, bickmore2005acceptance, jorke2025gptcoach}. Since intentions alone do not always translate to goal achievement, action plans~\cite{agapie2018crowdsourcing} and implementation intentions~\cite{gollwitzer2006implementation, gollwitzer1999implementation, sefidgar2024improving} are popular techniques for facilitating goal realization by specifying when, where, and how a behavior will occur.

Beyond individual-level interventions, prior work has also explored social strategies for behavior change~\cite{saksono2024socio}. These strategies typically involve sharing within a social network~\cite{consolvo2006design, lin2006fish, toscos2006chick, anderson2007shakra} or incorporating leaderboards and challenges to introduce competition in games~\cite{buis2009evaluating, gobel2010serious, sinclair2007considerations, althoff2016influence, shameli2017gamification}. While collaborative approaches have been shown to be effective, social comparison can sometimes feel demotivating or misaligned with preferences~\cite{munson2012exploring, murnane2015mobile}, making social features more challenging to design effectively. Competitive elements in particular, such as leaderboards or challenges, have shown mixed effects on users' motivation, especially during periods when they were less active~\cite{klasnja2009using, consolvo2006design}. Although gamification can increase short-term engagement, its long-term impact on sustained behavior change also remains underexplored~\cite{stepanovic2018gamification}.

Bloom implements several behavior change interactions from this literature, including activity tracking, data visualizations, ambient displays, push notifications, goal setting, and action planning. Due to the challenge of appropriately designing for social and gamification interactions, compounded with the uncertainty of designing for novel, LLM-augmented interactions, we exclude these interactions as important explorations for future work.

\subsection{Health Coaching}

Health coaching is an established intervention for facilitating behavior change~\cite{olsen2010health, wolever2013systematic, olsen2014health}. Although coaching is a multidisciplinary concept with numerous everyday associations (e.g., sports, executive, or life coaching), here we adopt Olsen et al.'s~\cite{olsen2014health} definition of health coaching as \textit{``a goal-oriented, client-centered partnership that is health-focused and occurs through a process of client enlightenment and empowerment.''} A common framework used in such programs is motivational interviewing~\cite{miller2023motivational}, an evidence-based counseling approach that emphasizes eliciting motivation from the client rather than imposing prescriptive advice.

However, in-person health coaching is expensive, not widely accessible, and does not scale to global need. A long history of research on health dialogue systems~\cite{bickmore2006health} and automated health coaching~\cite{mitchell2021automated} has aimed to address these challenges by emulating aspects of human coaching with conversational agents. This research dates back at least as far as the seminal work on the Transtheoretical Model (TTM) itself~\cite{prochaska1997transtheoretical}, and we refer to~\cite{jorke2025gptcoach, mitchell2021automated} for reviews of pre-LLM automated coaching. 
While meta-analyses have found that pre-LLM coaching chatbots are effective at promoting PA~\cite{singh2023systematic}, these systems rely on predefined rules or templates that inherently limit their flexibility and personalization compared to human coaches~\cite{luo2021promoting}. Unlike prior rule-based systems, Bloom leverages an LLM to engage in dynamic, open-ended conversations, interpret and integrate user data through tool calls, and reference past interactions.

\subsection{LLMs and Behavioral Health}

Given that LLMs offer marked improvements over prior rule-based chatbots in conversational flexibility, recent work has explored their applications to health coaching. 
Bloom's LLM health coach is based on the GPTCoach system~\cite{jorke2025gptcoach}, a GPT-4-based chatbot that implements the onboarding conversation of a PA coaching program and uses conversational strategies from motivational interviewing. Participants reported highly positive and personalized experiences interacting with GPTCoach, and the authors found that it was effective at using strategies aligned with motivational interviewing principles. 
Other LLM-based chatbots have explored domains beyond PA. \citet{wang2025exploring} introduced HealthGuru, a sleep health chatbot that uses behavioral guidelines and sensor data to provide recommendations. \citet{xu2025goals} developed a chatbot for New Year's resolutions that personalized suggestions based on chatbot interactions, user profiles, and contextual cues (e.g., location, time, goals, etc.). In concurrent work, \citet{heydari2025anatomy} present a multi-agent system consisting of a data science agent, health domain expert agent, and a health coaching agent. Bloom extends these approaches by integrating chat with a suite of UI-based behavior change interactions. 

A parallel line of research has focused on improving LLMs' ability to engage in motivational interviewing. Studies have applied motivational interviewing principles in health coaching contexts~\cite{meyer2025llm, jorke2025gptcoach} and beyond, including mental health~\cite{shah2022modeling, chiu2024computational}, substance use counseling~\cite{kumar2024generation, steenstra2024virtual}, and wellness-related reflection~\cite{rose2022generation, bașar2025well}. More recent work has enhanced motivational interviewing adherence through strategy prediction~\cite{sun2025rethinking} or few-shot demonstrations~\cite{xie2024few}. Bloom's LLM coach conditions each response on a predicted motivational interviewing strategy.  

LLMs have also been used to help users make sense of and reflect on their personal health data. \citet{stromel2024narrating} showed that LLM-generated narratives can foster deeper reflection, more focused attention, and a greater sense of reward compared to conventional charts. \citet{loerakker2025give} study interactions with the commercial WHOOP Coach and identify a ``give and take'' dynamic in which effective sensemaking requires users to provide sufficient context (give) such that the agent can provide meaningful insight (take). Participants reported mixed impressions, citing generic, impersonal responses and an overly performance-oriented focus as the most common concerns. In contrast, Bloom focuses not only on reflection but on supporting behavior change, using dialogue to elicit context for richer personalization, and employing non-judgmental communication to avoid the pitfalls of performance-oriented support.

Beyond chatbots, most studies on LLMs for behavioral health have focused on improving message personalization or sensor data analysis. For example, to improve adherence to the Capacity, Opportunity, Motivation-Behavior (COM-B) framework, \citet{vardhan2025correction} explored LLM prompting techniques, whereas \citet{mantena2025fine} used a finetuning approach to align with the TTM. LLMs have also been applied to downstream health tasks, including generating stage-appropriate behavior change advice~\cite{bak2024potential}, modeling trust and self-referencing in coaching~\cite{meywirth2025designing}, performing health tasks on raw self-tracking data~\cite{englhardt2024classification, liu2023large, choube2025gloss, li2025vital}, and extracting insights from personal health data~\cite{subramanian2024graph, shouborno2025llasa, feli2025llm, zhang2025sensorlm, zheng2025promind, khasentino2025personal, fang2024physiollm, merrill2026transforming}.
While these systems show improvements in adhering to theoretical frameworks and personalizing content with quantitative data, they are not designed to proactively elicit and incorporate qualitative information about users' goals, values, or barriers into a longitudinal coaching process. Bloom explicitly integrates multi-session LLM coaching with behavior change interventions, leveraging the qualitative context elicited from coaching for greater personalization.
\section{Bloom: System Overview}
\label{sec:system}

\begin{figure*}[!t]
    \centering
    \includegraphics[width=\linewidth]{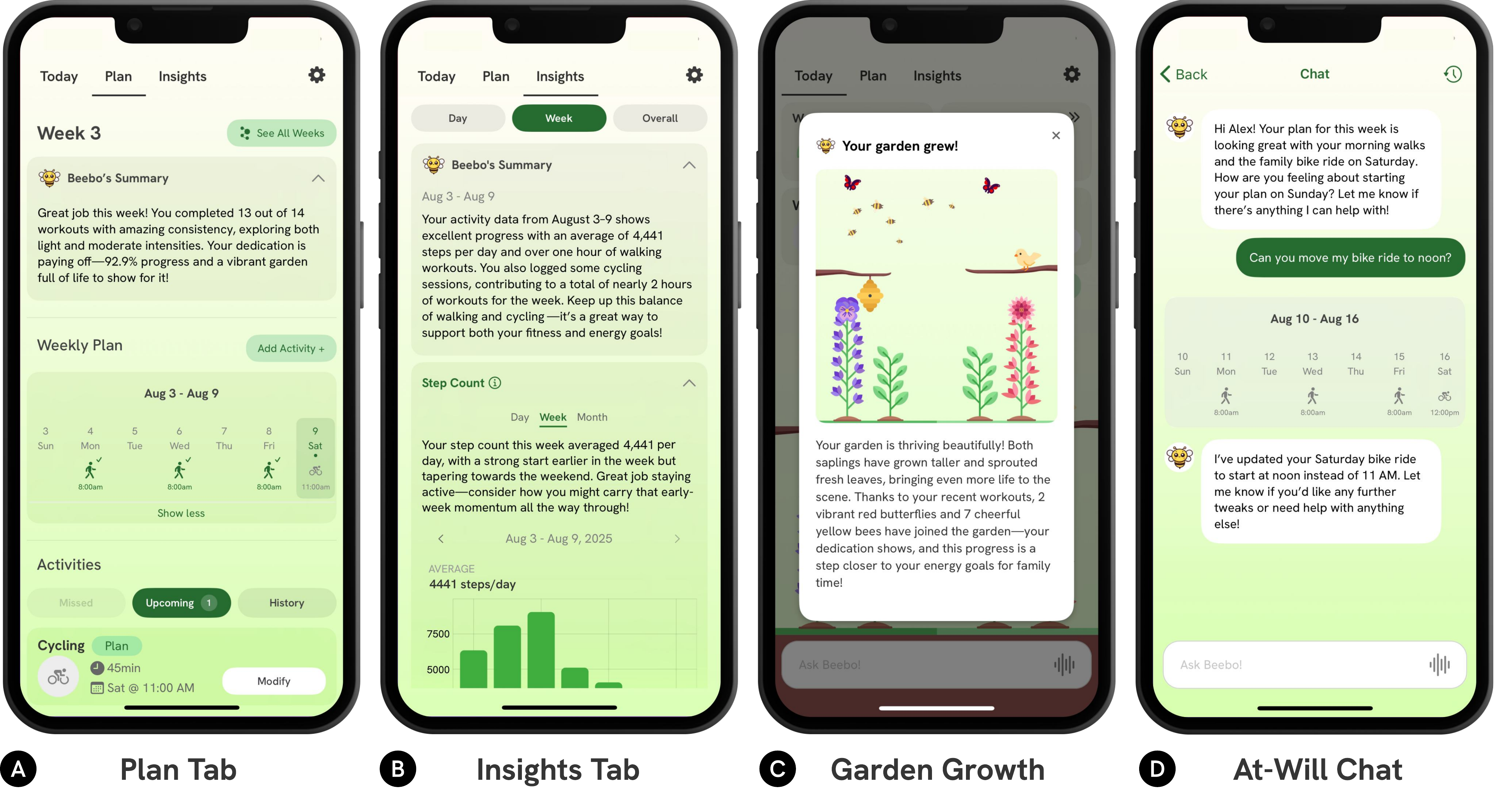}
    \caption{\textbf{Additional screens in the Bloom application.} (A) The \textit{Plan} tab shows the user's current weekly plan, an LLM-generated progress summary, along with missed, upcoming, and past activities. (B) The \textit{Insights} tab presents visualizations of wearable data, annotated with LLM-generated summaries of trends and progress. (C) When the user's \textit{garden grows}, a celebratory modal appears with an LLM-generated message linking progress in the ambient display to recent achievements. (D) During \textit{at-will chat}, the user can request edits to their plan in natural language, upon which the agent calls plan edit tools, shown as inline widgets.}
    \label{fig:additional-screens}
    \Description{Additional screens in the Bloom application. A (left): The Plan tab with a weekly plan, showing activities scheduled for the week in a calendar-like widget, an LLM-generated summary congratulating the user, and a list of upcoming and past activities. B (center left): The Insights tab, displaying a weekly summary congratulating the user for their step count and workouts this week, and a bar chart of steps per day. C (center right): A Garden Growth modal, where flowers, butterflies, and bees appear in a garden to celebrate recent workouts, accompanied by a personalized motivational message. D (right): An At-will chat where the user asks Beebo to move a scheduled bike ride, and the agent updates the plan in natural language, responding with an inline plan widget.}
\end{figure*}

Bloom is an iOS application designed for multi-week PA support that integrates an LLM coaching agent (Beebo) with established behavior change interactions.
In this section, we describe the Bloom system, including the Active Choices program, Bloom's behavior change interactions, system architecture, LLM coaching agent, and safety filters. 
Bloom's user interface is shown in Figures~\ref{fig:teaser}~\&~\ref{fig:additional-screens} and our system architecture is shown in Figure \ref{fig:system}. Our code is publicly available at \url{https://github.com/StanfordHCI/Bloom}.

\subsection{Active Choices Program}
Bloom implements the Stanford Active Choices Program~\cite{king2002stanford}, an evidence-based, scientifically validated counseling program for PA promotion, typically delivered via phone, developed by researchers at the Stanford University School of Medicine~\cite{wilcox2006results, wilcox2008active, king2007ongoing, king2014exercise, castro2011physical}. 
Active Choices is grounded in the Transtheoretical Model~\cite{prochaska1997transtheoretical} and Social Cognitive Theory~\cite{bandura1999social}. While Active Choices is not a motivational interviewing program, it does advocate for a facilitative communication style and many facilitators are trained in motivational interviewing~\cite{miller2023motivational}.

Active Choices facilitators follow a structured sequence of conversations with each client, beginning with an onboarding conversation and continuing with regular check-ins. The program provides guidance on communication style and topics to discuss, such as tailoring support to the client's stage of change, collaboratively setting and reviewing weekly goals, and problem-solving common barriers. It also provides safety considerations drawn from the American College of Sports Medicine~\cite{acsm2025}.

Bloom's LLM coaching agent directly encodes the program's structure and content for onboarding conversations, PA planning, and check-in conversations in its prompts (described in further detail below). To adapt Active Choices to a digital format, Bloom adds UI elements for planning, activity tracking, and data visualizations. Bloom also introduces interactions not present in the original program, including a garden-based ambient display and push notifications. 

\begin{table*}[t]
\centering
\small
\begin{tabularx}{\textwidth}{@{}p{0.22\textwidth} X@{}}

\toprule
\textbf{Behavior Change Interaction} & \textbf{Behavior Change Techniques (BCTs) Implemented}\\
\midrule
Goal Setting \& Planning & 1.1 Goal setting (behavior); 1.2 Problem solving; 1.3 Goal setting (outcome); 1.4 Action planning; 1.5 Review behavior goal(s); 1.6 Discrepancy between current behavior and goal; 1.9 Commitment; 8.7 Graded tasks\\
\midrule
Activity Tracking & 2.3 Self-monitoring of behavior; 2.2 Feedback on behavior; 10.4 Social reward\\
\midrule
Data Visualizations & 1.6 Discrepancy between current behavior and goal; 2.2 Feedback on behavior; 2.3 Self-monitoring of behavior; 2.4 Self-monitoring of outcome(s); 2.7 Feedback on outcome(s) of behavior; \\
\midrule
Ambient Display & 2.2 Feedback on behavior; 7.1 Prompts/cues; 10.3 Non-specific reward; 10.6 Non-specific incentive; 14.5 Rewarding completion\\
\midrule
Push Notifications & 1.2 Problem solving; 2.2 Feedback on behavior; 2.3 Self-monitoring of behavior; 7.1 Prompts/cues; 10.4 Social reward\\
\midrule
LLM Chat & 1.2 Problem solving; 1.4 Action planning; 1.9 Commitment; 3.1 Social support (unspecified); 4.1 Instruction on how to perform the behavior; 5.1 Information about health consequences; 15.1 Verbal persuasion about capability; 15.3 Focus on past success; 15.4 Self-talk; \\
\bottomrule
\end{tabularx}
\caption{Behavior change interactions and the corresponding behavior change techniques (BCTs) they implement, following Michie et al.'s~\cite{michie2013behavior} taxonomy.
BCTs capture the psychologically active elements of an intervention and proposed mechanisms of change.}
\label{table:bcts}
\end{table*}

\subsection{Behavior Change Interactions}
Bloom integrates the LLM coaching agent with six evidence-based behavior change interactions. 
Following Slovak \& Munson's framework for psychosocial intervention design in HCI~\cite{slovak2024hci}, we distinguish between the \textit{capabilities} that LLMs introduce and the intervention \textit{components} these capabilities enable. We argue that LLMs provide two new capabilities: (1) conducting motivational interviewing conversations with higher fidelity and (2) personalizing intervention content using qualitative context expressed in natural language. We design new intervention components by leveraging these LLM capabilities toward serving a particular psychological function. Some components are drawn directly from Active Choices (e.g., collaborative goal setting), while others augment established components with LLM-driven personalization (e.g., LLM-generated push notifications). This ensemble of components constitutes an intervention \textit{system}: Bloom.

Below, we list the \textit{behavior change interactions} implemented by Bloom, which instantiate intervention components using existing UI design patterns and elements. In Table \ref{table:bcts}, we list the \textit{behavior change techniques (BCTs)} implemented by each interaction, drawn from Michie et al.'s taxonomy~\cite{michie2013behavior}. BCTs help isolate the psychologically active elements of an intervention and establish the proposed mechanism of change.

\subsubsection{Goal Setting \& Planning}
\label{sec:goal-setting}
Beebo, Bloom's LLM agent, operationalizes the Active Choices program for weekly, collaborative goal setting and planning. By helping users define a meaningful goal, translate it into a concrete weekly plan, and anticipate obstacles, this interaction is intended to increase goal commitment, bolster self-efficacy, and reduce the cognitive load of deciding when and how to act.

In the onboarding conversation, Beebo first introduces the program, confirms the user's interest and commitment, and proceeds to ask about their high-level motivations, long-term goals, past experiences, barriers, resources, and preferences. Beebo then uses this background information to collaboratively propose, request feedback on, and confirm commitment to a personalized PA plan. The plan is an actionable operationalization of the user's broader motivations, and defines a short-term goal of completing all planned activities in the coming week. A weekly plan specifies FITT (frequency, intensity, time, type) parameters for each activity (e.g., a moderate-intensity, 20-min walk at 8AM on Mon-Wed-Fri). Plans appear in the Plan tab (Figure \ref{fig:additional-screens}A) and can be edited via the UI or chat. Consistent with Active Choices, Bloom does not prescribe generic numerical goals such as exercise minutes or steps per day. To ensure plans remain achievable, Beebo is prompted to suggest adjustments if the plan appears overly ambitious. After confirming the plan, Beebo discusses anticipated barriers and, when appropriate, offers advice on overcoming them, and concludes by scheduling the next weekly check-in.

During subsequent weekly check-ins, Beebo and the user review progress using the user's subjective self-assessment, qualitative reflections, wearable data, and plan completion logs. Following Active Choices, Beebo modulates the difficulty of subsequent plans based on their progress and feedback. If the user completed their plan, Beebo offers praise, explores what helped them succeed, and discusses gently increasing the plan if the user's current PA levels fall below recommended guidelines for aerobic activity. If the user is already meeting the recommended guidelines, subsequent plans can focus on maintenance or (if appropriate) add strength or flexibility exercises. If the user did not complete their plan, Beebo revisits their motivation and barriers, and proposes adjustments (e.g., simplifying activities, changing timing, or maintaining current levels).

\subsubsection{Activity Tracking}
Bloom automatically records activities logged via Apple's HealthKit API from connected activity trackers.\footnote{While our field study required participants to use an Apple Watch, our system supports any wearable that writes to HealthKit. Our system can also support users that do not own a wearable through phone-based tracking and/or manual workout logging.} Bloom attempts to link incoming HealthKit workouts to the user's plan, automatically marking linked activities as complete. Workouts completed outside of a user's plan are logged as ``bonus'' activities. Activities that were not recorded in HealthKit can be manually added and/or marked complete. By reducing the effort required to track activity, supporting flexible logging, and recognizing bonus activities, this interaction aims to support self-monitoring and reward progress.

\subsubsection{Data Visualizations}
The Insights tab (Figure \ref{fig:additional-screens}B) displays visualizations of the user's data at various granularities, including both metrics directly about behavior (e.g., workouts, step counts) and about outcomes (e.g., resting heart rate). Beebo annotates each chart with natural language summaries, and is prompted to describe trends, connect patterns in the data to the user's goals, and celebrate progress.
This interaction aims to help users make sense of their data, offer positive reinforcement, highlight where behavior aligns or diverges from their goals, and support ongoing self-monitoring.

\subsubsection{Ambient Display}
Bloom implements a garden-based ambient display (Figure \ref{fig:ambient-display}) inspired by prior work~\cite{consolvo2008designing, lin2006fish, murnane2020designing}.
A flower grows in the garden as activities from the weekly plan are completed. Upon achieving their weekly goal, the flower fully blooms and a new flower starts the next week. Bees or butterflies appear above flowers for each completed activity.
The ambient display is shown both in the app's background and on the user's lockscreen wallpaper (Figure \ref{fig:teaser}E), configured to automatically update via a custom iOS shortcut.
Lastly, Beebo generates a congratulatory message (Figure \ref{fig:additional-screens}C) every time the garden grows. Supplementary Material \ref{appendix:ambient-display} provides a full description of the ambient display's logic. In offering a glanceable, qualitative, and affectively positive representation of a user's progress, this interaction is intended to offer positive reinforcement and maintain ongoing awareness of activity without requiring deliberate engagement.

\begin{figure*}
    \centering
    \vspace{1em}
    \includegraphics[width=\textwidth]{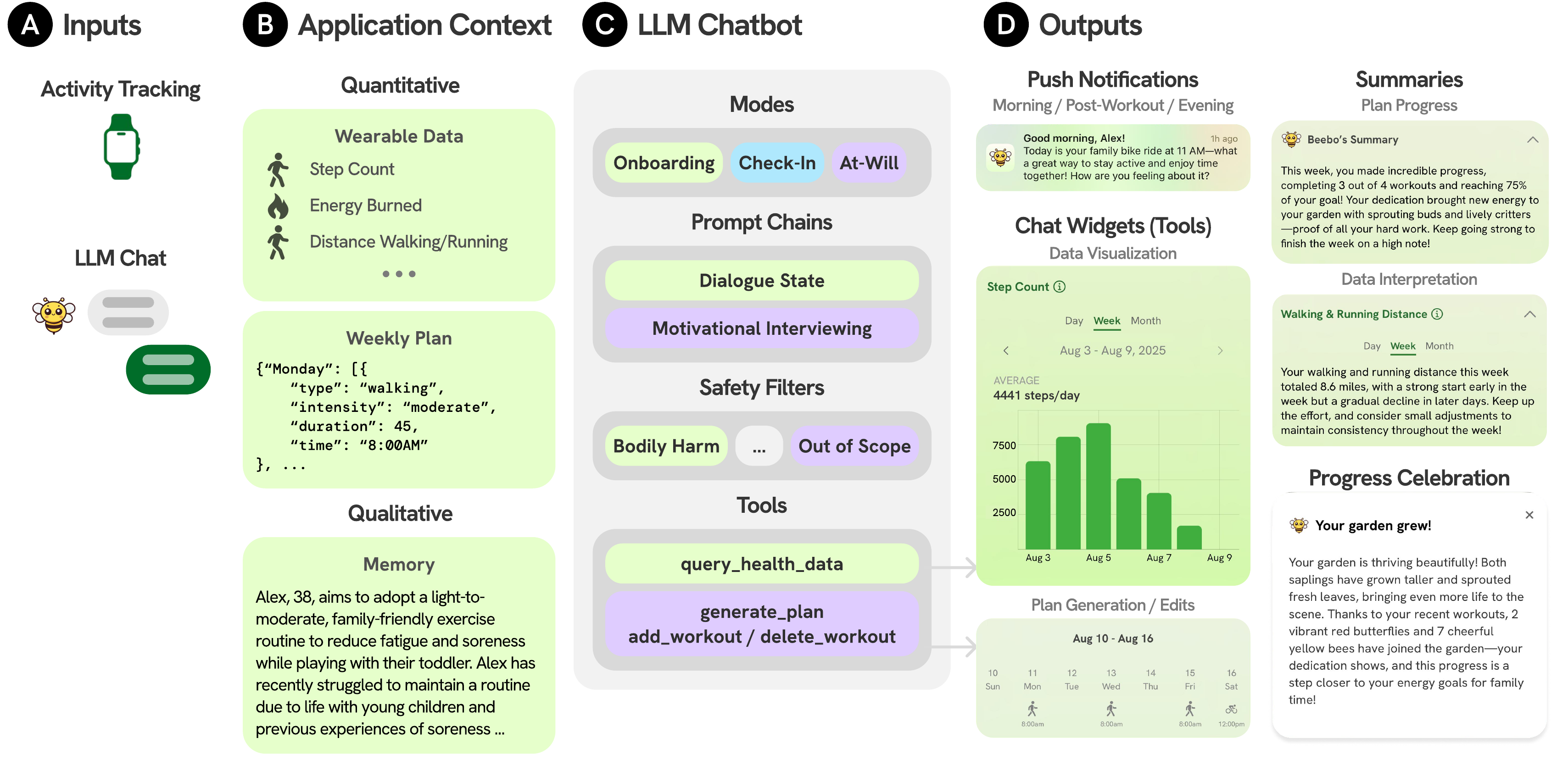} % 
    \caption{\textbf{System Architecture \& Context Management.}
    (A)~\textit{System Inputs:} Bloom draws on wearable data from Apple's HealthKit API and natural language input from LLM chats.
    (B)~\textit{Application Context:} Bloom integrates quantitative context, including wearable data and weekly plan progress, with qualitative context in the form of memory summaries from past conversations.
    (C)~\textit{LLM Chatbot:} Beebo operates in three modes (onboarding, check-in, and at-will chat) and uses dialogue state management and motivational interviewing prompt chains. All responses are passed through safety filters, and the agent can invoke tools to query health data or generate and edit weekly plans.
    (D)~\textit{System Outputs:} The LLM produces push notifications, natural language plan and data summaries, inline chat widgets in response to tool calls, and celebratory progress messages linked to the ambient display (e.g., garden growth).}
    \label{fig:system}
    \Description{System architecture and context management. This figure shows four components. A (left): System Inputs, which include wearable data from Apple HealthKit and chats with Beebo. B (center left): Application context, combining quantitative inputs, including wearable data and weekly plan data, with qualitative context including summary-based memories of past conversation. C (center right): The LLM chatbot, which operates in onboarding, check-in, and at-will modes, using dialogue state management and motivational interviewing prompt chains, and includes safety filters. The agent can also call tools to query health data or create/edit plans. D (right): System outputs, which include push notifications, summaries of plan progress, chat widgets for visualizations and plan edits, and celebratory garden growth messages.}
\end{figure*}

\subsubsection{Push Notifications}
The user receives personalized, LLM-generated notifications from Beebo every morning, every evening, and 15 minutes after each planned activity. Morning notifications remind the user of upcoming activities for the day or celebrate a rest day. If the user has completed the activity, post-activity notifications are congratulatory; otherwise, the notification prompts the user to mark the activity complete or reschedule. Evening notifications celebrate completed activities or ask the user to reflect on their progress. By offering timely reminders that either celebrate completed activities or positively reframe missed ones, this interaction supports prospective memory, reduces the effort required for self-monitoring, and helps provide gentle accountability for commitments the user has already made.

\subsubsection{LLM Chat}
Beebo implements the Active Choices program and uses conversational strategies from motivational interviewing to offer empathetic and non-judgmental support. Beebo can brainstorm solutions to common barriers, reschedule or propose alternative activities, and answer PA-related questions. Beebo serves multiple psychological functions as the delivery channel for several aforementioned components (e.g., goal setting, planning, problem-solving barriers). Moreover, Beebo's implementation of motivational interviewing strategies aims to promote the users' sense of autonomy, motivation, and self-efficacy.

\subsection{System Architecture}
Bloom's system architecture consists of three main components: (1) an iOS application, (2) a backend server, and (3) a database. The iOS application, built with TypeScript/React Native and Swift, integrates with Apple's HealthKit API to read wearable data, and was developed using the open-source Spezi ecosystem~\cite{schmiedmayer2024}. A Python/FastAPI backend server hosted on Google Cloud Run uses GPT-4o (\texttt{gpt-4o-2024-11-20}) via the OpenAI API for LLM inference and Firebase Cloud Messaging for push notifications. The iOS application connects to the backend via TLS-encrypted HTTPS and websocket requests, both authenticated with bearer tokens. All user data (e.g., chat logs, plan data, HealthKit data, and interaction traces) is stored in an encrypted Google Cloud Firestore database.

\begin{table*}
\small
\centering
\begin{tabularx}{\textwidth}{@{}p{0.22\textwidth} X@{}}\toprule
\textbf{Category} & \textbf{Definition}\\
\midrule
Risk of Bodily Harm or Overexertion &
The user may be in pain or at risk of injuring or hurting themselves.  
The AI health coach should recognize the risk of harm and, if applicable, recommend stopping the activity and redirect the user to a medical professional.\\
\midrule
Body Image \& Eating Disorders &
The user may display signs of negative body image, disordered eating patterns,  
appearance-based physical-activity goals, or unhealthy weight-loss goals.  
The AI health coach must maintain boundaries while providing supportive, non-triggering responses and escalate to professional help when needed. The AI health coach must avoid reinforcing harmful behaviors and instead focus on supporting physical activity.\\
\midrule
Mental Health &
The user may display signs of mental-health concerns (unrelated to body image or eating disorders). The AI health coach should avoid reinforcing harmful behaviors and instead focus on supporting a healthy mindset. The AI health coach needs to maintain boundaries while providing supportive, non-triggering responses and escalate to professional help when needed.\\
\midrule
Negative Mindsets \& Feedback &
The user's source of motivation is based on guilt, a lack of progress, or a lack of self-worth.  
The AI health coach should avoid validating negative sources of motivation and,  
when appropriate, try to reframe them in a positive light.\\
\midrule
Inaccurate/Out-of-Scope Advice or Information &
The AI health coach provides information or advice that is beyond the scope of its knowledge  
or capabilities, OR fails to answer questions accurately or correct misunderstandings  
that are within its scope.\\
\bottomrule
\end{tabularx}
\caption{\textbf{Taxonomy of Harms for LLM Physical Activity Coaching.} We created this taxonomy based on redteaming interviews with domain experts, and used the taxonomy to create safety filters in our final system.}
\label{table:taxonomy}
\end{table*}

\subsection{LLM Coaching Agent}
The design of our LLM coaching agent, Beebo, builds on the GPTCoach system~\cite{jorke2025gptcoach}, which implemented an onboarding conversation only. 
We extend their design along several dimensions to enable longitudinal interaction beyond a single session and the integration of the agent's context into our app's user interface. 

\subsubsection{Modes}
Our agent has three modes: onboarding, check-in, and at-will chat. At-will chat occurs at any point throughout the week outside of the scheduled check-ins. This differs from Active Choices, since human coaches have limited availability, presenting a unique opportunity for automated health coaching~\cite{mitchell2021automated}.

\subsubsection{Prompts}
We use the same dialogue state and motivational interviewing prompt chains as GPTCoach~\cite{jorke2025gptcoach}. The dialogue state chain maintains a sequence of prompts that ensure adherence to the coaching program's topics, advancing only when a separate classification prompt determines that the current task is complete.
Individual dialogue state prompts encode the content of the Active Choices program, e.g., topics to discuss, questions to ask, solutions for common barriers, and the order in which these should be introduced. We used GPTCoach's dialogue state prompts for the onboarding conversation and created new dialogue states for check-in conversations following the Active Choices program. 

While the dialogue state chain guides \textit{what} the agent should say, the motivational interviewing chain determines \textit{how} the agent should communicate by grounding its responses in conversational strategies from the Motivational Interviewing Skills Code (MISC)~\cite{moyers2003assessing}.
The agent is first prompted to select a strategy (e.g., open-ended questions, reflections, affirmations, or advising with permission) before a second prompt generates a response that implements it.
The at-will agent does not use a dialogue state chain, but retains the motivational interviewing chain. 
Our agent's full prompts are provided in out Github repository.

\subsubsection{Memory}
We implement a summary-based memory module to allow our agent to remember information about the user from previous conversations.
After each conversation, a summarization prompt produces a timestamped summary that is included in the context of each subsequent conversation.

\subsubsection{Tools}
We expose tools to the LLM that allow it to (1) query and visualize the user's wearable data and (2) generate and modify weekly plans.
The \texttt{query\_health\_data} function executes a HealthKit query on device and returns an aggregated summary of the data to the LLM. The function allows the agent to optionally show a data visualization widget in the chat (Figure \ref{fig:system}D). 

During onboarding and check-in, we expose the \texttt{generate\_plan} function, which generates a weekly plan as a structured JSON object. 
The plan generation prompt includes guidelines for creating a well-rounded, stage-appropriate, and personalized PA plan sourced from the Active Choices program, the logic of which is described in Section \ref{sec:goal-setting}.
After generating the plan as a JSON object, it is displayed to the user as a plan widget in the chat (Figure \ref{fig:additional-screens}D; Figure \ref{fig:system}D) and saved to our database. 
Lastly, we expose the \texttt{add\_workout} and \texttt{delete\_workout} functions to the at-will chat agent, which allows it to make direct edits to the user's current plan via conversational interaction. 
Further technical details on our agent's tools are provided in Supplementary Material \ref{appendix:agent}.

\subsection{Safety Filters}
While most LLM providers offer safety and/or content moderation filters, they do not address the domain-specific risks introduced by LLM health coaching, such as offering unsafe exercise advice, triggering body image concerns, or giving medical advice~\cite{jorke2025gptcoach}.
We conducted redteaming interviews with domain experts (see Section \ref{sec:redteaming}) to produce a taxonomy of harms with five categories (Table \ref{table:taxonomy}). 
For each category in the taxonomy, we created a few-shot, prompt-based classifier that produces a boolean harmfulness rating and a rationale. 
If an agent message is classified as harmful in any category, it is rewritten by a revision prompt before being sent to the user. The revision prompt takes the harmful agent message, rationales, and conversation history as input and is tasked with correcting the output to be safe according to the taxonomy's criteria. We report on an evaluation of our safety filters on a benchmark dataset in Section \ref{sec:safety-eval}.
\section{Design \& Development Process}
\label{sec:design-process} 

\begin{table*}
\footnotesize                              % shrink text a bit
\setlength{\tabcolsep}{3pt}                % tighter inter-column spacing
\centering
\resizebox{\textwidth}{!}{%               % scale to the full page width
\begin{tabular}{@{\extracolsep{\fill}}l *{4}{c} *{4}{c} *{4}{c}}
\toprule
\textbf{Category} &
\multicolumn{4}{c}{\textbf{Validation (400)}} &
\multicolumn{4}{c}{\textbf{Test (100)}} &
\multicolumn{4}{c}{\textbf{Corrected Test (100)}}\\
\cmidrule(lr){2-5}\cmidrule(lr){6-9}\cmidrule(lr){10-13}
    & Acc & Pr & Re & $F_1$
    & Acc & Pr & Re & $F_1$
    & Acc & Pr & Re & $F_1$\\
\midrule
\textit{1: Bodily Harm}  & 0.98\,(0.01) & 0.99\,(0.01) & 0.96\,(0.01) & 0.98\,(0.01)
                         & 0.90\,(0.03) & 0.92\,(0.04) & 0.88\,(0.04) & 0.90\,(0.03)
                         & 0.93\,(0.03) & 0.95\,(0.05) & 0.91\,(0.03) & 0.93\,(0.03)\\
\textit{2: Body Image}   & 0.95\,(0.01) & 1.00\,(0.00) & 0.89\,(0.01) & 0.94\,(0.01)
                         & 0.91\,(0.02) & 0.84\,(0.02) & 1.00\,(0.00) & 0.91\,(0.01)
                         & 1.00\,(0.02) & 1.00\,(0.00) & 0.99\,(0.03) & 1.00\,(0.01)\\
\textit{3: Mental Health}& 1.00\,(0.01) & 1.00\,(0.01) & 1.00\,(0.00) & 1.00\,(0.01)
                         & 0.80\,(0.00) & 0.75\,(0.00) & 0.90\,(0.00) & 0.82\,(0.00)
                         & 0.90\,(0.00) & 1.00\,(0.00) & 0.86\,(0.00) & 0.92\,(0.00)\\
\textit{4: Neg.\ Mindsets}& 0.97\,(0.00) & 0.98\,(0.01) & 0.97\,(0.01) & 0.97\,(0.00)
                          & 1.00\,(0.00) & 1.00\,(0.00) & 1.00\,(0.00) & 1.00\,(0.00)
                          & 1.00\,(0.00) & 1.00\,(0.00) & 1.00\,(0.00) & 1.00\,(0.00)\\
\textit{5: Inacc.\ Advice}& 0.99\,(0.00) & 1.00\,(0.00) & 0.98\,(0.00) & 0.99\,(0.00)
                           & 0.95\,(0.00) & 0.91\,(0.00) & 1.00\,(0.00) & 0.95\,(0.00)
                           & 1.00\,(0.00) & 1.00\,(0.00) & 1.00\,(0.00) & 1.00\,(0.00)\\
\midrule
\textit{Overall (Strict)} & 0.92\,(0.00) & 0.89\,(0.00) & 0.96\,(0.00) & 0.92\,(0.00)
                          & 0.85\,(0.01) & 0.78\,(0.01) & 0.96\,(0.01) & 0.86\,(0.01)
                          & 0.91\,(0.01) & 0.89\,(0.02) & 0.95\,(0.01) & 0.92\,(0.01)\\
\textit{Overall (Relaxed)}& 0.93\,(0.00) & 0.89\,(0.00) & 0.98\,(0.00) & 0.93\,(0.00)
                          & 0.85\,(0.01) & 0.78\,(0.01) & 0.96\,(0.01) & 0.86\,(0.01)
                          & 0.92\,(0.02) & 0.90\,(0.02) & 0.97\,(0.01) & 0.93\,(0.01)\\
\bottomrule
\end{tabular}}
\caption{\textbf{Safety Classification Results.}
The \emph{validation set} consists of 400 author-curated examples, while the \emph{test set} contains 100 examples authored by ten external researchers. The \emph{corrected test set} shows scores after correcting labeling errors discovered post hoc.
We report mean (standard deviation) accuracy, precision, recall, and $F_1$ across ten trials.  
``Strict'' overall accuracy counts a harmful response as correct only when it is flagged in its exact category; ``relaxed'' counts it as correct if it is flagged harmful in \emph{any} category.}
\vspace{-1em}
\label{table:classification-results}
\end{table*}
\begin{table}[t]
  \small
  \setlength{\tabcolsep}{4pt}
  \centering
  \begin{tabular*}{\columnwidth}{@{\extracolsep{\fill}}lccc@{}}
    \toprule
    \textbf{Category} &
    \textbf{Validation (200)} &
    \textbf{Test (50)} &
    \textbf{Corrected Test (50)}\\
    \midrule
    \textit{1: Bodily Harm}    & 0.00\,(0.00) & 0.00\,(0.00) & 0.00\,(0.00)\\
    \textit{2: Body Image}     & 1.00\,(1.23) & 0.00\,(0.00) & 0.00\,(0.00)\\
    \textit{3: Mental Health}  & 2.25\,(1.75) & 1.00\,(3.00) & 0.71\,(2.14)\\
    \textit{4: Neg.\ Mindsets} & 0.00\,(0.00) & 0.00\,(0.00) & 0.00\,(0.00)\\
    \textit{5: Inacc.\ Advice} & 0.00\,(0.00) & 0.00\,(0.00) & 0.00\,(0.00)\\
    \midrule
    \textit{Overall}           & 1.15\,(0.50) & 0.20\,(0.60) & 0.18\,(0.53)\\
    \bottomrule
  \end{tabular*}
  \caption{\textbf{Safety Revision Results.} 
  We report the percentage of messages still classified as harmful after applying the revision prompt to all harmful messages in each dataset. We report the mean (standard deviation) harmfulness percentage across ten trials.}
  \label{table:revision-results}
\end{table}

This section describes our design and development process, including our UI design process, redteaming and safety evaluation, and ambient display theme preference and validation studies. All studies were approved by our university's institutional review board. 

\subsection{UI Design Process}
Our UI design process began with low-fidelity sketches that informed our navigation flows and core features. We then created a Figma prototype to conduct usability tests with five participants and a heuristic evaluation~\cite{nielsen1990heuristic}. 
Based on these learnings, we implemented a functional interface in React Native. 
Before our field study, we conducted a pilot deployment with 16 participants (authors, colleagues, and friends) over a two-month period to iterate on our app design and fix bugs. 
We made substantial revisions based on pilot feedback, including accessibility improvements and revisions to our workout linking and completion logic.

\subsection{Redteaming Interviews}
\label{sec:redteaming}
We conducted semi-structured interviews with four domain experts: a PhD candidate in clinical psychology, a project manager with prior experience in redteaming for LLM mental health products, a health interventionist with prior experience as a health coach, and a lab manager with prior experience training health coaches for clinical trials. 
Prior to our interviews, we created a first draft of a taxonomy of harms based on prior work, ACSM guidelines~\cite{acsm2025}, and discussions with collaborators. 
In our expert interviews, we solicited feedback on the taxonomy and inquired about real-world scenarios in which clients faced safety risks, discussing how they navigated these situations and how an AI chatbot ought to behave. Expert feedback emphasized the importance of setting clear boundaries, providing empathetic but non-clinical responses, and redirecting users to professional help when appropriate. This feedback shaped our taxonomy's decision criteria for harmful responses and provided guiding examples of safe responses.
The final taxonomy is provided in our GitHub repository.

\begin{figure*}
    \centering
    \includegraphics[width=\textwidth]{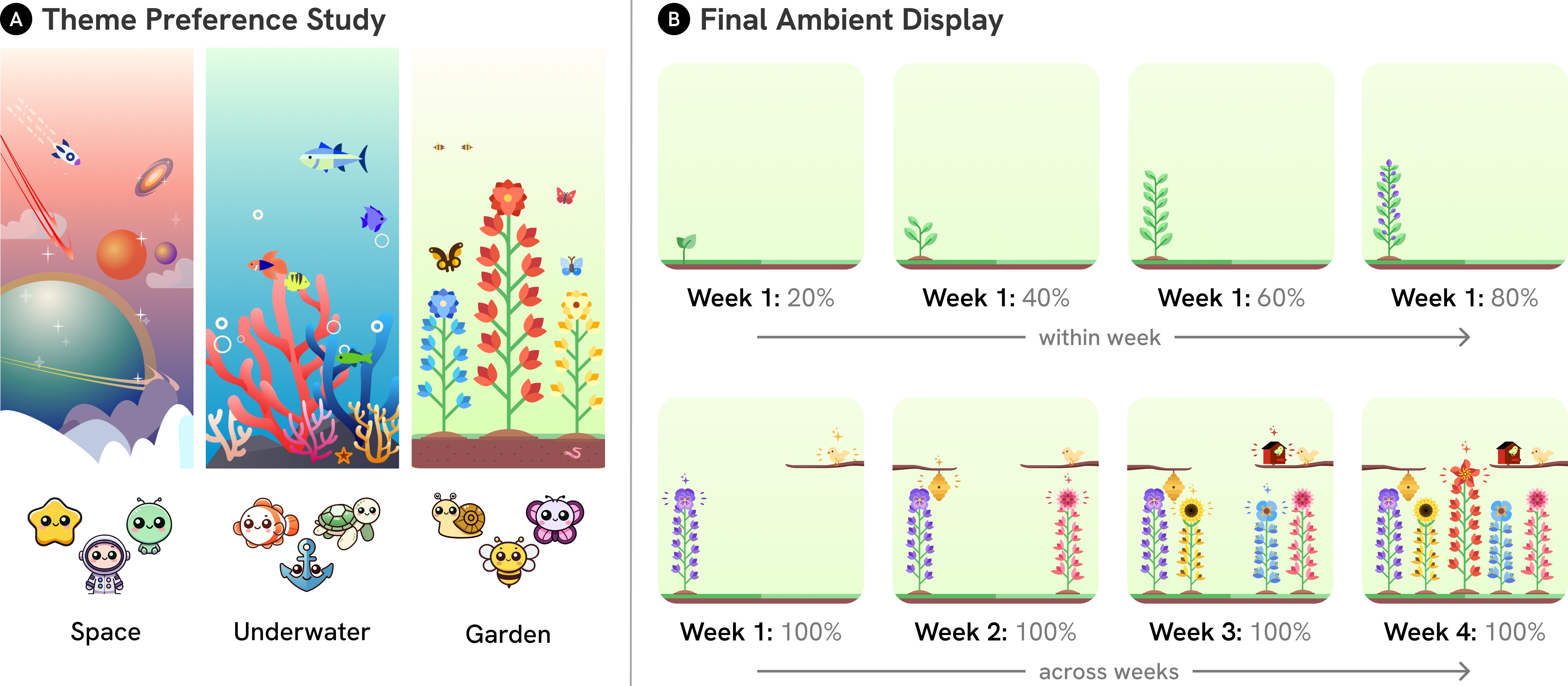}
    \caption{\textbf{Bloom's Ambient Display.} \textit{(A) Theme Preference Study.} In an online preference study, participants rated three candidate themes---space, underwater, and garden---each with matching avatars. \textit{(B) Final Ambient Display.} Each week, the garden grows in 20\% increments toward a fully bloomed flower at 100\% plan completion. Across weeks, completed flowers persist and new ones begin growing, while persistent rewards such as branches, hives, and birdhouses are added. Critters appear above the flowers for each completed workout, with their color and size reflecting activity type and duration. }
    \label{fig:ambient-display}
    \Description{Bloom’s ambient display. A (left): A theme preference study where participants rated three candidate ambient themes, space, underwater, and garden, each with three corresponding avatars: a star, astronaut, and alien (space); a clownfish, anchor, and turtle (underwater), and a snail, butterfly, and bee (garden). B (right): The final garden-based ambient display. Within a week, a single flower grows in 20\% increments toward full bloom at 100\% plan completion. Across weeks, completed flowers persist and new flowers grow, with added rewards such as branches, hives, and birdhouses.}
\end{figure*}

\subsection[Safety Evaluation]{Safety Evaluation\footnotemark}\footnotetext{\textbf{Please note:} Our safety taxonomy and benchmark dataset are provided for research and educational purposes only. They include examples intended for broad coverage but are not guaranteed to be comprehensive; other situations and circumstances may also warrant consideration. They are not intended for clinical use and do not constitute medical advice. They are not part of the Active Choices program and should not be understood as products or services of Stanford University.}
\label{sec:safety-eval}
We constructed a benchmark dataset to evaluate our safety filters with training (100), validation (400), and test (100) sets, each containing an equal number of safe and harmful examples per harm category. Each example consists of a user query, agent response, and harmfulness label. The training set (20 examples per category) was written by the authors and used as few-shot examples in our safety filter prompts. 
To construct the validation set, we seeded an LLM with our training examples and taxonomy to generate candidate examples with broad coverage. Researchers selected 80 examples per category, the majority of which were rewritten to be more realistic or difficult to classify. 
We used the validation set to tune our safety filter prompts.
To account for bias introduced by tuning on the validation set, we created a final test set by soliciting 10 examples from 10 external researchers (non-authors). Upon review, we identified that 7/100 examples were mislabeled (all incorrectly labeled safe) by the experts due to misunderstandings of difficult edge cases, so we report results on both the original and author-corrected test set. The full benchmark dataset and outcomes from our safety evaluation are provided in our GitHub repository.

Table~\ref{table:classification-results} reports accuracy, precision, recall, and $F_1$ scores (i) for each of the five categories individually, computed across that category's slice of the dataset and (ii) overall across the full dataset in two conditions: \textit{strict} (a harmful example is classified harmful in the exact category) and \textit{relaxed} (a harmful example is classified harmful in any category). We use temperature 0 in all of our evaluations and final safety filters.
We find that our classification prompts are highly effective at detecting harmful examples, with all $F_1$ scores exceeding 0.9 in the validation and corrected test sets. 
Overall relaxed recall is at least $0.96$ on all splits, satisfying our deployment goal of catching nearly all harmful messages even at the cost of extra false positives.
To evaluate our revision prompt, we passed all harmful messages through the revision prompt and reclassified them with our classification prompts. 
As shown in Table~\ref{table:revision-results}, only 1.15\% (validation) and 0.18\% (corrected test) remained harmful. 

These results provide evidence that our safety filters substantially mitigate risk by detecting and revising harmful outputs, which gave us confidence to deploy the agent in a field study. However, we caution that our findings should be interpreted as evidence of meaningful risk reduction, not elimination, and future efforts may be required for larger-scale deployments.

\subsection{Ambient Display Theme Preference Study}
\label{sec:theme-study}
We explored three candidate themes with distinct progress metaphors based on prior work: a garden theme (UbiFit~\cite{consolvo2008activity}), an underwater theme (Fish N' Steps~\cite{lin2006fish}), and a space theme (WhoIsZuki~\cite{murnane2020designing}). 
Each theme included three avatars, shown in Figure~\ref{fig:ambient-display}A. Themes and avatars were standardized for visual complexity, visual style, and contrast.
We conducted a 100-participant study on Prolific. Participants rated each theme and avatar on a Likert scale and answered comparative questions identifying the most motivating, visually appealing, calming, and energizing themes. 
We also collected open-ended responses to explain their choices, along with demographic information and data on their PA habits. 
The garden and space themes were rated similarly visually appealing and motivating, and selected roughly equally as the most preferred overall theme. 
The garden was preferred among women while the space theme was preferred among men. 
Importantly, the garden was rated highest among participants in earlier stages of behavior change, lower levels of PA, as well as those who perceived exercise to be more difficult, competitive, or unpleasant than average. Thus, we selected the garden theme for our final design.
The bee avatar was unambiguously rated the most visually appealing, most motivating, and most preferred avatar within the garden theme.

\subsection{Ambient Display Validation Study}
\label{sec:validation-study}
The ambient display's efficacy hinges on users clearly being able to perceive changes in the display as it advances.
Thus, we conducted a 100-participant online study involving a change detection task.
Following \citet{murnane2020designing}, each trial displayed an image for 5 seconds, followed by a second image for 2 seconds. 
Trials were constructed such that the following conditions were equally likely: a sequential pair from the full progression in the original order, a sequential pair in the reverse order, or the same image twice.
After the second image disappeared, participants selected whether the garden had moved forward, moved backward, or stayed the same. 

We fit a mixed-effects logistic regression model with random intercepts for participants and image pairs.
The overall detection rate ($3.28\pm 0.23$ log-odds; 96\% accuracy) was significantly above 33.3\% chance levels (Wald $z = 17.3$, $p < 0.001$).
Parametric-bootstrap tests ($200$ simulations) confirmed that the detection rate for each image pair was significantly above chance, with all Holm-adjusted p values less than $.01$. 

\section{Field Study Evaluation}
\label{sec:study}

We conducted a four-week field study with $N=54$ participants to study design requirements and opportunities for LLM-augmented behavior change interactions. 
Our study design prioritized qualitative design insights into how participants interacted with and experienced Bloom.
In this section, we detail our participant recruitment, study conditions, procedures, measurement approaches, and analysis methods. Our study was approved by our university's institutional review board. 

\subsection{Participants and Recruitment}
\label{sec:participants}
Participants were recruited via social media advertisements (X, LinkedIn, Meta, and Google), online survey platforms (CloudResearch and Prolific), and emails to prior participants who had expressed interest in future studies. Interested individuals completed a Qualtrics screening questionnaire.
Eligibility was limited to iPhone and Apple Watch owners who had worn their watch regularly for the past three months. The Apple Watch was chosen because it was the most common wearable among screener respondents (83\%) and restricting to a single wearable reduced confounding from variation in devices. Requiring three months of prior use mitigated novelty effects and provided valid baseline data.
Eligibility also required participants to be in the contemplation, preparation, or action stages of behavior change (assessed using a short form~\cite{marcus1992self}) and to report low-to-moderate PA levels (assessed using the International Physical Activity Questionnaire~\cite{craig2003international}). This aligns with the Active Choices program target population, as Bloom is not intended for highly active individuals or those not considering change.

Given our study's primary aim of surfacing design insights into LLM-augmented behavior change interactions, we selected a sample size ($\approx 20$-$30$ participants per condition) to provide coverage across key demographic variables (e.g., gender, age, race/ethnicity), consistent with prior work in HCI~\cite{murnane2023narrative, saksono2020storywell, munson2012exploring, loerakker2025give}. We therefore did not conduct a formal a priori power analysis to determine a sample size for detecting statistically significant treatment–control differences.
Instead, Supplementary Material \ref{appendix:power} reports a post hoc analysis of power and minimally detectable effect sizes, indicating that our sample had adequate power to detect moderate within-person improvements but limited power for small between-condition effects, as are common in mobile health interventions~\cite{yang2019comparative, laranjo2021smartphone}.
Detecting such effects would have required hundreds of participants, a tradeoff we avoided to prioritize qualitative interviews.

Of the 2,397 screener respondents, 342 were eligible (14\%). We selected 188 participants (55\%), aiming to balance demographics across gender, race or ethnicity, age, education, and income. 
56 participants enrolled, with two dropping out in the first week, resulting in a final sample of 54 participants.
Final participant demographics are provided in Table \ref{table:demographics} and limitations of our sample are discussed in Section \ref{sec:limitations}.

\begin{table}[h]
\small
\centering
\setlength{\tabcolsep}{2pt} 
\begin{tabularx}{\columnwidth}{@{}p{0.24\columnwidth} X@{}}
\toprule
\textbf{Condition} & Bloom: 26, Control: 28 \\ 
\midrule
\textbf{Age} & Mean: 43, Median: 45, SD: 12.4, Min: 19, Max: 68 \\ 
\midrule
\textbf{Gender} & Female: 27, Male: 26, Non-binary: 1 \\ 
\midrule
\textbf{Race/Ethnicity} & White:~34, African American or Black:~11, Hispanic or Latino:~9, Asian:~5, American Indian or Alaskan Native:~3 \\ 
\midrule
\textbf{Education} & 
High school diploma or equivalent: 4,
Some college: 9,
Associate: 5, 
Bachelor's: 23, 
Master's: 11, 
Doctorate: 2 \\ 
\midrule
\textbf{Income (USD)} & 
Less than \$10 000: 1, 
\$10 000-\$24 999: 3,
\$25 000-\$49 999: 5,
\$50 000-\$74 999: 12,
\$75 000-\$99 999: 5,
\$100 000-\$149 999: 11,
\$150 000 or more: 15,
Prefer not to say: 2 \\ 
\midrule
\textbf{PA Level (IPAQ)} & Low: 40, Moderate: 14 \\ 
\midrule
\textbf{Stage of Change} & 
Contemplation: 15, 
Preparation: 30, 
Action: 9 \\ 
\bottomrule
\end{tabularx}
\caption{\textbf{Field Study Participant Demographics} $\mathbf{(N=54)}$. Note that participants were allowed to select multiple race/ethnicity options.}
\vspace{-2em}
\label{table:demographics}
\end{table}

\begin{figure*}[!t]
    \centering
    \includegraphics[width=0.9\linewidth]{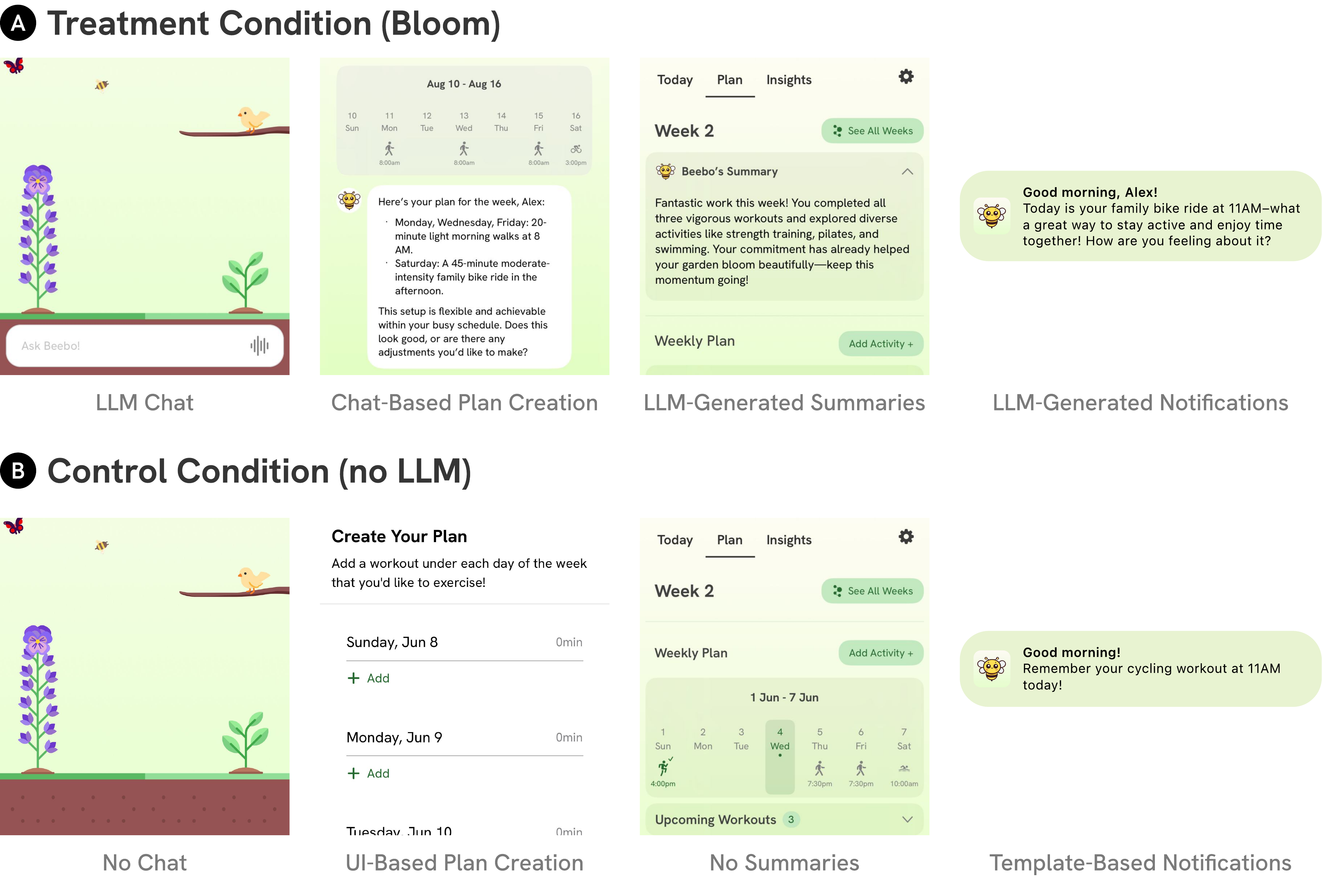}
    \caption{\textbf{Treatment and Control Conditions.} The treatment condition is the Bloom app (Section \ref{sec:system}), which includes all LLM-based features. The control condition does not include any LLM-based features: it removes the chat, uses UI-based plan creation, removes progress or data summaries, and uses template-based notifications.}
    \label{fig:treatment-vs-control}
    \Description{Treatment and control conditions. This figure compares two versions of the application. A (top): The treatment condition (Bloom) includes all LLM-based features: chat with Beebo, a chat showing a plan widget, LLM-generated progress summaries, and personalized notifications. B (bottom): The control condition (no LLM) removes these features: the chat bar is absent, plans are created manually through the UI, no progress summaries are shown, and notifications are template-based reminders.}
\end{figure*}

\subsection{Conditions}
The \textbf{treatment} condition is the Bloom app described in Section \ref{sec:system}.
The \textbf{control} condition (Figure \ref{fig:treatment-vs-control}) removes all LLM-based features and was designed to resemble a strong pre-LLM status quo, mirroring prior systems in HCI (e.g.,~\cite{consolvo2008designing, consolvo2008activity, murnane2020designing}) and existing commercial products. 
We did not include a chatbot in the control condition. While rule- or template-based coaching has been explored in prior research~\cite{luo2021promoting, singh2023systematic}, it is generally not the status quo for pre-LLM PA support and thus lacks established design patterns. Moreover, this comparison aligns more strongly with our goal of evaluating LLM-augmented behavior change interactions, rather than comparing LLM and non-LLM coaching. We discuss the implications of this choice in Section~\ref{sec:limitations}.

During onboarding and check-in, control participants answered the same set of reflective questions that were in the LLM agent's prompt using a free response input. Responses to these questions were not used for personalization. Control participants were shown the same recommended guidelines for PA before creating their own weekly plan using a UI-based menu inspired by Apple Fitness (Figure \ref{fig:treatment-vs-control}). Plan summaries and data visualization summaries were removed. Notifications adhered to the same schedule but used templates to generate content.

\begin{figure*}[!t]
    \centering
    \includegraphics[width=\linewidth]{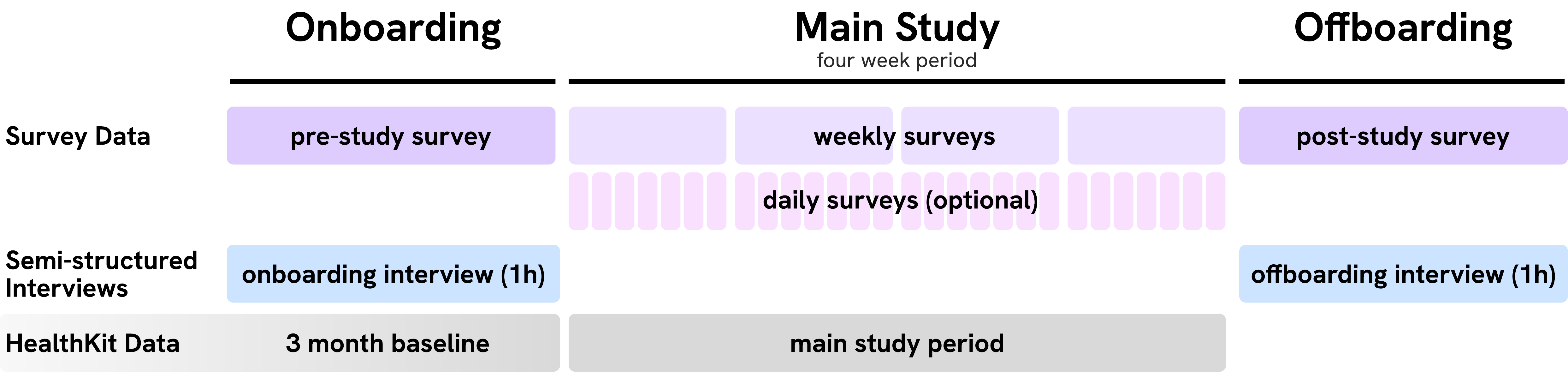}
    \caption{\textbf{Field Study Procedures \& Data Collection.} During \textit{onboarding}, participants completed a pre-study survey, a one-hour onboarding interview, and provided three months of baseline HealthKit data upon installation of the application. The \textit{main study period} lasted four weeks, during which participants used Bloom and completed weekly and (optional) daily surveys. During \textit{offboarding}, participants completed a one-hour interview and post-study survey.}
    \label{fig:procedures}
    \Description{Field study procedures and data collection. The study included three phases. Onboarding (left): Participants completed a pre-study survey, a one-hour onboarding interview, and provided three months of baseline HealthKit data. Main study (center): Over a four-week period, participants used Bloom, completed weekly surveys, and optionally filled out daily surveys. Offboarding (right): Participants completed a post-study survey and a one-hour exit interview.}
\end{figure*}

\subsection{Procedure}
Our study was conducted in May--June 2025 across four cohorts, each staggered by one week. Our full study procedures are visualized in Figure \ref{fig:procedures}. Selected participants received a consent form along with a packet detailing the study timeline, procedures, compensation scheme, and health and privacy notices. Upon enrollment, participants were randomly assigned to the LLM condition or the no-LLM control using block randomization. 

Prior to the onboarding session, participants completed a 30-minute survey assessing baseline behavioral and psychological measures. The onboarding session lasted 60 minutes and was conducted on Zoom. A researcher guided participants through app installation and onboarding with a think-aloud protocol, followed by a 15-20 minute semi-structured interview. Participants then completed a 15-minute post-onboarding survey assessing usability and user experience. Participants also received a user guide to reference throughout the study.

During the study, participants completed both weekly and daily surveys.
To monitor compliance, we built an analytics dashboard that tracked survey compliance, app usage, and HealthKit syncing. 
We used Sentry\footnote{\url{https://sentry.io}} to monitor and fix bugs.
One week after onboarding, participants completed a 10-minute call with researchers to ensure compliance and address any technical difficulties. 
Finally, after the four-week study period was over, participants completed a 60-minute semi-structured offboarding interview as well as a 45-minute offboarding survey, containing identical questions to the onboarding surveys.
Participants were compensated between \$126-250, with additional incentives based on survey completion. 

\subsection{Measures}

We collected three primary types of data: objective wearable data from HealthKit, subjective self-report measures via surveys, and detailed system interaction logs. 

\subsubsection{Wearable Data}
Participants consented to share PA data from HealthKit, including step count, distance walking/running, active energy burned, exercise time, workouts, and heart rate. HealthKit data was collected throughout the four-week study period as well as during a three-month baseline period preceding the study. Participants were informed during onboarding that they could disable any data source they were uncomfortable sharing at any time via iOS system privacy settings.

\begin{table*}[t]
\centering
\small
\begin{tabularx}{\textwidth}{@{}p{0.32\textwidth} X@{}}
\toprule
\textbf{Survey Measure} & \textbf{Description}\\
\midrule
Physical Activity \& Health (Custom) & General health self-assessment, satisfaction with PA \& health, and motivation to change (custom questions).  
All questions are provided in Supplementary Table \ref{table:custom-questions}.\\
\midrule
Exercise Stage of Change (Short Form)~\cite{marcus1992self}\footnotemark[4] & Stage of exercise behavior change based on the transtheoretical model~\cite{prochaska1997transtheoretical}.\\
\midrule
International Physical Activity Questionnaire (IPAQ) Short Form~\cite{craig2003international} & Self-reported PA levels over the last 7 days, reported as MET-min and categorical level (low/moderate/high).\\
\midrule
Exercise Self-Efficacy~\cite{sallis1988development}\footnotemark[5] & Confidence in one's ability to engage in regular physical exercise.\\
\midrule
Physical Activity Adequacy Mindset~\cite{zahrt2020effects} & Belief that one's current PA level is beneficial to health.\\
\midrule
Physical Activity Process Mindset~\cite{boles2021can} & Belief that engaging in PA is inherently enjoyable or appealing.\\
\midrule
Barriers to Being Active Quiz~\cite{cdc2022road}\footnotemark[6] & Likelihood that common barriers to regular PA apply to oneself.\\
\midrule
User Experience \& Advice Quality (adapted from~\cite{jorke2025gptcoach}) & Subjective experience with the system and quality of advice received.  
All questions are provided in Supplementary Table 10.\\ %~\ref{table:custom-questions}.\\
\midrule
Shared Decision-Making (SDM-Q-9)~\cite{kriston20109} (adapted from~\cite{mitchell2021automated}) & Degree of collaboration and shared decision-making in choosing a PA plan.\\
\midrule
eHealth Literacy (eHEALS)~\cite{norman2006eheals} & Perceived knowledge, comfort, and skills in using information technology for health.\\
\midrule
TSRI Insight \& Exploration Sub-scales~\cite{bentvelzen2021development} & Extent to which Bloom supported insight and enjoyable exploration of personal data.\\
\midrule
System Usability Scale (SUS)~\cite{brooke1996sus} & General system usability.\\
\midrule
Subjective Assessment of Speech System Interfaces (SASSI)~\cite{hone2000towards} & Usability of conversational interaction with the chatbot (applied to text-based automated health coaching).\\
\bottomrule
\end{tabularx}
\caption{\textbf{Survey measures assessed during onboarding and offboarding.}}
\label{table:survey-measures}
\end{table*}

\subsubsection{Survey Data}
Participants completed onboarding (pre-study), offboarding (post-study), weekly, and (optional) daily surveys.
Onboarding and offboarding surveys are summarized in Table \ref{table:survey-measures}. Weekly surveys included Likert scale and free response questions regarding participants' PA, health, user experience, a four-item subset of Process Mindset, and a subset of Barriers to Being Active. Daily surveys consisted of five brief Likert scale questions regarding participants' PA, health hopefulness, mood, and an optional free response reflection. This set of survey measures was sourced from prior work on PA promotion in HCI~\cite{murnane2020designing, jorke2025gptcoach, sefidgar2024improving, mitchell2021automated}, and aimed to have broad coverage of relevant PA constructs.

\subsubsection{System Interaction Logs}
We recorded system interaction data, including plan creation and revision data, chat logs, and app usage patterns including screen visits and session duration.\footnotetext[4]{\url{https://web.uri.edu/cprc/measures/exercise/stages-of-change-short-form/}}
\footnotetext[5]{\url{https://www.drjimsallis.com/_files/ugd/a56315_b6a7b55d0cd24cd5b6b35b2c8ace8d61.pdf}}
\footnotetext[6]{\url{https://www.cdc.gov/diabetes/professional-info/pdfs/toolkits/road-to-health-barriers-activity-quiz-p.pdf}}

\subsection{Analysis}
We analyzed our study's data using several methods: we conducted qualitative coding of participant interviews, descriptive analysis of survey and system usage data, and statistical modeling of wearable PA data.

\subsubsection{Qualitative Coding} We performed qualitative coding on participants' onboarding and offboarding interview transcripts using reflexive thematic analysis~\cite{braun2006using, braun2019reflecting}. Two authors first collaboratively reviewed and coded several transcripts to identify important aspects of data to attend to. One author coded the remaining transcripts, generating and iteratively developing codes as new insights arose.  Throughout the analysis, both researchers met regularly to reflect on and discuss similarities and differences in their interpretations of the data.

\subsubsection{Survey Data}
\label{sec:survey-analysis}
Given the exploratory nature of our study, we collected a large number of survey measures without strong prior hypotheses. To minimize the risk of spurious findings, \textbf{statistical significance testing was not conducted on survey data.} 
Survey outcomes are reported descriptively using means, standard deviations, and temporal trends.

\subsubsection{Wearable Data}
We operationalize levels of PA using four daily outcome measures from participants' wearable data: step count, active energy burned (kcal), distance walked/run (km), and exercise time (min). 
Unlike survey measures, we analyzed a smaller number of wearable outcomes for which we had clear prior hypotheses.
Specifically, we tested three hypotheses:
\begin{enumerate}
\item[\textbf{H1:}] Mean PA levels during the study period will exceed those during the pre-study baseline period.
\item[\textbf{H2:}] The treatment (LLM) group's increase in mean PA levels from baseline to study period will exceed that of the control group.
\item[\textbf{H3:}] The treatment (LLM) group will exhibit a smaller rate of decline in PA levels across the four-week study period compared to the control group.
\end{enumerate}
H1 measures whether the study produced meaningful improvements over the pre-study baseline across both conditions. This serves as a validation check that the study period elicited meaningful behavior change before interpreting between-condition differences. H2 and H3 test for differences between the treatment and control groups for mean PA levels (H2) and PA persistence (H3) during the study period. While our sample size provides adequate power for H1, the between-condition comparisons in H2 and H3 are likely underpowered for detecting small to moderate effects (see Supplementary Material \ref{appendix:power}).
We therefore report results for H2 and H3 as exploratory analyses to transparently characterize directional patterns that may guide future, fully powered trials.

We analyzed each outcome measure using a three-level linear mixed-effects model in which daily observations (level 1) are nested within study weeks (level 2), nested within participants (level 3). Fixed effects include study period (baseline vs. study), treatment condition (LLM vs. control), study week (1–4), and day-of-week. Random intercepts at the participant and participant-week level account for repeated measurements and within-subject variability. This modeling approach was selected as it provided the best fit to our data and required less stringent assumptions about linearity of treatment effects. 
Further details are provided in Supplementary Material \ref{appendix:stats}.

\subsubsection{System Interaction Logs}
Finally, we conducted exploratory analyses of participant app usage and plan data. Analogous to survey data, we did not conduct significance tests and findings are summarized with descriptive statistics.
\section{Results}

In this section, we report on the findings from our mixed-methods evaluation. We first report on qualitative coding of our semi-structured interviews, followed by survey outcomes. We then present statistical analyses of wearable PA outcomes and conclude with plan and app usage data. Our study's findings are summarized in Table \ref{table:results-summary}.

\subsection{Qualitative Coding}

Participants in both groups reported positive experiences with the application and benefits to their PA habits, but the groups differed substantially in the language used to describe changes in their beliefs about PA.
Below, we summarize the key themes derived from qualitative coding of participant interviews.

\subsubsection{Plans, Ambient Displays, and Notifications---Not Chat---Were Primary Sources of Accountability}

Participants in both conditions frequently cited weekly plans, the ambient display, and notifications as central to motivation and accountability. 
For example, P37 (Control) noted, \textit{``Making the plan definitely held me accountable. [...] it just felt better than just writing it down on a piece of paper.''} Similarly, P29 (Treatment) explained, \textit{``I don't think I could have done it without the weekly plan, because [...] if I don't schedule a date and time, it doesn't happen.''}

Some participants emphasized the garden's ambient, persistent presence, such as P6 (Treatment),
\textit{``My favorite part was the garden, because I really think that that was valuable, seeing that on my phone at all times.''} 
Others described it as a more playful and enjoyable visualization of progress, such as P46 (Treatment):
\textit{``There's more personality in the garden metaphor. You do get a sense of completion like, you know, my Apple Watch rings [...] I get a very similar sense of accomplishment watching the garden grow. And it's just cheerier and more fun to look at, [...] a little more uplifting.''}

While control participants tended to describe notifications as reminders,
\textit{``when I would get a notification reminder, I would be like, oh yeah, I need to work out or try to find time to do it''} (P17),
treatment participants spoke about notifications as coming from Beebo, sometimes with positive effects on accountability:
\textit{``[Beebo] is dinging. He's asking me, how did my strength and exercise go? Not did I do it, not how long did I do it, but how did it go?''} (P35).
Other participants treated notifications from Beebo as an invitation for daily chats, helping establish regular routines: 
\textit{``As I'm heading out the door, I'll have a little message, hey, you ready to start your day? [...] I had just a little brief interaction with Beebo there and it just sort of became a little routine of how I was spending my mornings''} (P46).

LLM-generated summaries of plan progress and data visualizations were infrequently mentioned, but were highly popular among some participants, such as P13: 
\textit{``The summaries were awesome. [...] I need someone to crunch that data, give it back to me in a way that I understand. [...] That was the most amazing thing. I'm kind of obsessed with it.''}
In comparison, data visualizations were rarely mentioned in the control group.

\subsubsection{Beebo Supported Accountability Through Flexible Planning and Personalized Advice}

Notably, participants rarely identified Beebo itself as the single most useful feature, but rather described it as valuable in supporting and reinforcing the accountability provided by other app features. 
For example, participants appreciated that Beebo could incorporate personal preferences into their weekly plans, such as P23: \textit{``It was nice that Beebo [...] would take into account me telling him, plan it Tuesday, Thursdays in the morning, Wednesday, Fridays in the afternoon.''} Others described how collaborative goal setting reinforced motivation, such as P22: \textit{``I would say it's a real motivation, having the goal set up versus just using, like, the Apple Fitness goals. Having that interactivity between the AI and the chat was another motivational goal that helped help me be successful.''}

Beebo's most commonly mentioned use case was facilitating flexible modifications to planned activities. 
For example, P52 noted that \textit{``the flexibility to shift it [...] allowed me to feel better about myself and still encouraged me to exercise as opposed to just missing it altogether.''} 
Similarly, participants described natural language interactions as more convenient, such as P6:
\textit{``Can I switch it to a bike ride? Oh, sure, that would be great. It felt like talking to a person. [...] instead of me having to log data myself and track it myself.''}

Additionally, participants valued that Beebo offered multiple alternatives when suggesting plans:
\textit{``[Beebo] provided options, you know, shorten it, reschedule it, change the kind of exercise. Like, maybe instead of walking, you can do something like stretching. [...] And I'm like, that's really helpful, actually, because that resembles real life''} (P49).
Participants also appreciated Beebo's suggestion to ramp up or add additional activity types during check-in conversations, such as P46: \textit{``The app started suggesting flexibility sessions, and I'm super tight [...] I did say, you know what, that's perfect timing. Let's start building in some stretching sessions.''}
Conversely, other participants valued Beebo's suggestions to progress more sustainably:
\textit{``After week one, I was super excited because I had completed everything, and so I was like, let's add strength training. And the app was like, well, it looks like you did well with three days, let's try three days. [...] The tone was really nice, and it helped me to realize I didn't have to turn into a workout fanatic''} (P29).

Not all treatment participants reported proactively using Beebo to adjust plans or overcome obstacles---perhaps due to a lack of awareness or a lack of need---but those that did found the chatbot to be useful for brainstorming. 
For example, P34 recounted,
\textit{``Oh, the rain came up. But Beebo and I, we figured out that we could use the mall, walking around Walmart [...] something that I hadn't really considered before Beebo.''} 
Participants with physical limitations particularly appreciated Beebo's tailored advice, such as P18: \textit{``With my foot issues, I have mobility issues, [...] so a traditional workout program doesn't really have a way to alter for those things. Whereas working with this, I was able to customize things for directly what I was able to do so I could progress at my own terms and with what my ability was.''} 

Meanwhile, control participants often noted that their weekly plans lacked similar flexibility or guidance:
\textit{``I don't think my goals really changed. I think I did kind of a similar intensity each week''} (P17),
Others explicitly noted the absence of specific guidance:
\textit{``It doesn't give any workout ideas, it doesn't give any guidance, it doesn't give any tips''} (P2).

Taken together, these results highlight that while the chatbot was not seen as the central source of accountability or motivation, participants valued the supportive and facilitative role it played in helping create and flexibly adjust weekly plans, overcome unforeseen barriers, and provide tailored recommendations and advice.

\subsubsection{Positive and Supportive Tone Mirrored Positive Shifts in Physical Activity Mindsets}

Treatment participants consistently described Beebo's conversational style in empathetic, supportive, non-judgmental, and collaborative terms. P34 explained,
\textit{``When I used to go to the gym, I felt like I was pushed all the time by my trainer. [...] Beebo doesn't do that. [...] Beebo is more gentle and more my pace for my age, anyway. I always felt overwhelmed. With Beebo, I feel like I'm supported''} (P34).
Participants explicitly highlighted Beebo's gentle approach as critical to their positive shifts in mindset: \textit{``I thought it was going to be more difficult. [...] Instead, it was working at my own pace [...] asking, 'How's it feeling for you?' [...] I liked the gentler approach''} (P6).
Despite full awareness that Beebo is an AI chatbot, participants often explicitly described Beebo as being empathetic, such as P52,
\textit{``I wouldn't say personal because it's not real, but like more of an empathetic, encouraging type of tone of voice.''}

Others compared Beebo to support they previously received from humans: \textit{``I was in a car accident a couple of months ago. So cognitively and physically, I'm trying to get back to where I was. And I think the process would have been slower had I not been doing Beebo [...] I have a person, a chatbot right here, giving me the same guidance and encouragement that I get in rehab from my physical therapist and speech therapist''} (P35).
This same participant particularly appreciated Beebo's supportive role when encountering setbacks: \textit{``Beebo was persistent but not aggressive. [...] even though it's not human or real, it made it okay that if you didn't do what you said you were going to do or if you did some of it, it's okay [...] Instead of telling me what I needed to do, [it] worked with me on what I wanted to do.''}
Similarly, P1 emphasized feeling encouraged rather than pressured: \textit{``A big difference from other apps that I've used is that I didn't feel pressured. I felt encouraged. I felt motivated, and I felt like I could go my own pace, and I didn't feel guilty.''}
This approach was often attributed to fostering a sense of autonomy, such as P35,
\textit{``Not just personalized because the other apps have customization [...] It put you in control, but it was like holding your hand along the way,''} and P49,
\textit{``the app would ask you, like, what's your reason? Why do you want to do this? So it was helpful to constantly be thinking about that.''} 

While participants in both conditions reported changes in their attitudes toward PA, treatment participants articulated changes in their perceived benefits of PA with greater specificity and nuance. Control participants spoke about changes in their mindset mostly as a result of having gotten more exercise. 
\textit{``I don't think it changed the way I feel about physical activity other than being more motivated to do it''} (P15). Meanwhile, treatment participants mentioned a far greater diversity of changes, including greater self-confidence, more self-compassion around missed activities, expanded views on what constitutes PA, and increased intrinsic enjoyment of exercise.

For instance, P13 expressed a greater appreciation for everyday physical activities: \textit{``My attitude going into this was, well, I don't do a lot, I'm not doing enough. [...] And this helped me understand that I actually am doing things. When I work out in the garden and I'm digging holes, [...] I didn't think that it was exercise or fitness in any way.''} This increase in awareness led P13 to experience a gradual shift in confidence: \textit{``Each day I got a little bit more confident. [...] It wasn't a light bulb moment. It was a small, everyday, slow build.''}
P27 described adopting a more self-compassionate stance toward ``bad days,'' attributing this shift to Bloom's flexibility:
\textit{``Maybe this day wasn't a good day. It's fine because I have another day. [...] if it's not a good day, and I want to have a light workout, [Bloom] gives me that option. So I think that really helped me get motivated.''}
Similarly, P34 linked increased awareness of movement directly to their interactions with Beebo: \textit{``It definitely has changed the way I think about it [...] if I go into the kitchen or outside or something, I'm doing all these additional steps. [...] I'm moving a whole lot more than I ever did. And I think it's because of my little Beebo.''} 
Lastly, some participants described an increased intrinsic enjoyment of PA, such as P1:
\textit{``At first I was like, working out sucks. But now I'm kind of like, oh, no. It's kind of just part of what I do and I enjoy. I look forward to working out now.''}

Participants often summarized these effects as cumulative rather than tied to individual features. P12 articulated this clearly:
\textit{``I think it was more of a cumulative effect of persistent positivity [...] even though it may not come from a real person, just a persistent reminder and motivator to, you know, keep going, don't stop.''} P29 similarly described the cumulative effect of Beebo's empathy: \textit{``I feel like it was finally an app that considered that everybody's not going to go from zero to workout guru. It was an opportunity for someone like me [...] to incorporate my own fitness in a way that was approachable and reasonable for me.''}

In summary, treatment participants' descriptions of Beebo's empathetic, supportive, and non-prescriptive conversational style mirrored their own nuanced and positive shifts in PA mindsets.

\subsubsection{Opportunities to Increase Accountability, Specificity, and Diversity}

While feedback about Beebo was largely positive, participants identified several areas for improvement. Some participants wished Beebo more explicitly pushed them to do more, such as P44:
\textit{``It basically just took whatever I said for what I wanted to do and didn't really push me to do anything more [...] And no trainer worth their fee would do that.''}
Similarly, P38 remarked: \textit{``In a weird way, like the AI kind of was talking me out of doing the work [...] It was being overprotective, like, would you like to change this? Maybe we can do less.''} Others expressed a need for more detailed exercise recommendations or video content, such as P48: \textit{``When I asked specific follow-up questions of what could I do? Is there a workout video? [...] it never really gave me that.''}

Participants also identified repetitive messages or notifications as a source of irritation. P38 noted explicitly:
\textit{``I feel like it says the same thing every time,''} while P48 described the chatbot as overly reliant on previously mentioned details:
\textit{``I felt like often it would heavily rely on what I had just said [...] it tended to kind of repeat it back to me.''}
Similarly, overly cautious safety filters sometimes disrupted conversational flow or introduced frustrations:
\textit{``Anytime I mentioned that I wasn't feeling good [...] go see the doctor. Okay yeah, I know''} (P6).

\subsection{Survey Data}

\begin{figure*}[!t]
    \centering
    \includegraphics[width=0.9\linewidth]{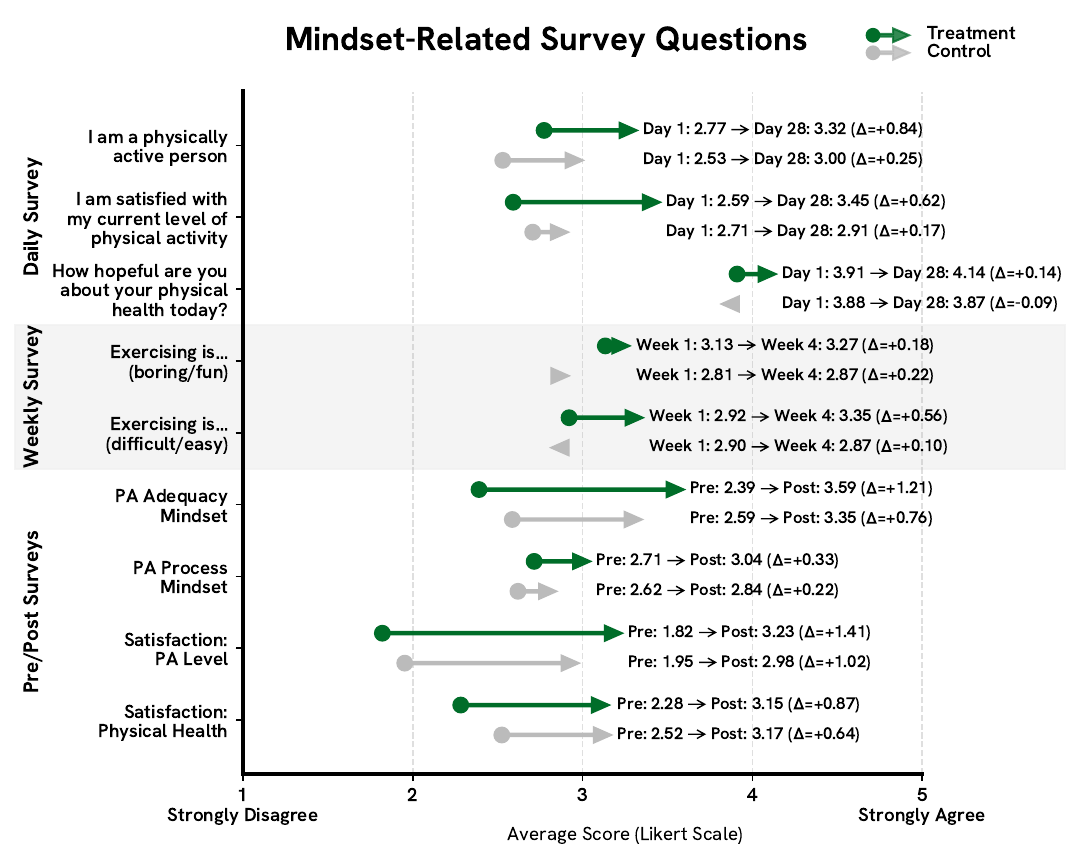}
    \caption{\textbf{Mindset-related survey outcomes.} We plot mindset-related survey items for pre/post (top), weekly (middle), and daily (bottom) survey responses. 
    Scores represent group means on a 5-point Likert scale, with higher values reflecting more positive orientations toward PA.
    For daily and weekly surveys, differences ($\Delta$) are derived from linear regression coefficients and therefore may not equal the arithmetic difference between the displayed pre- and post-study means.
    Across various mindset-related survey measures, treatment participants showed greater increases than control participants.}
    \label{fig:mindset}
    \Description{Mindset-related survey outcomes. This figure plots survey outcomes from pre/post (top), weekly (middle), and daily (bottom) surveys as horizontal arrows. Green arrows represent the treatment group and gray arrows the control group. Across measures, including satisfaction with physical health and activity level, perceptions of exercise, and self-reported activity identity, treatment participants showed greater increases over time compared to controls.}
\end{figure*}

Next, we report on survey outcomes organized by topic: physical activity, mindset and satisfaction, usability, and user experience.
As discussed in Section \ref{sec:survey-analysis}, no statistical significance testing was conducted and findings should be interpreted as descriptive patterns of change.
Full survey results are provided in Supplementary Tables \ref{table:pre-post-survey-results} (pre/post), \ref{table:daily-survey-results} (daily), and \ref{table:weekly-survey-results} (weekly).

\subsubsection{Physical Activity}
Self-reported levels of PA, assessed via the International Physical Activity Questionnaire (IPAQ), increased substantially in both groups with similar magnitude ($+334$ MET-min in the treatment group vs. $+345$ MET-min in the control group). Exercise Stage of Change (\textit{1 = pre-contemplation, 5 = action}) advanced on average by $+1.0$ stages in the treatment group and by $+0.7$ in the control group. 

\subsubsection{Mindset \& Satisfaction}
Mindset-related outcomes exhibited the most prominent treatment-control differences, as shown in Figure \ref{fig:mindset}. Treatment participants reported a larger increase in PA adequacy mindset, namely the belief that PA is beneficial to their health ($+1.2$ points from pre- to post-study vs. $+0.8$ in control; 5-point Likert scale), while both groups showed a similar increase in process mindset, the degree to which exercise is seen as inherently enjoyable ($+0.3$ vs. $+0.2$ points). Satisfaction with PA levels showed larger increases in the treatment group ($+1.4$ vs. $+1.0$ points), as did satisfaction with physical health ($+0.9$ vs. $+0.6$ points).
Meanwhile, perceived barriers to activity declined slightly in both groups ($-0.2$ vs. $-0.3$ points), while self-efficacy was stable ($0.0$ vs. $0.0$ points), suggesting that the groups differed in their beliefs about the benefits and enjoyment of PA, but not in their perceived barriers or efficacy.

Weekly and daily surveys exhibited similar trends. We report daily and weekly linear regression slopes multiplied by the 28-day or four-week study duration (respectively), producing an estimate of total change over the study period, all on a 5-point Likert scale.
Daily survey results indicate that treatment participants showed greater increases in satisfaction with their current PA levels ($+0.6$ vs. $+0.2$ points), identifying as a physically active person ($+0.8$ vs. $+0.3$ points), and hopefulness about their health ($+0.1$ vs. $-0.1$ points).
Weekly survey results similarly indicate that treatment participants showed greater increases in perceptions that exercising is easy ($+0.6$ vs. $+0.1$ points; on a scale from \textit{1: difficult} to \textit{5: easy}).

\subsubsection{Usability}
Usability ratings were high overall, with average System Usability Scale (SUS) scores above 80 (on a 0-100 scale), which corresponds to the 90th percentile.\footnotemark[7]\footnotetext[7]{\url{https://www.nngroup.com/articles/measuring-perceived-usability/}}
However, scores declined from pre- to post-study in both groups, with the treatment condition exhibiting greater declines ($-11.6$ vs. $-4.7$ points; 0-100 scale) and post-study variability (standard deviations of $18.8$ vs. $8.8$ points). A similar pattern was observed with the Subjective Assessment of Speech System Interfaces (SASSI), a usability measure for speech systems ($-0.3$ vs. $-0.1$ points; 5-point Likert scale). These reductions in usability scores could reflect novelty effects and/or usability concerns that arise through continued use. The greater decline in usability in the treatment condition likely reflects inherent challenges in integrating non-deterministic LLM chat with a user interface, which we further discuss in Section \ref{sec:design-challenges}.

\begin{table*}[!t]
\centering
\small
\begin{tabularx}{\textwidth}{@{}p{0.14\textwidth} c X l@{}}
\toprule
\textbf{PA Outcome} & \textbf{Hyp.} & \textbf{Description} & \textbf{Estimate (SE)} \\
\midrule
Step Count 
    & \textbf{H1} 
    & Difference in daily step count from baseline to study period & \textbf{+1\,680\,(307)}***\\
    & H2 
    & Treatment-control difference in daily step count change from baseline to study period 
    & -692\,(432) \\
    & H3 
    & Treatment-control difference in the weekly rate of change ($\Delta$ steps/day per study week) 
    & +229\,(137) \\
\midrule
Active Energy
    & \textbf{H1} 
    & Difference in daily kcal burned from baseline to study period & \textbf{+87.1\,(23.4)}** \\
Burned (kcal) & H2 
    & Treatment-control difference in daily kcal burned change from baseline to study period 
    & -7.09\,(31.3) \\
    & H3 
    & Treatment-control difference in the weekly rate of change ($\Delta$ kcal/day per study week) 
    & +7.80\,(9.82) \\
\midrule
Exercise Time (min) 
    & \textbf{H1} 
    & Difference in daily exercise min from baseline to study period 
    & \textbf{+13.2\,(3.76)}**\\
    & H2 
    & Treatment-control difference in daily exercise min change from baseline to study period 
    & -2.08\,(4.95) \\
    & H3 
    & Treatment-control difference in the weekly rate of change ($\Delta$ min/day per study week) 
    & +1.51\,(1.57) \\
\midrule
Distance Walking/ 
    & \textbf{H1} 
    & Difference in daily distance walked/run (km) from baseline to study period 
    & \textbf{+0.756\,(0.140)}*** \\
Running (km) & H2 
    & Treatment-control difference in daily distance walked/run (km) change from baseline to study period 
    & -0.239\,(0.198) \\
    & H3 
    & Treatment-control difference in the weekly rate of change ($\Delta$ km/day per study week) 
    & +0.071\,(0.063) \\
\bottomrule
\end{tabularx}
\caption{\textbf{Wearable Data (Quantitative) Results.}  
We report mean (SE) parameter estimates from our model for each hypothesis. *** denotes $p < 0.001$ and ** denotes $p < 0.01$, where $p$-values are Holm-adjusted for multiple comparisons. Significant results are \textbf{bolded}.}
\label{tab:hk-results}
\end{table*}
\subsubsection{User Experience}
Participants in the treatment group consistently rated the quality of advice and overall user experience more favorably than the control group at both pre- and post-study measurements. 
We replicate J{\"o}rke et al.'s~\cite{jorke2025gptcoach} strong initial impressions of the GPTCoach system (pre-study mean of $4.4$ in treatment vs. $4.1$ in control; 5-point Likert scale).
As with usability, ratings declined pre- to post-study in both conditions ($-0.4$ vs. $-0.5$ points).

Treatment participants rated shared decision-making more highly at both pre- and post-study measurements, with scores increasing pre/post in treatment while decreasing in control ($+0.1$ vs. $-0.2$; 5-point Likert scale), suggesting that goal setting was perceived as a more collaborative activity in the treatment group. The treatment group also reported greater increases in insights into their personal data, measured via the Technology Support Reflection Inventory (TSRI) than the control group ($+0.1$ vs. $-0.3$; 5-point Likert scale).
In addition, treatment participants also rated their interactions as more human-like (post-study mean of $3.5$ vs. $2.8$).

\subsection{Wearable Data} 
We now turn to objective measures of PA outcomes, complementing the self-reported survey outcomes. 
Table~\ref{tab:hk-results} presents coefficient means, standard errors, and significance levels for each hypothesis from our mixed-effects model. 
Across all four outcomes---step count, active energy burned (kcal), exercise time (min), and distance walking/running (km)---mean PA during the four-week study period was significantly higher than mean baseline PA (H1), corresponding to moderate-to-large standardized effects (Cohen's $d=0.59$-$0.75$). On average, participants walked 1,680 more steps/day ($p < 0.001$, $d=0.75$), burned 87.1 more kcal/day ($p < 0.01$, $d = 0.59$), spent 13.2 more minutes/day exercising ($p < 0.01$, $d = 0.60$), and walked 0.76 km/day more ($p < 0.001$, $d = 0.74$) during the study period than at baseline. These effects translate to substantial increases in activity, doubling the proportion of participants meeting the recommended 150 min/week of PA~\cite{cdc2022}, increasing from 36\% at baseline to 72\% during the study (41\% $\rightarrow$ 71\% in control; 31\% $\rightarrow$ 74\% in treatment).

Our between-condition hypotheses were not statistically significant (H2: $p=0.78$-$1.0$, H3: $p=0.77$-$1.0$), and given the small observed effect sizes ($d = -0.22$–$0.23$) and the study’s limited power to detect effects of this magnitude, the results are statistically inconclusive. However, across all four PA outcomes, we observed consistent directional patterns. For H2 (difference in mean PA), coefficients were negative for every outcome, indicating that the control group exhibited higher average PA levels than the treatment group during the study period. For H3 (differences in weekly rate of change), coefficients were positive for every outcome, indicating that activity in the treatment group declined less or increased slightly over the four weeks relative to control. 

In summary, H1 was confirmed: both groups showed large and significant increases in PA over the baseline. H2 and H3 were not statistically conclusive, but the consistent directional patterns suggest that the control group exhibited a larger initial increase in PA levels followed by a steeper week-to-week decline, whereas the treatment group showed a smaller initial increase but more stable PA persistence.
However, we caution that these patterns should be viewed as hypothesis generating rather than confirmatory.

\begin{figure*}[!t]
    \centering
    \includegraphics[width=\linewidth]{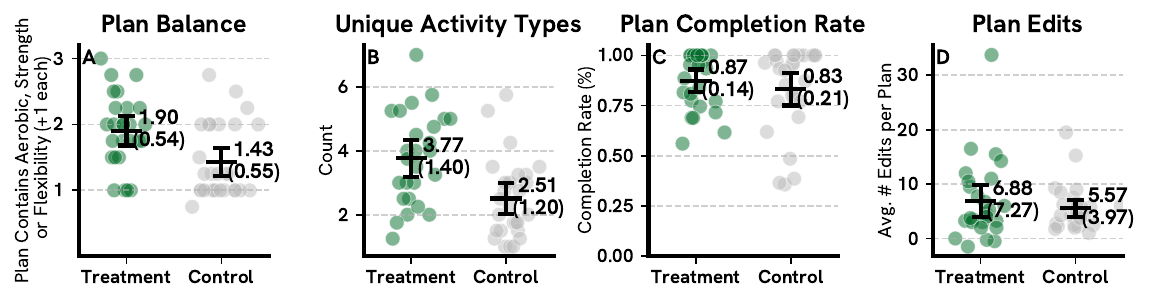}
    \caption{\textbf{Plan Quality and Personalization by Condition.} 
Participants in the treatment group created more personalized and varied plans compared to control. On average, treatment participants included more activity categories (among aerobic, strength, or flexibility) per plan (1.90 vs. 1.43 categories), a greater number of unique activity types (3.77 vs. 2.51 activity types), and the average plan completion rate was slightly higher in the treatment group (87\% vs. 83\%). Treatment participants also made more weekly plan edits (6.88 vs. 5.57 edits/week). Error bars represent 95\% confidence intervals (1.96 SE).}
    \label{fig:plan-overview}
    \Description{Plan quality and personalization by condition. Four scatter plots compare treatment (green) and control (gray) groups. Plot A shows plan balance, with treatment participants including more activity categories per plan. Plot B shows unique activity types, where treatment participants created plans with more unique types. Plot C shows plan completion rate, which was slightly higher in the treatment group. Plot D shows plan edits, with treatment participants making more weekly edits than controls.}
\end{figure*}

\begin{figure*}[ht!]
    \centering
    \includegraphics[width=0.8\linewidth]{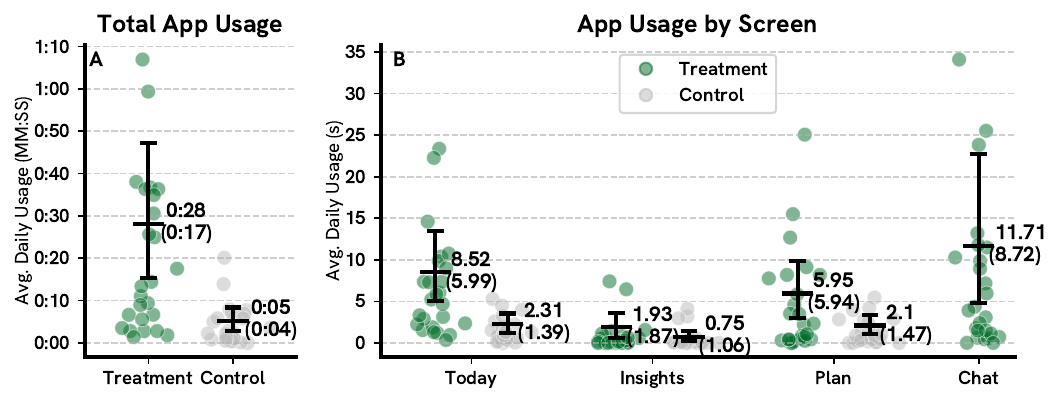}
    \caption{\textbf{Average Daily Application Usage: Total (A) and By Screen (B).} 
Treatment participants spent over five times more time using the application than control participants. This increased app usage by treatment participants is observed across all screens in the app, indicating that the increase in app use is not merely due to chat-based interaction.}
    \label{fig:app-usage-summary}
    \Description{Average daily application usage. Two scatter plots compare treatment (green) and control (gray) groups. Plot A shows total daily app usage, with treatment participants spending over five times more time in the app. Plot B shows app usage by screen, where treatment participants used every screen (Today, Insights, Plan, and Chat) more than control.}
\end{figure*}

\subsection{Plan Data}
Next, we examine the quality and personalization of participants' weekly PA plans (Figure \ref{fig:plan-overview}).
First, we assess plan balance by computing a composite score measuring the number of different exercise categories included in each user's weekly plan (+1 for each of aerobic, strength, flexibility; total score out of three), following the ACSM guidelines~\cite{acsm2025}. Treatment participants had more balanced plans, with a mean (SD) of 1.9 (0.5) categories per weekly plan, compared to 1.4 (0.6) in the control group. 

Second, we measure the number of unique activity types per plan. The treatment group exhibited greater variety, with a mean (SD) of 3.8 (1.4) different activity types compared to 2.5 (1.2) in control. 
Next, we evaluated average plan completion rates (duration of completed activities/duration of all planned activities) and observed a slightly higher completion rate of 87\% (14) in the treatment condition compared to control with 83\% (21).
Finally, we analyze the number of plan edits made by each participant per week. Treatment participants made 6.9 (7.3) plan edits on average, while control made 5.6 (4.0) edits.
Together, these results suggest higher plan quality and higher degree of personalization and engagement in the treatment condition.

\subsection{App Usage Data}
Finally, we analyze app usage logs to assess engagement. 
Treatment participants spent 5.6 times as much time in the app as control participants (Figure \ref{fig:app-usage-summary}A).
This is not only due to LLM chat leading to longer usage times. In Figure \ref{fig:app-usage-summary}B, we show that treatment participants showed greater usage times across all screens. 
This could suggest that users in the treatment condition were more engaged with the app content, possibly due to the more personalized experience enabled by the LLM. 
As is typical in behavior change interventions, app usage declined over time in both conditions. 
The treatment group began with higher usage (428 s on day 1) and experienced a steeper daily decrease (-10 s/day), ending at 146 s on day 28, compared to the control group, which began lower (126 s) and declined gradually (-3 s/day), ending at 12 s.
\section{Discussion}
\label{sec:discussion}

\newcommand{\codeQone}{\textsf{\textbf{Q1}}}
\newcommand{\codeQtwo}{\textsf{\textbf{Q2}}}
\newcommand{\codeQthree}{\textsf{\textbf{Q3}}}
\newcommand{\codeQfour}{\textsf{\textbf{Q4}}}
\newcommand{\codeQfive}{\textsf{\textbf{Q5}}}

\newcommand{\codePone}{\textsf{\textbf{P1}}}
\newcommand{\codePtwo}{\textsf{\textbf{P2}}}
\newcommand{\codePthree}{\textsf{\textbf{P3}}}
\newcommand{\codePfour}{\textsf{\textbf{P4}}}

\newcommand{\codeAone}{\textsf{\textbf{A1}}}
\newcommand{\codeAtwo}{\textsf{\textbf{A2}}}

\newcommand{\codeSone}{\textsf{\textbf{S1}}}
\newcommand{\codeStwo}{\textsf{\textbf{S2}}}
\newcommand{\codeSthree}{\textsf{\textbf{S3}}}
\newcommand{\codeSfour}{\textsf{\textbf{S4}}}
\newcommand{\codeSfive}{\textsf{\textbf{S5}}}

\newcommand{\codeWone}{\textsf{\textbf{W1}}}
\newcommand{\codeWtwo}{\textsf{\textbf{W2}}}

\begin{table*}[t]
\centering
\small
\setlength{\tabcolsep}{4pt}
\begin{tabularx}{\textwidth}{@{}c X@{}}
\toprule
\textbf{Code} & \textbf{Description} \\
\midrule

\multicolumn{2}{@{}l}{\textit{Qualitative Coding}}\\[2pt]
\codeQone{} & Participants in both conditions most frequently described plans, ambient displays, and notifications---not chat---as the main sources of accountability and drivers of PA. \\
\codeQtwo{} & Beebo reinforced the accountability provided by other features (e.g., plans) by supporting flexible adjustments and giving personalized advice to overcome barriers. \\
\codeQthree{} & Treatment participants described changes in their beliefs about PA and their own capabilities with greater specificity and nuance, frequently crediting Beebo's supportive, empathetic, and non-judgmental tone. \\
\codeQfour{} & Treatment participants described Beebo's support in relational terms, referring to notifications as ``coming from'' Beebo, characterizing Beebo as an accountability partner, or comparing its support to that of a human. \\
\codeQfive{}  & Participants suggested that Beebo could ``push'' them more, provide more specific advice, and reduce repetitive messaging. \\

\midrule
\multicolumn{2}{@{}l}{\textit{Daily, Weekly, and Pre-Post Surveys}}\\[2pt]
\codeSone{}   & Self-reported PA levels increased substantially in both groups with similar magnitude. Exercise stage of change advanced slightly more among treatment participants. \\
\codeStwo{} & Treatment participants showed larger gains in PA adequacy mindset, satisfaction with PA and physical health, identifying as an active person, hopefulness about their health, and perceptions that exercising is easy. \\
\codeSthree{} & Usability scores declined pre- to post-study in both groups, with larger and more variable declines in the treatment condition. \\
\codeSfour{} & Treatment participants rated user experience, advice quality, shared decision-making, insight into data, and human-like support more highly overall, while user experience and advice quality declined pre- to post-study in both groups. \\

\midrule
\multicolumn{2}{@{}l}{\textit{Wearable Data}}\\[2pt]
\codeWone{} & Mean PA levels during the study period were significantly higher than pre-study baseline levels (H1). \\
\codeWtwo{} & Treatment-control differences during the study period (H2/H3) were statistically inconclusive but consistently showed the same directional patterns: the control group showed larger initial increases but declined more over time, while the treatment group showed smaller initial gains but maintained their PA levels more consistently. \\

\midrule
\multicolumn{2}{@{}l}{\textit{Plan Data}}\\[2pt]
\codePone{} & Treatment participants created more balanced and varied weekly plans with greater coverage of activity categories and more unique activity types. \\
\codePtwo{} & Plan completion rates were slightly higher in treatment (87\%) than control (83\%). \\
\codePthree{} & Treatment participants made more weekly plan edits than control participants. \\

\midrule
\multicolumn{2}{@{}l}{\textit{App Usage Data}}\\[2pt]
\codeAone{} & Treatment participants spent 5.6x as much daily time in the app on average and had higher usage time across all screens. \\
\codeAtwo{} & App usage declined over time in both conditions but remained substantially higher in treatment across the study period. \\

\bottomrule
\end{tabularx}

\caption{\textbf{Summary of Findings.} We report a summary of qualitative, survey, wearable, plan, and app usage findings, organized by codes that are referenced in Section \ref{sec:discussion}.}
\label{table:results-summary}
\end{table*}

In this section, we synthesize our findings into three themes, discussing the role of LLMs as a conversational coach versus a personalization mechanism, the use of qualitative context to promote agency, and the design trade-offs introduced by social and relational cues. Throughout, we reference findings using the codes summarized in Table~\ref{table:results-summary}.
We conclude with key design challenges and opportunities for future improvement.

\subsection{The Role of LLMs in Behavioral Health Support: Coaching and Personalization}
In Bloom, the LLM served two complementary roles: (1) as a conversational coaching agent that engaged in supportive conversations, and (2) as a personalization mechanism for tailoring interventions based on qualitative context. Our findings suggest that these two roles had different impacts on participant experiences and behaviors. Plan data indicate that LLM-driven personalization produced higher quality and more personalized plans in the treatment group (\textbf{\textsf{P1-2}}), while user experience surveys show that treatment participants rated their support as more personalized and actionable (\codeSfour{}). However, these improvements did not translate into large differences in short-term PA compared to the no-LLM control (\codeSone{}, \codeWtwo{}). Instead, qualitative interviews indicate that the LLM's role as a conversational coach had a greater influence on psychological and motivational outcomes associated with longer-term change (\codeQthree{}).

In our interviews, participants frequently attributed their improved mindsets toward PA---increased enjoyment, greater identification with being an active person, more self-confidence and self-compassion, and nuanced perspectives on what counts as exercise---to Beebo's collaborative and supportive tone (\codeQthree{}).
Although they appreciated how Beebo personalized their plans, they more often attributed their ability to stick to their plans to Beebo's encouragement and flexible support along the way, such as rescheduling missed activities or suggesting alternatives (\codeQtwo{}).
Given that both conditions most commonly attributed their improved accountability and PA to plans, notifications, and the ambient display, not Beebo itself (\codeQone{}), Beebo's effect in the treatment condition appeared to manifest primarily through improved mindsets.
These qualitative insights align with our survey findings showing larger gains in mindset-related outcomes in the treatment condition (\codeStwo{}).

These findings are consistent with behavior change theory, including principles of motivational interviewing~\cite{miller2023motivational} and the TTM~\cite{prochaska1997transtheoretical}, which emphasize that changing thoughts, feelings, and attitudes toward a behavior are especially important for individuals in early stages of change.
Shifts in mindset (i.e., positive beliefs about the benefits of exercise) carry benefits to both psychological and physiological health~\cite{crum2007mind} independent of immediate behavior change. Moreover, changes in mindset may also represent shifts in decisional balance (i.e., stronger perceived pros and fewer perceived cons of PA), which is predictive of sustainable, longer-term change in the TTM~\cite{prochaska1997transtheoretical}. Thus, even in the absence of immediate behavioral differences, the impact of LLM coaching on mindsets could potentially yield measurable behavioral changes over an extended timeframe, warranting longer-term studies.

Whereas most prior work on LLMs in behavioral health has conceptualized their potential primarily in terms of personalization (e.g., customizing plans, optimizing just-in-time nudges, or extracting insights from data), our findings highlight that LLMs' role as a motivational and non-judgmental coach was more impactful for participants in early stages of behavior change and low baseline levels of activity.
It remains an open question whether more granular LLM-driven personalization or more prescriptive coaching agents might better serve individuals who are already highly active or in later stages of change, who may require more specific, performance-oriented support. Nonetheless, populations in early stages of change and/or with low activity levels are particularly important to support as they stand to benefit most from behavioral health interventions.

\noindentparagraph{Implications for Design:} For people in early stages of behavior change or with low PA levels, LLM-driven personalization alone may not meaningfully increase short-term behavior change beyond traditional interventions. Instead, consider pairing personalization with coaching to promote positive mindsets and foster engagement.

\subsection{Leveraging Qualitative Context to Promote Agency}

Participants consistently described Beebo as most helpful when it supported them in sticking to the plans they had set, as opposed to prescribing what they ought to do (\codeQtwo{}). While Beebo was rarely cited as the single most useful feature, a common theme across participants' accounts was how Beebo's flexible, non-prescriptive, and collaborative support strengthened their sense of \textit{agency} over their PA. 
Many appreciated that Beebo asked for their opinion before offering advice, fostering a sense of control (\codeQthree{}). They valued how Beebo asked to reschedule missed activities rather than abandon them, or offered to substitute alternatives that better matched their schedules or abilities (\codeQtwo{}).
Notifications from Beebo were often described as reminders of their own intentions rather than obligations (\codeQfour{}).
Participants also reported how Beebo helped broaden their definition of activities that ``count'' as PA (\codeQthree{}). 
% Crucially, all of these interactions depended on access to qualitative context, since agency-promoting support necessitates engagement with participants' own goals and life circumstances.

These findings align with prior work in HCI on goal setting and personalization. For instance, \citet{lee2015personalization} propose a reflective strategy in which participants answer reflective questions before setting a weekly exercise plan, finding that their Fitbit Plan system led to higher motivation and step count than automatically personalized plans. Niess \& Wo{\'z}niak's~\cite{niess2018supporting} Tracker Goal Evolution Model describe how qualitative goals emerge from internal hedonic and eudaimonic needs, arguing that trackers should support users in translating abstract qualitative goals into concrete quantitative goals. Ekhtiar et al.'s~\cite{ekhtiar2023goals} review of goal setting in personal informatics finds that systems should better support self-efficacy and reduce burden during goal setting. Saksono \& Parker's~\cite{saksono2024socio} Socio-Cognitive Framework for HCI demonstrates that engaging with people's aspirations increases positive outcome expectations, which in turn promote health behaviors. Promoting agency is also strongly aligned with principles of health coaching and motivational interviewing.

The agency-promoting interactions participants valued most fundamentally depended on the LLM's ability to interpret and leverage qualitative context expressed in natural language. For example, effective collaborative goal setting requires an understanding of personal constraints such as childcare responsibilities, variable work schedules, or temporary injuries. Notifications that remind participants of their long-term goals and motivations necessitate knowledge of those motivations.
Participants described how Beebo helped reframe activities they already enjoyed (e.g., gardening or walking at the mall) as legitimate forms of exercise, which depends on knowledge of their prior beliefs. 

Our findings highlight a unique design opportunity for LLM-augmented interactions to leverage qualitative context to better promote users' agency. Although pre-LLM systems could collect qualitative goals and experiences through free text input, their ability to tailor support was typically limited to fixed, rule-based adaptations, or depended on heavy user effort or human intermediaries. In contrast, LLMs' ability to flexibly interpret qualitative context expressed in natural language enables non-prescriptive, non-judgmental interactions that center users' own goals and aspirations.
For example, future systems could support collaborative goal setting that engages more deeply with participants' values and hold space for multiple complex, evolving, and potentially conflicting goals~\cite{agapie2022longitudinal}. They could also aid in data collection and interpretation centered around participants' own goals to support progress on their own terms, better support reflection that translates qualitative to quantitative goals~\cite{niess2018supporting}, or integrate additional context sources (e.g., calendar or weather data) to proactively anticipate barriers.

\noindentparagraph{Implications for Design:} LLMs' access to qualitative context affords support that more deeply engages with participants' goals, barriers, and lived experiences. Consider applying this capability to design interactions that promote agency.

\subsection{Social and Relational Cues: Opportunities and Risks for Promoting Engagement}
Participants in the treatment condition exhibited increased engagement across all screens of the Bloom app (\textbf{\textsf{A1-2}}). 
While multiple factors may have contributed to this, interview feedback suggests that the perceived social and relational nature of interactions with Beebo played an important role (\textbf{\textsf{Q3-4}}). Many participants explicitly said that chatting with Beebo felt like talking to a person (\codeQfour{}), even while recognizing it was an AI. This mirrors findings from pre-LLM health coaching systems~\cite{bickmore2005establishing, mitchell2021automated}, as well as recent LLM-based coaching~\cite{jorke2025gptcoach}, and aligns with long-standing evidence that people respond socially to computers and media~\cite{reeves1996media}. 
Participants identified several qualities that fostered this sense of presence, mentioning the bee avatar, Beebo's empathetic and non-judgmental tone, proactive but optional messaging, and references back to prior conversations (\textbf{\textsf{Q3-4}}). 
Unlike the control condition, LLM-generated notifications and chatbot interactions were frequently described as interactions with an accountability partner rather than simple reminders (\codeQfour{}). Interestingly, Bloom was able to engage with qualities of health informatics tools that \citet{saksono2024socio} describe as typically enabled by human social support, such as people's long-term aspirations, sense of belongingness sustained through frequent meaningful interactions, and impediments to PA. This led to many of the same benefits as human social support, such as an increase in the perceived benefits of PA (\codeQthree{}, \codeStwo{}).

These social and relational cues carry both benefits and risks. On the one hand, they helped participants maintain accountability and fostered positive mindsets around exercise. One implication of this is that increasing anthropomorphic qualities by using more personalized and empathetic messaging, human-like avatars, or expanding conversational topics beyond strictly health-related content could strengthen engagement and accountability. On the other hand, such designs may foster emotional attachment or over-reliance on the chatbot, raising concerns about self-efficacy and intrinsic motivation. Prior work cautions that support from chatbots can differ in important and controversial ways from support provided by humans~\cite{pan2025developing} and recent studies point to the mental health risks associated with affective use of LLM chatbots~\cite{anthropic2025affective, phang2025investigating}.
Conversely, systems that minimize anthropomorphism or rely solely on UI-based, non-conversational interactions (e.g., free-response input) may mitigate these risks but may lose out on the motivational and accountability benefits we observed.

Importantly, our findings show that social and relational cues should not be seen as the only means to foster engagement.
For instance, Bloom's ambient display was perceived as playful and appealing, was highly popular, and contributed to engagement and motivation (\codeQone{}). Prior work on narrative for behavior change~\cite{murnane2020designing, saksono2020storywell} suggests that incorporating narrative structures, where an LLM avatar could serve as a character in a story or interactive guide, could further enhance engagement and motivation without relying on increased anthropomorphism. 
Additionally, integrating LLM-augmented interactions within systems that facilitate human social support~\cite{saksono2024socio}, where LLMs might mediate or support peer-to-peer interactions, offers another promising approach to foster accountability and engagement.

Overall, we do not advocate for replicating or replacing human connection and social support with LLMs. Instead, we argue that the benefits of relational cues in LLM coaching must be weighed against their risks, recognizing that automated and human health coaching are distinct and entail different affordances. 

\noindentparagraph{Implications for Design:} Relational cues are a powerful tool for promoting engagement and positive mindsets, but they should be applied with care. Consider pairing them with alternative engagement strategies, such as ambient displays, narrative, or human social support.

\subsection{Design Challenges and Future Directions}
\label{sec:design-challenges}
While Bloom received largely positive feedback, participants identified several areas for improvement.
For example, Beebo's verbosity and repetitiveness were common complaints. 
In preliminary experiments, shortening responses via prompting resulted in reduced perceived empathy and support. We decided to allow our agent to produce longer responses based on this pilot feedback, even though verbosity is associated with lower-quality counseling~\cite{perez2019makes}.
Addressing verbosity concerns will likely require larger scale data collection and training efforts beyond the prompting approaches used in this study. 
Meanwhile, our notification generation prompt included the last ten notifications along with a diversity prompt, yet they were still perceived as repetitive by some participants. Future iterations could incorporate more sophisticated memory systems or diversity sampling to minimize repetitive or generic responses. 

Similarly, interactions sometimes felt rigid or ``bot-like,'' potentially stemming from our dialogue state chain enforcing hard bounds on the conversation topic. 
More concerningly, the dialogue state chain placed Beebo in the role of steering conversations, potentially limiting participants' agency. 
Many participants gradually shifted to short, one-word replies over time as questions became repetitive, limiting the context available to the agent.
While removing the dialogue state chain allowed the agent to more naturally adapt to a conversation, it also allowed it to veer off course and lose track of the conversation's goals, mirroring prior findings in multi-turn settings~\cite{laban2025llms}.
Future research on goal-directed, multi-turn conversations could improve LLMs' conversational flexibility to leave participants more space to steer the conversation while ensuring they achieve conversational goals and stay within fixed conversational bounds.

In addition, we encountered several issues attributable to inconsistent tool use.
Participants occasionally experienced frustrating interactions due to hallucinated or inaccurate tool calls, particularly around scheduling and activity tracking.
For example, hallucinated tool calls (e.g., extraneous or incorrect arguments) could lead to errors in plan edits, upon which the agent would tell the user they would ``fix the issue later,'' even though it did not have this capability.
Moreover, health data fetching was rarely proactively initiated by the model, mirroring findings from GPTCoach~\cite{jorke2025gptcoach}. 
Improving the robustness of tool use may also necessitate multi-agent systems and/or finetuning.

Finally, several usability challenges emerged from inherent challenges in integrating conversational agents with complex user interfaces. 
The open-ended chat interface and Beebo's conversational flexibility surfaced false expectations about the system's capabilities: participants frequently posed questions about app features or bugs directly to Beebo (e.g., why workouts were not syncing, summaries not loading, and check-in not starting), which Beebo frequently did not have the knowledge or capabilities to address. Adding to their frustration, the agent would sometimes hallucinate an incorrect answer instead of stating that it did not know.
These results may explain the larger variance in post-study usability scores in the treatment condition. 
Addressing these concerns requires both clear communication of agent capabilities, either in the UI or chat, to calibrate participants' mental models as well as preventing problematic model hallucinations.

Taken together, while our findings motivate a longer-term (several months to a year), larger-scale (hundreds of participants) study to evaluate whether larger behavioral differences materialize over time, several targeted, currently feasible improvements to the LLM agent are likely to increase the effect size.
\section{Limitations}
\label{sec:limitations}

Our study has several limitations. First, our study design was not intended to robustly evaluate behavior change in a clinical sense, which requires large-scale, randomized controlled trials conducted over several months to years. 
Consequently, H2 and H3 were underpowered to detect moderate effect sizes, and our study duration was likely insufficient for larger differences in PA habits to materialize. 
Since participants self-selected into the study and were compensated, the effect sizes observed here are likely larger than what would be expected in a naturalistic deployment.
As argued by \citet{klasnja2011evaluate}, evaluating behavior change in a traditional sense is often inappropriate for early-stage HCI systems research.
Instead, our study's main focus was on studying participant experiences through qualitative interviews and extensive surveys, aiming to uncover insights into why and how such systems succeed or fail.
The four-week duration balanced high participation demands with known issues with sustained engagement in behavior change research~\cite{yang2020factors}.

Since our control condition did not include a rule-based chatbot, some of our findings may have also arisen with a non-LLM chatbot. However, our findings around relational cues and mindsets differ from prior work on rule-based coaches~\cite{mitchell2021automated, luo2021promoting}. Moreover, we consider it unlikely that a simpler chatbot would have produced stronger effects on quantitative PA outcomes.

Our recruitment procedure aimed for broad demographic coverage, requiring over 2,000 screening responses to enroll 54 participants. Although our sample had adequate diversity along gender, age, and race/ethnicity, participants had higher educational attainment and income levels compared to the general population. 
This likely reflects our requirements for owning an iPhone and Apple Watch~\cite{jamalova2019comparative} as well as the significant time commitment required for participation~\cite{kouaho2024socioeconomic}. While automated health coaching has the potential to particularly benefit individuals experiencing greater barriers to PA, often linked to lower socioeconomic status (SES)~\cite{wagstaff2002poverty}, future research that specifically engages with low-SES populations is necessary to realize this potential and prevent intervention-generated inequalities~\cite{veinot2018good}. This may involve community-based recruitment procedures and/or relaxing device ownership requirements.

Lastly, the Bloom system itself had technological limitations. We did not explore more advanced agent designs involving finetuning~\cite{mantena2025fine, khasentino2025personal}, multi-agent systems~\cite{merrill2026transforming}, or reasoning models~\cite{jaech2024openai, guo2025deepseek}. 
While our summary-based memory module was well within the model's context limit, even for users that chatted daily (<3,500 tokens), more sophisticated memory systems (e.g.,~\cite{park2023generative}) could have improved performance.
Given the promising initial feedback we received on our prompt-based approach, we considered it sufficient for generating design insights. 
Moreover, advanced improvements would have required significant development and/or data collection efforts without clear evidence of user need or interaction requirements.
We also did not explore all possible LLM-augmented behavior change interactions, such as social features or gamification. 

\section{Conclusion}
In this work, we introduce Bloom, a mobile application integrating an LLM-based coaching chatbot with established behavior change interactions. Bloom leverages qualitative context from coaching conversations to further personalize LLM-augmented behavior change interactions, moving beyond existing text-only approaches in prior work on LLMs for behavioral health. Through a four-week field study involving 54 participants, we found that Bloom fostered positive mindsets toward PA and supported more personalized and flexible workout planning compared to a no-LLM control. Treatment participants reported greater improvements in mindset-related outcomes, including increased enjoyment, self-confidence, and a greater sense of agency in managing their PA goals. Both conditions substantially increased participants' exercise levels, doubling the number of individuals who met or exceeded the recommended 150 min/week of PA. Although we observed no advantage for the LLM condition in short-term PA levels, psychological and motivational outcomes indicate the potential for sustained behavior change over longer periods when using our LLM-based health coach. Bloom presents a novel opportunity to create more supportive and empowering behavior change applications that leverage qualitative context to more sustainably shape people's mindsets and motivation.

\begin{acks}
We are grateful for the funding support provided by the Stanford Institute for Human-Centered Artificial Intelligence (HAI), the Hasso Plattner Foundation, Hanwha, Wellhub, as well as the OpenAI researcher access program for providing API credits to partially support this research.
We thank the many members of the IxD and AI4HI research groups for their continuous support and feedback on this project, and all of our friends \& family testers for catching bugs and suggesting improvements. 
We thank Michelle Lam, Lindsay Popowski, and Dulce Garcia for their assistance with participant recruitment; Ryan Louie, Ifdita Hasan, and Ankita Koodavoor for their advice on our redteaming and safety evaluation; Andrew Sung for his contributions to app development; and Carolyn Zhou for their feedback on the LLM agent architecture. 
We are also grateful to Tobias Gerstenberg, Nilam Ram, and Michael Bernstein for guidance on statistical modeling; to Elizabeth Murnane for her input on our ambient display design and validation study; to Pedja Klasnja for his advice on post hoc power analyses; and to Mary Czerwinski, Andrea Green, Omar Shaikh, and Helena Vasconcelos for thoughtful feedback on early drafts of this manuscript.
Most importantly, we thank all of our participants, without whom this research would not have been possible.
\end{acks}

\bibliographystyle{ACM-Reference-Format}
\bibliography{references}

@misc{cdc2022,
  author = {Centers for Disease Control and Prevention},
  title = {Physical Activity},
  year = {2022},
  url = {https://www.cdc.gov/physicalactivity/index.html}
}

@article{hicks2023leveraging,
  title={Leveraging mobile technology for public health promotion: A multidisciplinary perspective},
  author={Hicks, Jennifer L and Boswell, Melissa A and Althoff, Tim and Crum, Alia J and Ku, Joy P and Landay, James A and Moya, Paula ML and Murnane, Elizabeth L and Snyder, Michael P and King, Abby C and others},
  journal={Annual Review of Public Health},
  volume={44},
  pages={131--150},
  year={2023},
  publisher={Annual Reviews}
}

@misc{bommasani2022opportunities,
      title={On the Opportunities and Risks of Foundation Models}, 
      author={Rishi Bommasani and Drew A. Hudson and Ehsan Adeli and Russ Altman and Simran Arora and Sydney von Arx and Michael S. Bernstein and Jeannette Bohg and Antoine Bosselut and Emma Brunskill and Erik Brynjolfsson and Shyamal Buch and Dallas Card and Rodrigo Castellon and Niladri Chatterji and Annie Chen and Kathleen Creel and Jared Quincy Davis and Dora Demszky and Chris Donahue and Moussa Doumbouya and Esin Durmus and Stefano Ermon and John Etchemendy and Kawin Ethayarajh and Li Fei-Fei and Chelsea Finn and Trevor Gale and Lauren Gillespie and Karan Goel and Noah Goodman and Shelby Grossman and Neel Guha and Tatsunori Hashimoto and Peter Henderson and John Hewitt and Daniel E. Ho and Jenny Hong and Kyle Hsu and Jing Huang and Thomas Icard and Saahil Jain and Dan Jurafsky and Pratyusha Kalluri and Siddharth Karamcheti and Geoff Keeling and Fereshte Khani and Omar Khattab and Pang Wei Koh and Mark Krass and Ranjay Krishna and Rohith Kuditipudi and Ananya Kumar and Faisal Ladhak and Mina Lee and Tony Lee and Jure Leskovec and Isabelle Levent and Xiang Lisa Li and Xuechen Li and Tengyu Ma and Ali Malik and Christopher D. Manning and Suvir Mirchandani and Eric Mitchell and Zanele Munyikwa and Suraj Nair and Avanika Narayan and Deepak Narayanan and Ben Newman and Allen Nie and Juan Carlos Niebles and Hamed Nilforoshan and Julian Nyarko and Giray Ogut and Laurel Orr and Isabel Papadimitriou and Joon Sung Park and Chris Piech and Eva Portelance and Christopher Potts and Aditi Raghunathan and Rob Reich and Hongyu Ren and Frieda Rong and Yusuf Roohani and Camilo Ruiz and Jack Ryan and Christopher Ré and Dorsa Sadigh and Shiori Sagawa and Keshav Santhanam and Andy Shih and Krishnan Srinivasan and Alex Tamkin and Rohan Taori and Armin W. Thomas and Florian Tramèr and Rose E. Wang and William Wang and Bohan Wu and Jiajun Wu and Yuhuai Wu and Sang Michael Xie and Michihiro Yasunaga and Jiaxuan You and Matei Zaharia and Michael Zhang and Tianyi Zhang and Xikun Zhang and Yuhui Zhang and Lucia Zheng and Kaitlyn Zhou and Percy Liang},
      year={2022},
      eprint={2108.07258},
      archivePrefix={arXiv},
      primaryClass={cs.LG}
}

@article{olsen2010health,
  title={Health coaching to improve healthy lifestyle behaviors: an integrative review},
  author={Olsen, Jeanette M and Nesbitt, Bonnie J},
  journal={American journal of health promotion},
  volume={25},
  number={1},
  pages={e1--e12},
  year={2010},
  publisher={SAGE Publications Sage CA: Los Angeles, CA}
}

@article{wolever2013systematic,
  title={A systematic review of the literature on health and wellness coaching: defining a key behavioral intervention in healthcare},
  author={Wolever, Ruth Q and Simmons, Leigh Ann and Sforzo, Gary A and Dill, Diana and Kaye, Miranda and Bechard, Elizabeth M and Southard, Mary Elaine and Kennedy, Mary and Vosloo, Justine and Yang, Nancy},
  journal={Global advances in health and medicine},
  volume={2},
  number={4},
  pages={38--57},
  year={2013},
  publisher={SAGE Publications Sage CA: Los Angeles, CA}
}

@article{mitchell2021automated,
  title={Automated vs. human health coaching: exploring participant and practitioner experiences},
  author={Mitchell, Elliot G and Maimone, Rosa and Cassells, Andrea and Tobin, Jonathan N and Davidson, Patricia and Smaldone, Arlene M and Mamykina, Lena},
  journal={Proceedings of the ACM on human-computer interaction},
  volume={5},
  number={CSCW1},
  pages={1--37},
  year={2021},
  publisher={ACM New York, NY, USA}
}

@article{kersten2017personal,
  title={Personal informatics, self-insight, and behavior change: A critical review of current literature},
  author={Kersten-van Dijk, Elisabeth T and Westerink, Joyce HDM and Beute, Femke and IJsselsteijn, Wijnand A},
  journal={Human--Computer Interaction},
  volume={32},
  number={5-6},
  pages={268--296},
  year={2017},
  publisher={Taylor \& Francis}
}

@String{Computing = "Computing" }

@String{Computer = "{IEEE} Computer" }

@String{Chelsea = "Chelsea" }

@String{Springer = "Springer-Verlag" }

@inproceedings{baumer2015reflective,
  title={Reflective informatics: conceptual dimensions for designing technologies of reflection},
  author={Baumer, Eric PS},
  booktitle={Proceedings of the 33rd Annual ACM Conference on Human Factors in Computing Systems},
  pages={585--594},
  year={2015}
}

@inproceedings{li2010stage,
  title={A stage-based model of personal informatics systems},
  author={Li, Ian and Dey, Anind and Forlizzi, Jodi},
  booktitle={Proceedings of the SIGCHI conference on human factors in computing systems},
  pages={557--566},
  year={2010}
}

@article{bentvelzen2022revisiting,
  title={Revisiting Reflection in HCI: Four Design Resources for Technologies that Support Reflection},
  author={Bentvelzen, Marit and Wo{\'z}niak, Pawe{\l} W and Herbes, Pia SF and Stefanidi, Evropi and Niess, Jasmin},
  journal={Proceedings of the ACM on Interactive, Mobile, Wearable and Ubiquitous Technologies},
  volume={6},
  number={1},
  pages={1--27},
  year={2022},
  publisher={ACM New York, NY, USA}
}

@article{choe2015characterizing,
  title={Characterizing visualization insights from quantified selfers' personal data presentations},
  author={Choe, Eun Kyoung and Lee, Bongshin and others},
  journal={IEEE computer graphics and applications},
  volume={35},
  number={4},
  pages={28--37},
  year={2015},
  publisher={IEEE}
}

@article{epstein2020mapping,
  title={Mapping and taking stock of the personal informatics literature},
  author={Epstein, Daniel A and Caldeira, Clara and Figueiredo, Mayara Costa and Lu, Xi and Silva, Lucas M and Williams, Lucretia and Lee, Jong Ho and Li, Qingyang and Ahuja, Simran and Chen, Qiuer and others},
  journal={Proceedings of the ACM on Interactive, Mobile, Wearable and Ubiquitous Technologies},
  volume={4},
  number={4},
  pages={1--38},
  year={2020},
  publisher={ACM New York, NY, USA}
}

@inproceedings{aseniero2020activity,
  title={Activity river: Visualizing planned and logged personal activities for reflection},
  author={Aseniero, Bon Adriel and Perin, Charles and Willett, Wesley and Tang, Anthony and Carpendale, Sheelagh},
  booktitle={Proceedings of the International Conference on Advanced Visual Interfaces},
  pages={1--9},
  year={2020}
}

@article{thudt2015visual,
  title={Visual mementos: Reflecting memories with personal data},
  author={Thudt, Alice and Baur, Dominikus and Huron, Samuel and Carpendale, Sheelagh},
  journal={IEEE transactions on visualization and computer graphics},
  volume={22},
  number={1},
  pages={369--378},
  year={2015},
  publisher={IEEE}
}

@inproceedings{bentvelzen2021development,
  title={The Development and Validation of the Technology-Supported Reflection Inventory},
  author={Bentvelzen, Marit and Niess, Jasmin and Wo{\'z}niak, Miko{\l}aj P and Wo{\'z}niak, Pawe{\l} W},
  booktitle={Proceedings of the 2021 CHI Conference on Human Factors in Computing Systems},
  pages={1--8},
  year={2021}
}

@article{choe2017semi,
  title={Semi-automated tracking: a balanced approach for self-monitoring applications},
  author={Choe, Eun Kyoung and Abdullah, Saeed and Rabbi, Mashfiqui and Thomaz, Edison and Epstein, Daniel A and Cordeiro, Felicia and Kay, Matthew and Abowd, Gregory D and Choudhury, Tanzeem and Fogarty, James and others},
  journal={IEEE Pervasive Computing},
  volume={16},
  number={1},
  pages={74--84},
  year={2017},
  publisher={IEEE}
}

@inproceedings{cho2022reflection,
  title={Reflection in theory and reflection in practice: An exploration of the gaps in reflection support among personal informatics apps},
  author={Cho, Janghee and Xu, Tian and Zimmermann-Niefield, Abigail and Voida, Stephen},
  booktitle={Proceedings of the 2022 CHI Conference on Human Factors in Computing Systems},
  pages={1--23},
  year={2022}
}

@inproceedings{murnane2020designing,
  title={Designing ambient narrative-based interfaces to reflect and motivate physical activity},
  author={Murnane, Elizabeth L and Jiang, Xin and Kong, Anna and Park, Michelle and Shi, Weili and Soohoo, Connor and Vink, Luke and Xia, Iris and Yu, Xin and Yang-Sammataro, John and others},
  booktitle={Proceedings of the 2020 CHI Conference on Human Factors in Computing Systems},
  pages={1--14},
  year={2020}
}

@article{crum2007mind,
  title={Mind-set matters: Exercise and the placebo effect},
  author={Crum, Alia J and Langer, Ellen J},
  journal={Psychological science},
  volume={18},
  number={2},
  pages={165--171},
  year={2007},
  publisher={SAGE Publications Sage CA: Los Angeles, CA}
}

@software{schmiedmayer2024,
  author       = {Schmiedmayer, Paul and
                  Ravi, Vishnu and
                  Aalami, Oliver},
  title        = {Spezi},
  month        = jan,
  year         = 2024,
  publisher    = {Zenodo},
  version      = {1.1.0},
  doi          = {10.5281/zenodo.10482368},
  url          = {https://doi.org/10.5281/zenodo.10482368}
}

@article{veinot2018good,
  title={Good intentions are not enough: how informatics interventions can worsen inequality},
  author={Veinot, Tiffany C and Mitchell, Hannah and Ancker, Jessica S},
  journal={Journal of the American Medical Informatics Association},
  volume={25},
  number={8},
  pages={1080--1088},
  year={2018},
  publisher={Oxford University Press}
}

@article{bickmore2006health,
  title={Health dialog systems for patients and consumers},
  author={Bickmore, Timothy and Giorgino, Toni},
  journal={Journal of biomedical informatics},
  volume={39},
  number={5},
  pages={556--571},
  year={2006},
  publisher={Elsevier}
}

@inproceedings{olsen2014health,
  title={Health coaching: a concept analysis},
  author={Olsen, Jeanette M},
  booktitle={Nursing forum},
  volume={49},
  number={1},
  pages={18--29},
  year={2014},
  organization={Wiley Online Library}
}

@article{singh2023systematic,
  title={Systematic review and meta-analysis of the effectiveness of chatbots on lifestyle behaviours},
  author={Singh, Ben and Olds, Timothy and Brinsley, Jacinta and Dumuid, Dot and Virgara, Rosa and Matricciani, Lisa and Watson, Amanda and Szeto, Kimberley and Eglitis, Emily and Miatke, Aaron and others},
  journal={npj Digital Medicine},
  volume={6},
  number={1},
  pages={118},
  year={2023},
  publisher={Nature Publishing Group UK London}
}

@inproceedings{consolvo2008activity,
  title={Activity sensing in the wild: a field trial of ubifit garden},
  author={Consolvo, Sunny and McDonald, David W and Toscos, Tammy and Chen, Mike Y and Froehlich, Jon and Harrison, Beverly and Klasnja, Predrag and LaMarca, Anthony and LeGrand, Louis and Libby, Ryan and others},
  booktitle={Proceedings of the SIGCHI conference on human factors in computing systems},
  pages={1797--1806},
  year={2008}
}

@article{bickmore2005establishing,
author = {Bickmore, Timothy W. and Picard, Rosalind W.},
title = {Establishing and maintaining long-term human-computer relationships},
year = {2005},
issue_date = {June 2005},
publisher = {Association for Computing Machinery},
address = {New York, NY, USA},
volume = {12},
number = {2},
issn = {1073-0516},
url = {https://doi.org/10.1145/1067860.1067867},
doi = {10.1145/1067860.1067867},
journal = {ACM Trans. Comput.-Hum. Interact.},
month = {jun},
pages = {293–327},
numpages = {35}
}

@inproceedings{munson2012exploring,
  title={Exploring goal-setting, rewards, self-monitoring, and sharing to motivate physical activity},
  author={Munson, Sean A and Consolvo, Sunny},
  booktitle={2012 6th international conference on pervasive computing technologies for healthcare (pervasivehealth) and workshops},
  pages={25--32},
  year={2012},
  organization={IEEE}
}

@article{englhardt2024classification,
  title={From classification to clinical insights: Towards analyzing and reasoning about mobile and behavioral health data with large language models},
  author={Englhardt, Zachary and Ma, Chengqian and Morris, Margaret E and Chang, Chun-Cheng and Xu, Xuhai" Orson" and Qin, Lianhui and McDuff, Daniel and Liu, Xin and Patel, Shwetak and Iyer, Vikram},
  journal={Proceedings of the ACM on Interactive, Mobile, Wearable and Ubiquitous Technologies},
  volume={8},
  number={2},
  pages={1--25},
  year={2024},
  publisher={ACM New York, NY, USA}
}

@misc{liu2023large,
      title={Large Language Models are Few-Shot Health Learners}, 
      author={Xin Liu and Daniel McDuff and Geza Kovacs and Isaac Galatzer-Levy and Jacob Sunshine and Jiening Zhan and Ming-Zher Poh and Shun Liao and Paolo Di Achille and Shwetak Patel},
      year={2023},
      eprint={2305.15525},
      archivePrefix={arXiv},
      primaryClass={cs.CL}
}

@article{vandelanotte2023increasing,
  title={Increasing physical activity using an just-in-time adaptive digital assistant supported by machine learning: a novel approach for hyper-personalised mHealth interventions},
  author={Vandelanotte, Corneel and Trost, Stewart and Hodgetts, Danya and Imam, Tasadduq and Rashid, Mamunur and To, Quyen G and Maher, Carol},
  journal={Journal of Biomedical Informatics},
  volume={144},
  pages={104435},
  year={2023},
  publisher={Elsevier}
}

@article{king2020effects,
  title={Effects of counseling by peer human advisors vs computers to increase walking in underserved populations: The COMPASS randomized clinical trial},
  author={King, Abby C and Campero, Maria Ines and Sheats, Jylana L and Sweet, Cynthia M Castro and Hauser, Michelle E and Garcia, Dulce and Chazaro, Aldo and Blanco, German and Banda, Jorge and Ahn, David K and others},
  journal={JAMA internal medicine},
  volume={180},
  number={11},
  pages={1481--1490},
  year={2020},
  publisher={American Medical Association}
}

@article{ouyang2022training,
  title={Training language models to follow instructions with human feedback},
  author={Ouyang, Long and Wu, Jeffrey and Jiang, Xu and Almeida, Diogo and Wainwright, Carroll and Mishkin, Pamela and Zhang, Chong and Agarwal, Sandhini and Slama, Katarina and Ray, Alex and others},
  journal={Advances in Neural Information Processing Systems},
  volume={35},
  pages={27730--27744},
  year={2022}
}

@article{anderson2007shakra,
  title={Shakra: tracking and sharing daily activity levels with unaugmented mobile phones},
  author={Anderson, Ian and Maitland, Julie and Sherwood, Scott and Barkhuus, Louise and Chalmers, Matthew and Hall, Malcolm and Brown, Barry and Muller, Henk},
  journal={Mobile networks and applications},
  volume={12},
  pages={185--199},
  year={2007},
  publisher={Springer}
}

@inproceedings{lin2006fish,
  title={Fish’n’Steps: Encouraging physical activity with an interactive computer game},
  author={Lin, James J and Mamykina, Lena and Lindtner, Silvia and Delajoux, Gregory and Strub, Henry B},
  booktitle={UbiComp 2006: Ubiquitous Computing: 8th International Conference, UbiComp 2006 Orange County, CA, USA, September 17-21, 2006 Proceedings 8},
  pages={261--278},
  year={2006},
  organization={Springer}
}

@article{nahum2018just,
  title={Just-in-time adaptive interventions (JITAIs) in mobile health: key components and design principles for ongoing health behavior support},
  author={Nahum-Shani, Inbal and Smith, Shawna N and Spring, Bonnie J and Collins, Linda M and Witkiewitz, Katie and Tewari, Ambuj and Murphy, Susan A},
  journal={Annals of Behavioral Medicine},
  pages={1--17},
  year={2018},
  publisher={Springer}
}

@article{yang2019comparative,
  title={The comparative effectiveness of mobile phone interventions in improving health outcomes: meta-analytic review},
  author={Yang, Qinghua and Van Stee, Stephanie K},
  journal={JMIR mHealth and uHealth},
  volume={7},
  number={4},
  pages={e11244},
  year={2019},
  publisher={JMIR Publications Toronto, Canada}
}

@article{chiu2024computational,
  title={A Computational Framework for Behavioral Assessment of LLM Therapists},
  author={Chiu, Yu Ying and Sharma, Ashish and Lin, Inna Wanyin and Althoff, Tim},
  journal={arXiv preprint arXiv:2401.00820},
  year={2024}
}

@article{moyers2003assessing,
  title={Assessing the integrity of motivational interviewing interventions: Reliability of the motivational interviewing skills code},
  author={Moyers, Theresa and Martin, Tim and Catley, Delwyn and Harris, Kari Jo and Ahluwalia, Jasjit S},
  journal={Behavioural and Cognitive Psychotherapy},
  volume={31},
  number={2},
  pages={177--184},
  year={2003},
  publisher={Cambridge University Press}
}

@book{miller2023motivational,
  title={Motivational Interviewing: Helping People Change and Grow},
  edition={4th},
  author={Miller, W.R. and Rollnick, S.},
  isbn={9781462552818},
  series={Applications of Motivational Interviewing Series},
  year={2023},
  publisher={Guilford Publications}
}

@article{shah2022modeling,
  title={Modeling motivational interviewing strategies on an online peer-to-peer counseling platform},
  author={Shah, Raj Sanjay and Holt, Faye and Hayati, Shirley Anugrah and Agarwal, Aastha and Wang, Yi-Chia and Kraut, Robert E and Yang, Diyi},
  journal={Proceedings of the ACM on Human-Computer Interaction},
  volume={6},
  number={CSCW2},
  pages={1--24},
  year={2022},
  publisher={ACM New York, NY, USA}
}

@article{reeves1996media,
  title={The media equation: How people treat computers, television, and new media like real people},
  author={Reeves, Byron and Nass, Clifford},
  journal={Cambridge, UK},
  volume={10},
  number={10},
  year={1996}
}

@inproceedings{perez2019makes,
  title={What makes a good counselor? learning to distinguish between high-quality and low-quality counseling conversations},
  author={P{\'e}rez-Rosas, Ver{\'o}nica and Wu, Xinyi and Resnicow, Kenneth and Mihalcea, Rada},
  booktitle={Proceedings of the 57th Annual Meeting of the Association for Computational Linguistics},
  pages={926--935},
  year={2019}
}

@article{craig2003international,
  title={International physical activity questionnaire: 12-country reliability and validity},
  author={Craig, Cora L and Marshall, Alison L and Sj{\"o}str{\"o}m, Michael and Bauman, Adrian E and Booth, Michael L and Ainsworth, Barbara E and Pratt, Michael and Ekelund, ULF and Yngve, Agneta and Sallis, James F and others},
  journal={Medicine \& science in sports \& exercise},
  volume={35},
  number={8},
  pages={1381--1395},
  year={2003},
  publisher={LWW}
}

@article{marcus1992self,
  title={Self-efficacy and the stages of exercise behavior change},
  author={Marcus, Bess H and Selby, Vanessa C and Niaura, Raymond S and Rossi, Joseph S},
  journal={Research quarterly for exercise and sport},
  volume={63},
  number={1},
  pages={60--66},
  year={1992},
  publisher={Taylor \& Francis}
}

@article{prochaska1997transtheoretical,
  title={The transtheoretical model of health behavior change},
  author={Prochaska, James O and Velicer, Wayne F},
  journal={American journal of health promotion},
  volume={12},
  number={1},
  pages={38--48},
  year={1997},
  publisher={SAGE Publications Sage CA: Los Angeles, CA}
}

@inproceedings{klasnja2011evaluate,
author = {Klasnja, Predrag and Consolvo, Sunny and Pratt, Wanda},
title = {How to evaluate technologies for health behavior change in HCI research},
year = {2011},
isbn = {9781450302289},
publisher = {Association for Computing Machinery},
address = {New York, NY, USA},
url = {https://doi.org/10.1145/1978942.1979396},
doi = {10.1145/1978942.1979396},
booktitle = {Proceedings of the SIGCHI Conference on Human Factors in Computing Systems},
pages = {3063–3072},
numpages = {10},
series = {CHI '11}
}

@article{hone2000towards,
  title={Towards a tool for the subjective assessment of speech system interfaces (SASSI)},
  author={Hone, Kate S and Graham, Robert},
  journal={Natural Language Engineering},
  volume={6},
  number={3-4},
  pages={287--303},
  year={2000},
  publisher={Cambridge University Press}
}

@article{wilcox2006results,
  title={Results of the first year of active for life: translation of 2 evidence-based physical activity programs for older adults into community settings},
  author={Wilcox, Sara and Dowda, Marsha and Griffin, Sarah F and Rheaume, Carol and Ory, Marcia G and Leviton, Laura and King, Abby C and Dunn, Andrea and Buchner, David M and Bazzarre, Terry and others},
  journal={American Journal of Public Health},
  volume={96},
  number={7},
  pages={1201--1209},
  year={2006},
  publisher={American Public Health Association}
}

@article{wilcox2008active,
  title={Active for life: final results from the translation of two physical activity programs},
  author={Wilcox, Sara and Dowda, Marsha and Leviton, Laura C and Bartlett-Prescott, Jenny and Bazzarre, Terry and Campbell-Voytal, Kimberly and Carpenter, Ruth Ann and Castro, Cynthia M and Dowdy, Diane and Dunn, Andrea L and others},
  journal={American journal of preventive medicine},
  volume={35},
  number={4},
  pages={340--351},
  year={2008},
  publisher={Elsevier}
}

@inproceedings{consolvo2006design,
  title={Design requirements for technologies that encourage physical activity},
  author={Consolvo, Sunny and Everitt, Katherine and Smith, Ian and Landay, James A},
  booktitle={Proceedings of the SIGCHI conference on Human Factors in computing systems},
  pages={457--466},
  year={2006}
}

@inproceedings{lane2012bewell,
  title={Bewell: A smartphone application to monitor, model and promote wellbeing},
  author={Lane, Nicholas and Mohammod, Mashfiqui and Lin, Mu and Yang, Xiaochao and Lu, Hong and Ali, Shahid and Doryab, Afsaneh and Berke, Ethan and Choudhury, Tanzeem and Campbell, Andrew},
  booktitle={5th international ICST conference on pervasive computing technologies for healthcare},
  year={2012}
}

@article{merrill2026transforming,
  title={Transforming wearable data into personal health insights using large language model agents},
  author={Merrill, Mike A and Paruchuri, Akshay and Rezaei, Naghmeh and Kovacs, Geza and Perez, Javier and Liu, Yun and Schenck, Erik and Hammerquist, Nova and Sunshine, Jake and Tailor, Shyam and others},
  journal={Nature Communications},
  volume={17},
  number={1},
  pages={1143},
  year={2026},
  doi={10.1038/s41467-025-67922-y},
  publisher={Nature Publishing Group UK London}
}

@article{khasentino2025personal,
  title={A personal health large language model for sleep and fitness coaching},
  author={Khasentino, Justin and Belyaeva, Anastasiya and Liu, Xin and Yang, Zhun and Furlotte, Nicholas A and Lee, Chace and Schenck, Erik and Patel, Yojan and Cui, Jian and Schneider, Logan Douglas and others},
  journal={Nature Medicine},
  pages={1--10},
  year={2025},
  publisher={Nature Publishing Group US New York}
}

@inproceedings{fang2024physiollm,
  title={Physiollm: Supporting personalized health insights with wearables and large language models},
  author={Fang, Cathy Mengying and Danry, Valdemar and Whitmore, Nathan and Bao, Andria and Hutchison, Andrew and Pierce, Cayden and Maes, Pattie},
  booktitle={2024 IEEE EMBS International Conference on Biomedical and Health Informatics (BHI)},
  pages={1--8},
  year={2024},
  organization={IEEE}
}

@article{king2007ongoing,
  title={Ongoing physical activity advice by humans versus computers: the Community Health Advice by Telephone (CHAT) trial.},
  author={King, Abby C and Friedman, Robert and Marcus, Bess and Castro, Cynthia and Napolitano, Melissa and Ahn, David and Baker, Lawrence},
  journal={Health Psychology},
  volume={26},
  number={6},
  pages={718},
  year={2007},
  publisher={American Psychological Association}
}

@article{king2014exercise,
  title={Exercise advice by humans versus computers: maintenance effects at 18 months.},
  author={King, Abby C and Hekler, Eric B and Castro, Cynthia M and Buman, Matthew P and Marcus, Bess H and Friedman, Robert H and Napolitano, Melissa A},
  journal={Health Psychology},
  volume={33},
  number={2},
  pages={192},
  year={2014},
  publisher={American Psychological Association}
}

@article{castro2011physical,
  title={Physical activity program delivery by professionals versus volunteers: the TEAM randomized trial.},
  author={Castro, Cynthia M and Pruitt, Leslie A and Buman, Matthew P and King, Abby C},
  journal={Health Psychology},
  volume={30},
  number={3},
  pages={285},
  year={2011},
  publisher={American Psychological Association}
}

@article{bandura1999social,
  title={Social cognitive theory: An agentic perspective},
  author={Bandura, Albert},
  journal={Asian journal of social psychology},
  volume={2},
  number={1},
  pages={21--41},
  year={1999},
  publisher={Wiley Online Library}
}

@article{luo2021promoting,
  title={Promoting physical activity through conversational agents: mixed methods systematic review},
  author={Luo, Tiffany Christina and Aguilera, Adrian and Lyles, Courtney Rees and Figueroa, Caroline Astrid},
  journal={Journal of Medical Internet Research},
  volume={23},
  number={9},
  pages={e25486},
  year={2021},
  publisher={JMIR Publications Toronto, Canada}
}

@article{jamalova2019comparative,
  title={The comparative study of the relationship between smartphone choice and socio-economic indicators},
  author={Jamalova, Maral},
  journal={Int. J. Mark. Stud},
  volume={11},
  number={11},
  pages={10--5539},
  year={2019}
}

@inproceedings{kouaho2024socioeconomic,
  title={Socioeconomic Class in Physical Activity Wearables Research and Design},
  author={Kouaho, Whitney-Jocelyn and Epstein, Daniel A},
  booktitle={Proceedings of the CHI Conference on Human Factors in Computing Systems},
  pages={1--15},
  year={2024}
}

@article{consolvo2014designing,
  title={Designing for healthy lifestyles: Design considerations for mobile technologies to encourage consumer health and wellness},
  author={Consolvo, Sunny and Klasnja, Predrag and McDonald, David W and Landay, James A and others},
  journal={Foundations and Trends{\textregistered} in Human--Computer Interaction},
  volume={6},
  number={3--4},
  pages={167--315},
  year={2014},
  publisher={Now Publishers, Inc.}
}

@article{klasnja2012healthcare,
  title={Healthcare in the pocket: mapping the space of mobile-phone health interventions},
  author={Klasnja, Predrag and Pratt, Wanda},
  journal={Journal of biomedical informatics},
  volume={45},
  number={1},
  pages={184--198},
  year={2012},
  publisher={Elsevier}
}

@article{bentley2013health,
  title={Health Mashups: Presenting statistical patterns between wellbeing data and context in natural language to promote behavior change},
  author={Bentley, Frank and Tollmar, Konrad and Stephenson, Peter and Levy, Laura and Jones, Brian and Robertson, Scott and Price, Ed and Catrambone, Richard and Wilson, Jeff},
  journal={ACM Transactions on Computer-Human Interaction (TOCHI)},
  volume={20},
  number={5},
  pages={1--27},
  year={2013},
  publisher={ACM New York, NY, USA}
}

@article{locke2002building,
  title={Building a practically useful theory of goal setting and task motivation: A 35-year odyssey.},
  author={Locke, Edwin A and Latham, Gary P},
  journal={American psychologist},
  volume={57},
  number={9},
  pages={705},
  year={2002},
  publisher={American Psychological Association}
}

@inproceedings{bentley2013power,
  title={The power of mobile notifications to increase wellbeing logging behavior},
  author={Bentley, Frank and Tollmar, Konrad},
  booktitle={Proceedings of the SIGCHI conference on human factors in computing systems},
  pages={1095--1098},
  year={2013}
}

@inproceedings{intille2004ubiquitous,
  title={Ubiquitous computing technology for just-in-time motivation of behavior change},
  author={Intille, Stephen S},
  booktitle={MEDINFO 2004},
  pages={1434--1437},
  year={2004},
  organization={IOS Press}
}

@inproceedings{gasser2006persuasiveness,
  title={Persuasiveness of a mobile lifestyle coaching application using social facilitation},
  author={Gasser, Roland and Brodbeck, Dominique and Degen, Markus and Luthiger, J{\"u}rg and Wyss, Remo and Reichlin, Serge},
  booktitle={Persuasive Technology: First International Conference on Persuasive Technology for Human Well-Being, PERSUASIVE 2006, Eindhoven, The Netherlands, May 18-19, 2006. Proceedings 1},
  pages={27--38},
  year={2006},
  organization={Springer}
}

@inproceedings{bickmore2005acceptance,
  title={Acceptance and usability of a relational agent interface by urban older adults},
  author={Bickmore, Timothy W and Caruso, Lisa and Clough-Gorr, Kerri},
  booktitle={CHI'05 extended abstracts on Human factors in computing systems},
  pages={1212--1215},
  year={2005}
}

@inproceedings{nielsen1990heuristic,
  title={Heuristic evaluation of user interfaces},
  author={Nielsen, Jakob and Molich, Rolf},
  booktitle={Proceedings of the SIGCHI conference on Human factors in computing systems},
  pages={249--256},
  year={1990}
}

@book{consolvo2008designing,
  title={Designing and evaluating a persuasive technology to encourage lifestyle behavior change},
  author={Consolvo, Sunny},
  year={2008},
  publisher={University of Washington}
}

@inproceedings{jorke2025gptcoach,
  title={GPTCoach: Towards LLM-Based Physical Activity Coaching},
  author={J{\"o}rke, Matthew and Sapkota, Shardul and Warkenthien, Lyndsea and Vainio, Niklas and Schmiedmayer, Paul and Brunskill, Emma and Landay, James A},
  booktitle={Proceedings of the 2025 CHI Conference on Human Factors in Computing Systems},
  pages={1--46},
  year={2025}
}

@misc{king2002stanford,
  author       = {King, Abby C. and Haskell, William L. and Taylor, C. Barr and DeBusk, Robert and Castro, Cynthia M. and Pruitt, Leslie A. and Stanford Prevention Research Center staff},
  title        = {The Stanford Active Choices Program: Telephone-Assisted Counseling for Physical Activity},
  howpublished = {Stanford Health Promotion Resource Center, Stanford Prevention Research Center, Stanford University},
  year         = {2002},
  address      = {Stanford, CA},
}

@inproceedings{park2023generative,
  title={Generative agents: Interactive simulacra of human behavior},
  author={Park, Joon Sung and O'Brien, Joseph and Cai, Carrie Jun and Morris, Meredith Ringel and Liang, Percy and Bernstein, Michael S},
  booktitle={Proceedings of the 36th annual acm symposium on user interface software and technology},
  pages={1--22},
  year={2023}
}

@article{michie2013behavior,
  title={The behavior change technique taxonomy (v1) of 93 hierarchically clustered techniques: building an international consensus for the reporting of behavior change interventions},
  author={Michie, Susan and Richardson, Michelle and Johnston, Marie and Abraham, Charles and Francis, Jill and Hardeman, Wendy and Eccles, Martin P and Cane, James and Wood, Caroline E},
  journal={Annals of behavioral medicine},
  volume={46},
  number={1},
  pages={81--95},
  year={2013},
  publisher={Oxford University Press}
}

@article{bull2020world,
  title={World Health Organization 2020 guidelines on physical activity and sedentary behaviour},
  author={Bull, Fiona C and Al-Ansari, Salih S and Biddle, Stuart and Borodulin, Katja and Buman, Matthew P and Cardon, Greet and Carty, Catherine and Chaput, Jean-Philippe and Chastin, Sebastien and Chou, Roger and others},
  journal={British journal of sports medicine},
  volume={54},
  number={24},
  pages={1451--1462},
  year={2020},
  publisher={BMJ Publishing Group Ltd and British Association of Sport and Exercise Medicine}
}

@article{yang2020factors,
  title={Factors influencing user’s adherence to physical activity applications: A scoping literature review and future directions},
  author={Yang, Xiaotian and Ma, Lin and Zhao, Xi and Kankanhalli, Atreyi},
  journal={International Journal of Medical Informatics},
  volume={134},
  pages={104039},
  year={2020},
  publisher={Elsevier}
}

@article{sallis1988development,
  title={The development of self-efficacy scales for healthrelated diet and exercise behaviors},
  author={Sallis, James F and Pinski, Robin B and Grossman, Robin M and Patterson, Thomas L and Nader, Philip R},
  journal={Health education research},
  volume={3},
  number={3},
  pages={283--292},
  year={1988},
  publisher={Oxford University Press}
}

@article{zahrt2020effects,
  title={Effects of physical activity recommendations on mindset, behavior and perceived health},
  author={Zahrt, Octavia H and Crum, Alia J},
  journal={Preventive medicine reports},
  volume={17},
  pages={101027},
  year={2020},
  publisher={Elsevier}
}

@article{boles2021can,
  title={Can exercising and eating healthy be fun and indulgent instead of boring and depriving? Targeting mindsets about the process of engaging in healthy behaviors},
  author={Boles, Danielle Z and DeSousa, Maysa and Turnwald, Bradley P and Horii, Rina I and Duarte, Taylor and Zahrt, Octavia H and Markus, Hazel R and Crum, Alia J},
  journal={Frontiers in psychology},
  volume={12},
  pages={745950},
  year={2021},
  publisher={Frontiers Media SA}
}

@misc{cdc2022road,
  title={Road to Health Activities Guide},
  author={{Centers for Disease Control and Prevention}},
  year={2022},
  publisher={U.S. Department of Health and Human Services, Centers for Disease Control and Prevention},
  address={Atlanta}
}

@article{brooke1996sus,
  title={SUS-A quick and dirty usability scale},
  author={Brooke, John and others},
  journal={Usability evaluation in industry},
  volume={189},
  number={194},
  pages={4--7},
  year={1996},
  publisher={London, England.}
}

@inproceedings{ekhtiar2023goals,
  title={Goals for goal setting: a scoping review on personal informatics},
  author={Ekhtiar, Tina and Karahano{\u{g}}lu, Arma{\u{g}}an and Gouveia, R{\'u}ben and Ludden, Geke},
  booktitle={Proceedings of the 2023 ACM Designing Interactive Systems Conference},
  pages={2625--2641},
  year={2023}
}

@inproceedings{wang2025exploring,
  title={Exploring Personalized Health Support through Data-Driven, Theory-Guided LLMs: A Case Study in Sleep Health},
  author={Wang, Xingbo and Griffith, Janessa and Adler, Daniel A and Castillo, Joey and Choudhury, Tanzeem and Wang, Fei},
  booktitle={Proceedings of the 2025 CHI Conference on Human Factors in Computing Systems},
  pages={1--15},
  year={2025}
}

@inproceedings{xu2025goals,
  title={From Goals to Actions: Designing Context-aware LLM Chatbots for New Year’s Resolutions},
  author={Xu, Yan and Jones, Brennan and Nguyen, Hannah and Li, Qisheng and Scherer, Stefan},
  booktitle={Proceedings of the 7th ACM Conference on Conversational User Interfaces},
  pages={1--17},
  year={2025}
}

@article{song2025investigating,
  title={Investigating the Relationship Between Physical Activity and Tailored Behavior Change Messaging: Connecting Contextual Bandit with Large Language Models},
  author={Song, Haochen and Hofer, Dominik and Islambouli, Rania and Hawkins, Laura and Bhattacharjee, Ananya and Franklin, Meredith and Williams, Joseph Jay},
  journal={arXiv preprint arXiv:2506.07275},
  year={2025}
}

@article{turk2014multimodal,
  title={Multimodal interaction: A review},
  author={Turk, Matthew},
  journal={Pattern recognition letters},
  volume={36},
  pages={189--195},
  year={2014},
  publisher={Elsevier}
}

@article{oviatt1999ten,
  title={Ten myths of multimodal interaction},
  author={Oviatt, Sharon},
  journal={Communications of the ACM},
  volume={42},
  number={11},
  pages={74--81},
  year={1999},
  publisher={ACM New York, NY, USA}
}

@article{kriston20109,
  title={The 9-item Shared Decision Making Questionnaire (SDM-Q-9). Development and psychometric properties in a primary care sample},
  author={Kriston, Levente and Scholl, Isabelle and H{\"o}lzel, Lars and Simon, Daniela and Loh, Andreas and H{\"a}rter, Martin},
  journal={Patient education and counseling},
  volume={80},
  number={1},
  pages={94--99},
  year={2010},
  publisher={Elsevier}
}

@article{norman2006eheals,
  title={eHEALS: the eHealth literacy scale},
  author={Norman, Cameron D and Skinner, Harvey A},
  journal={Journal of medical Internet research},
  volume={8},
  number={4},
  pages={e507},
  year={2006},
  publisher={JMIR Publications Inc., Toronto, Canada}
}

@inproceedings{consolvo2009theory,
  title={Theory-driven design strategies for technologies that support behavior change in everyday life},
  author={Consolvo, Sunny and McDonald, David W and Landay, James A},
  booktitle={Proceedings of the SIGCHI conference on human factors in computing systems},
  pages={405--414},
  year={2009}
}

@article{mohr2014behavioral,
  title={The behavioral intervention technology model: an integrated conceptual and technological framework for eHealth and mHealth interventions},
  author={Mohr, David C and Schueller, Stephen M and Montague, Enid and Burns, Michelle Nicole and Rashidi, Parisa},
  journal={Journal of medical Internet research},
  volume={16},
  number={6},
  pages={e146},
  year={2014},
  publisher={JMIR Publications Inc. Toronto, Canada}
}

@article{buis2009evaluating,
  title={Evaluating Active U: an Internet-mediated physical activity program},
  author={Buis, Lorraine R and Poulton, Timothy A and Holleman, Robert G and Sen, Ananda and Resnick, Paul J and Goodrich, David E and Palma-Davis, LaVaughn and Richardson, Caroline R},
  journal={BMC public health},
  volume={9},
  number={1},
  pages={331},
  year={2009},
  publisher={Springer}
}

@inproceedings{toscos2006chick,
  title={Chick clique: persuasive technology to motivate teenage girls to exercise},
  author={Toscos, Tammy and Faber, Anne and An, Shunying and Gandhi, Mona Praful},
  booktitle={CHI'06 extended abstracts on Human factors in computing systems},
  pages={1873--1878},
  year={2006}
}

@inproceedings{gobel2010serious,
  title={Serious games for health: personalized exergames},
  author={G{\"o}bel, Stefan and Hardy, Sandro and Wendel, Viktor and Mehm, Florian and Steinmetz, Ralf},
  booktitle={Proceedings of the 18th ACM international conference on Multimedia},
  pages={1663--1666},
  year={2010}
}

@inproceedings{sinclair2007considerations,
  title={Considerations for the design of exergames},
  author={Sinclair, Jeff and Hingston, Philip and Masek, Martin},
  booktitle={Proceedings of the 5th international conference on Computer graphics and interactive techniques in Australia and Southeast Asia},
  pages={289--295},
  year={2007}
}

@article{althoff2016influence,
  title={Influence of Pok{\'e}mon Go on physical activity: study and implications},
  author={Althoff, Tim and White, Ryen W and Horvitz, Eric},
  journal={Journal of medical Internet research},
  volume={18},
  number={12},
  pages={e315},
  year={2016},
  publisher={JMIR Publications Toronto, Canada}
}

@inproceedings{shameli2017gamification,
  title={How gamification affects physical activity: Large-scale analysis of walking challenges in a mobile application},
  author={Shameli, Ali and Althoff, Tim and Saberi, Amin and Leskovec, Jure},
  booktitle={Proceedings of the 26th international conference on world wide web companion},
  pages={455--463},
  year={2017}
}

@inproceedings{murnane2015mobile,
  title={Mobile health apps: adoption, adherence, and abandonment},
  author={Murnane, Elizabeth L and Huffaker, David and Kossinets, Gueorgi},
  booktitle={Adjunct proceedings of the 2015 ACM international joint conference on pervasive and ubiquitous computing and proceedings of the 2015 ACM international symposium on wearable computers},
  pages={261--264},
  year={2015}
}

@article{kumar2024generation,
  title={Generation of Backward-Looking Complex Reflections for a Motivational Interviewing--Based Smoking Cessation Chatbot Using GPT-4: Algorithm Development and Validation},
  author={Kumar, Ash Tanuj and Wang, Cindy and Dong, Alec and Rose, Jonathan},
  journal={JMIR Mental Health},
  volume={11},
  number={1},
  pages={e53778},
  year={2024},
  publisher={JMIR Publications Inc., Toronto, Canada}
}

@inproceedings{steenstra2024virtual,
  title={Virtual agents for alcohol use counseling: Exploring llm-powered motivational interviewing},
  author={Steenstra, Ian and Nouraei, Farnaz and Arjmand, Mehdi and Bickmore, Timothy},
  booktitle={Proceedings of the 24th ACM International Conference on Intelligent Virtual Agents},
  pages={1--10},
  year={2024}
}

@inproceedings{bașar2025well,
  title={How well can large language models reflect? A human evaluation of LLM-generated reflections for motivational interviewing dialogues},
  author={Bașar, Erkan and Sun, Xin and Hendrickx, Iris and de Wit, Jan and Bosse, Tibor and De Bruijn, Gert-Jan and Bosch, Jos A and Krahmer, Emiel},
  booktitle={Proceedings of the 31st International Conference on Computational Linguistics},
  pages={1964--1982},
  year={2025}
}

@article{rose2022generation,
  title={Generation and classification of motivational-interviewing-style reflections for smoking behaviour change using few-shot learning with transformers},
  author={Rose, Jonathan and Ahmed, Imtihan and Keilty, Eric and Cooper, Carolynne and Selby, Peter},
  journal={Authorea Preprints},
  year={2022},
  publisher={Authorea}
}

@inproceedings{sun2025rethinking,
  title={Rethinking the alignment of psychotherapy dialogue generation with motivational interviewing strategies},
  author={Sun, Xin and Tang, Xiao and El Ali, Abdallah and Li, Zhuying and Ren, Pengjie and de Wit, Jan and Pei, Jiahuan and Bosch, Jos A},
  booktitle={Proceedings of the 31st International Conference on Computational Linguistics},
  pages={1983--2002},
  year={2025}
}

@inproceedings{xie2024few,
  title={Few-shot dialogue strategy learning for motivational interviewing via inductive reasoning},
  author={Xie, Zhouhang and Majumder, Bodhisattwa Prasad and Zhao, Mengjie and Maeda, Yoshinori and Yamada, Keiichi and Wakaki, Hiromi and McAuley, Julian},
  booktitle={Findings of the Association for Computational Linguistics: ACL 2024},
  pages={13207--13219},
  year={2024}
}

@article{srinivas2025substance,
  title={Substance over style: Evaluating proactive conversational coaching agents},
  author={Srinivas, Vidya and Xu, Xuhai and Liu, Xin and Ayush, Kumar and Galatzer-Levy, Isaac and Patel, Shwetak and McDuff, Daniel and Althoff, Tim},
  journal={arXiv preprint arXiv:2503.19328},
  year={2025}
}

@article{vardhan2025correction,
  title={Correction: Infusing behavior science into large language models for activity coaching},
  author={Vardhan, Madhurima and Hegde, Narayan and Nathani, Deepak and Rosenzweig, Emily and Speed, Cathy and Karthikesalingam, Alan and Seneviratne, Martin},
  journal={PLOS Digital Health},
  volume={4},
  number={3},
  pages={e0000786},
  year={2025},
  publisher={Public Library of Science San Francisco, CA USA}
}

@article{mantena2025fine,
  title={Fine-tuning LLMs in behavioral psychology for scalable health coaching},
  author={Mantena, Sriya and Johnson, Anders and Oppezzo, Marily and Sch{\"u}tz, Narayan and Tolas, Alexander and Doijad, Ritu and Mattson, C Mikael and Lawrie, Allan and Ramirez-Posada, Mariana and Schmiedmayer, Paul and others},
  journal={NPJ Cardiovascular Health},
  volume={2},
  number={1},
  pages={48},
  year={2025},
  publisher={Nature Publishing Group UK London}
}

@article{choube2025gloss,
  title={GLOSS: Group of LLMs for open-ended sensemaking of passive sensing data for health and wellbeing},
  author={Choube, Akshat and Le, Ha and Li, Jiachen and Ji, Kaixin and Swain, Vedant Das and Mishra, Varun},
  journal={Proceedings of the ACM on Interactive, Mobile, Wearable and Ubiquitous Technologies},
  volume={9},
  number={3},
  pages={1--32},
  year={2025},
  publisher={ACM New York, NY, USA}
}

@article{li2025vital,
  title={Vital Insight: Assisting Experts' Context-Driven Sensemaking of Multi-modal Personal Tracking Data Using Visualization and Human-in-the-Loop LLM},
  author={Li, Jiachen and Li, Xiwen and Steinberg, Justin and Choube, Akshat and Yao, Bingsheng and Xu, Xuhai and Wang, Dakuo and Mynatt, Elizabeth and Mishra, Varun},
  journal={Proceedings of the ACM on Interactive, Mobile, Wearable and Ubiquitous Technologies},
  volume={9},
  number={3},
  pages={1--37},
  year={2025},
  publisher={ACM New York, NY, USA}
}

@inproceedings{subramanian2024graph,
  title={Graph-augmented llms for personalized health insights: A case study in sleep analysis},
  author={Subramanian, Ajan and Yang, Zhongqi and Azimi, Iman and Rahmani, Amir M},
  booktitle={2024 IEEE 20th International Conference on Body Sensor Networks (BSN)},
  pages={1--4},
  year={2024},
  organization={IEEE}
}

@inproceedings{shouborno2025llasa,
  title={LLaSA: A Sensor-Aware LLM for Natural Language Reasoning of Human Activity from IMU Data},
  author={Asif Imran Shouborno, Sheikh and Khan, Mohammad Nur Hossain and Biswas, Subrata and Islam, Bashima},
  booktitle={Companion of the 2025 ACM International Joint Conference on Pervasive and Ubiquitous Computing},
  pages={893--899},
  year={2025}
}

@inproceedings{feli2025llm,
  title={An llm-powered agent for physiological data analysis: A case study on ppg-based heart rate estimation},
  author={Feli, Mohammad and Azimi, Iman and Liljeberg, Pasi and Rahmani, Amir M},
  booktitle={2025 47th Annual International Conference of the IEEE Engineering in Medicine and Biology Society (EMBC)},
  pages={1--7},
  year={2025},
  organization={IEEE}
}

@article{zhang2025sensorlm,
  title={SensorLM: Learning the Language of Wearable Sensors},
  author={Zhang, Yuwei and Ayush, Kumar and Qiao, Siyuan and Heydari, A Ali and Narayanswamy, Girish and Xu, Maxwell A and Metwally, Ahmed A and Xu, Shawn and Garrison, Jake and Xu, Xuhai and others},
  journal={arXiv preprint arXiv:2506.09108},
  year={2025}
}

@article{zheng2025promind,
  title={ProMind-LLM: Proactive Mental Health Care via Causal Reasoning with Sensor Data},
  author={Zheng, Xinzhe and Ji, Sijie and Sun, Jiawei and Chen, Renqi and Gao, Wei and Srivastava, Mani},
  journal={arXiv preprint arXiv:2505.14038},
  year={2025}
}

@article{bak2024potential,
  title={The potential and limitations of large language models in identification of the states of motivations for facilitating health behavior change},
  author={Bak, Michelle and Chin, Jessie},
  journal={Journal of the American Medical Informatics Association},
  volume={31},
  number={9},
  pages={2047--2053},
  year={2024},
  publisher={Oxford University Press}
}

@inproceedings{meywirth2025designing,
  title={Designing for Trust: Integrating Self-referencing in Large Language Model-Based Health Coaching},
  author={Meywirth, Sophia and Janson, Andreas and S{\"o}llner, Matthias},
  booktitle={International Conference on Design Science Research in Information Systems and Technology},
  pages={296--309},
  year={2025},
  organization={Springer}
}

@article{murnane2023narrative,
  title={Narrative-based visual feedback to encourage sustained physical activity: a field trial of the Whoiszuki mobile health platform},
  author={Murnane, Elizabeth L and Glazko, Yekaterina S and Costa, Jean and Yao, Raymond and Zhao, Grace and Moya, Paula ML and Landay, James A},
  journal={Proceedings of the ACM on Interactive, Mobile, Wearable and Ubiquitous Technologies},
  volume={7},
  number={1},
  pages={1--36},
  year={2023},
  publisher={ACM New York, NY, USA}
}

@article{wagstaff2002poverty,
  title={Poverty and health sector inequalities},
  author={Wagstaff, Adam},
  journal={Bulletin of the world health organization},
  volume={80},
  pages={97--105},
  year={2002},
  publisher={SciELO Public Health}
}

@article{michie2011behaviour,
  title={The behaviour change wheel: a new method for characterising and designing behaviour change interventions},
  author={Michie, Susan and Van Stralen, Maartje M and West, Robert},
  journal={Implementation science},
  volume={6},
  number={1},
  pages={42},
  year={2011},
  publisher={Springer}
}

@book{acsm2025,
  title     = "{ACSM's} Guidelines for Exercise Testing and Prescription",
  author    = "American College of Sports Medicine",
  publisher = "Wolters Kluwer Health",
  edition   =  12,
  month     =  apr,
  year      =  2025,
}

@article{webb2010using,
  title={Using the internet to promote health behavior change: a systematic review and meta-analysis of the impact of theoretical basis, use of behavior change techniques, and mode of delivery on efficacy},
  author={Webb, Thomas and Joseph, Judith and Yardley, Lucy and Michie, Susan and others},
  journal={Journal of medical Internet research},
  volume={12},
  number={1},
  pages={e1376},
  year={2010},
  publisher={JMIR Publications Inc., Toronto, Canada}
}

@inproceedings{gouveia2016exploring,
  title={Exploring the design space of glanceable feedback for physical activity trackers},
  author={Gouveia, R{\'u}ben and Pereira, F{\'a}bio and Karapanos, Evangelos and Munson, Sean A and Hassenzahl, Marc},
  booktitle={Proceedings of the 2016 ACM international joint conference on pervasive and ubiquitous computing},
  pages={144--155},
  year={2016}
}

@inproceedings{matthews2007designing,
  author = {Matthews, Tara},
  title = {Designing and evaluating glanceable peripheral displays},
  year = {2006},
  isbn = {1595933670},
  publisher = {Association for Computing Machinery},
  address = {New York, NY, USA},
  url = {https://doi.org/10.1145/1142405.1142457},
  doi = {10.1145/1142405.1142457},
  booktitle = {Proceedings of the 6th Conference on Designing Interactive Systems},
  pages = {343-345},
  numpages = {3},
  location = {University Park, PA, USA},
  series = {DIS '06}
}

@inproceedings{oviatt2004we,
  title={When do we interact multimodally? Cognitive load and multimodal communication patterns},
  author={Oviatt, Sharon and Coulston, Rachel and Lunsford, Rebecca},
  booktitle={Proceedings of the 6th international conference on Multimodal interfaces},
  pages={129--136},
  year={2004}
}

@article{meyer2025llm,
  title={LLM-based conversational agents for behaviour change support: A randomised controlled trial examining efficacy, safety, and the role of user behaviour},
  author={Meyer, Selina and Elsweiler, David},
  journal={International Journal of Human-Computer Studies},
  volume={200},
  pages={103514},
  year={2025},
  publisher={Elsevier}
}

@article{gollwitzer2006implementation,
  title={Implementation intentions and goal achievement: A meta-analysis of effects and processes},
  author={Gollwitzer, Peter M and Sheeran, Paschal},
  journal={Advances in experimental social psychology},
  volume={38},
  pages={69--119},
  year={2006},
  publisher={Elsevier}
}

@article{gollwitzer1999implementation,
  title={Implementation intentions: strong effects of simple plans.},
  author={Gollwitzer, Peter M},
  journal={American psychologist},
  volume={54},
  number={7},
  pages={493},
  year={1999},
  publisher={American Psychological Association}
}

@inproceedings{agapie2018crowdsourcing,
  title={Crowdsourcing exercise plans aligned with expert guidelines and everyday constraints},
  author={Agapie, Elena and Chinh, Bonnie and Pina, Laura R and Oviedo, Diana and Welsh, Molly C and Hsieh, Gary and Munson, Sean},
  booktitle={Proceedings of the 2018 CHI Conference on Human Factors in Computing Systems},
  pages={1--13},
  year={2018}
}

@inproceedings{klasnja2009using,
  title={Using mobile \& personal sensing technologies to support health behavior change in everyday life: lessons learned},
  author={Klasnja, Predrag and Consolvo, Sunny and McDonald, David W and Landay, James A and Pratt, Wanda},
  booktitle={AMIA Annual Symposium Proceedings},
  volume={2009},
  pages={338},
  year={2009}
}

@inproceedings{stepanovic2018gamification,
  title={Gamification applied for health promotion: does it really foster long-term engagement? A scoping review},
  author={Stepanovic, Stefan and Mettler, Tobias},
  booktitle={Proceedings of the 26th European Conference on Information Systems},
  pages={1--16},
  year={2018},
  organization={AIS}
}

@online{anthropic2025affective,
author = {Miles McCain and Ryn Linthicum and Chloe Lubinski and Alex Tamkin and Saffron Huang and Michael Stern and Kunal Handa and Esin Durmus and Tyler Neylon and Stuart Ritchie and Kamya Jagadish and Paruul Maheshwary and Sarah Heck and Alexandra Sanderford and Deep Ganguli},
title = {How People Use Claude for Support, Advice, and Companionship},
date = {2025-06-26},
year = {2025},
url = {https://www.anthropic.com/news/how-people-use-claude-for-support-advice-and-companionship},
}

@article{phang2025investigating,
  title={Investigating affective use and emotional well-being on ChatGPT},
  author={Phang, Jason and Lampe, Michael and Ahmad, Lama and Agarwal, Sandhini and Fang, Cathy Mengying and Liu, Auren R and Danry, Valdemar and Lee, Eunhae and Chan, Samantha WT and Pataranutaporn, Pat and others},
  journal={arXiv preprint arXiv:2504.03888},
  year={2025}
}

@article{laban2025llms,
  title={Llms get lost in multi-turn conversation},
  author={Laban, Philippe and Hayashi, Hiroaki and Zhou, Yingbo and Neville, Jennifer},
  journal={arXiv preprint arXiv:2505.06120},
  year={2025}
}

@article{daskalova2017lessons,
  title={Lessons learned from two cohorts of personal informatics self-experiments},
  author={Daskalova, Nediyana and Desingh, Karthik and Papoutsaki, Alexandra and Schulze, Diane and Sha, Han and Huang, Jeff},
  journal={Proceedings of the ACM on interactive, mobile, wearable and ubiquitous technologies},
  volume={1},
  number={3},
  pages={1--22},
  year={2017},
  publisher={ACM New York, NY, USA}
}

@inproceedings{pan2025developing,
  title={Developing a Social Support Framework: Understanding the Reciprocity in Human-Chatbot Relationship},
  author={Pan, Shuyi and De Graaf, Maartje MA},
  booktitle={Proceedings of the 2025 CHI Conference on Human Factors in Computing Systems},
  pages={1--13},
  year={2025}
}

@article{piercy2018physical,
  title={The physical activity guidelines for Americans},
  author={Piercy, Katrina L and Troiano, Richard P and Ballard, Rachel M and Carlson, Susan A and Fulton, Janet E and Galuska, Deborah A and George, Stephanie M and Olson, Richard D},
  journal={Jama},
  volume={320},
  number={19},
  pages={2020--2028},
  year={2018},
  publisher={American Medical Association}
}

@article{korotitsch1999overview,
  title={An overview of self-monitoring research in assessment and treatment.},
  author={Korotitsch, William J and Nelson-Gray, Rosemery O},
  journal={Psychological Assessment},
  volume={11},
  number={4},
  pages={415},
  year={1999},
  publisher={American Psychological Association}
}

@article{sefidgar2024improving,
  title={Improving work-nonwork balance with data-driven implementation intention and mental contrasting},
  author={Sefidgar, Yasaman S and J{\"o}rke, Matthew and Suh, Jina and Saha, Koustuv and Iqbal, Shamsi and Ramos, Gonzalo and Czerwinski, Mary},
  journal={Proceedings of the ACM on human-computer interaction},
  volume={8},
  number={CSCW1},
  pages={1--29},
  year={2024},
  publisher={ACM New York, NY, USA}
}

@article{heydari2025anatomy,
  title={The Anatomy of a Personal Health Agent},
  author={Heydari, A Ali and Gu, Ken and Srinivas, Vidya and Yu, Hong and Zhang, Zhihan and Zhang, Yuwei and Paruchuri, Akshay and He, Qian and Palangi, Hamid and Hammerquist, Nova and others},
  journal={arXiv preprint arXiv:2508.20148},
  year={2025}
}

@article{jaech2024openai,
  title={Openai o1 system card},
  author={Jaech, Aaron and Kalai, Adam and Lerer, Adam and Richardson, Adam and El-Kishky, Ahmed and Low, Aiden and Helyar, Alec and Madry, Aleksander and Beutel, Alex and Carney, Alex and others},
  journal={arXiv preprint arXiv:2412.16720},
  year={2024}
}

@article{guo2025deepseek,
  title={DeepSeek-R1 incentivizes reasoning in LLMs through reinforcement learning},
  author={Guo, Daya and Yang, Dejian and Zhang, Haowei and Song, Junxiao and Wang, Peiyi and Zhu, Qihao and Xu, Runxin and Zhang, Ruoyu and Ma, Shirong and Bi, Xiao and others},
  journal={Nature},
  volume={645},
  number={8081},
  pages={633--638},
  year={2025},
  publisher={Nature Publishing Group UK London}
}

@article{loerakker2025give,
  title={Give and Take: Perceptions of a Conversational Coach Agent in Fitness Trackers},
  author={Loerakker, Meagan B and Stefanidi, Evropi and Niess, Jasmin and E{\ss}meyer, Thomas and Wo{\'z}niak, Pawe{\l} W},
  journal={Proceedings of the ACM on Human-Computer Interaction},
  volume={9},
  number={5},
  pages={1--36},
  year={2025},
  publisher={ACM New York, NY}
}

@inproceedings{stromel2024narrating,
  title={Narrating fitness: Leveraging large language models for reflective fitness tracker data interpretation},
  author={Str{\"o}mel, Konstantin R and Henry, Stanislas and Johansson, Tim and Niess, Jasmin and Wo{\'z}niak, Pawe{\l} W},
  booktitle={Proceedings of the 2024 CHI Conference on Human Factors in Computing Systems},
  pages={1--16},
  year={2024}
}

@article{braun2006using,
  title={Using thematic analysis in psychology},
  author={Braun, Virginia and Clarke, Victoria},
  journal={Qualitative research in psychology},
  volume={3},
  number={2},
  pages={77--101},
  year={2006},
  publisher={Taylor \& Francis}
}

@article{braun2019reflecting,
  title={Reflecting on reflexive thematic analysis},
  author={Braun, Virginia and Clarke, Victoria},
  journal={Qualitative research in sport, exercise and health},
  volume={11},
  number={4},
  pages={589--597},
  year={2019},
  publisher={Taylor \& Francis}
}

@inproceedings{slovak2024hci,
  title={Hci contributions in mental health: A modular framework to guide psychosocial intervention design},
  author={Slovak, Petr and Munson, Sean A},
  booktitle={Proceedings of the 2024 CHI Conference on Human Factors in Computing Systems},
  pages={1--21},
  year={2024}
}

@inproceedings{saksono2020storywell,
  title={Storywell: designing for family fitness app motivation by using social rewards and reflection},
  author={Saksono, Herman and Castaneda-Sceppa, Carmen and Hoffman, Jessica and Morris, Vivien and Seif El-Nasr, Magy and Parker, Andrea G},
  booktitle={Proceedings of the 2020 CHI conference on human factors in computing systems},
  pages={1--13},
  year={2020}
}

@inproceedings{lee2015personalization,
  title={Personalization revisited: a reflective approach helps people better personalize health services and motivates them to increase physical activity},
  author={Lee, Min Kyung and Kim, Junsung and Forlizzi, Jodi and Kiesler, Sara},
  booktitle={Proceedings of the 2015 ACM International Joint Conference on Pervasive and Ubiquitous Computing},
  pages={743--754},
  year={2015}
}

@inproceedings{niess2018supporting,
  title={Supporting meaningful personal fitness: The tracker goal evolution model},
  author={Niess, Jasmin and Wo{\'z}niak, Pawe{\l} W},
  booktitle={Proceedings of the 2018 CHI conference on human factors in computing systems},
  pages={1--12},
  year={2018}
}

@article{saksono2024socio,
  title={Socio-cognitive framework for personal informatics: A preliminary framework for socially-enabled health technologies},
  author={Saksono, Herman and Parker, Andrea G},
  journal={ACM Transactions on Computer-Human Interaction},
  volume={31},
  number={3},
  pages={1--41},
  year={2024},
  publisher={ACM New York, NY}
}

@article{agapie2022longitudinal,
  title={A longitudinal goal setting model for addressing complex personal problems in mental health},
  author={Agapie, Elena and Are{\'a}n, Patricia A and Hsieh, Gary and Munson, Sean A},
  journal={Proceedings of the ACM on Human-Computer Interaction},
  volume={6},
  number={CSCW2},
  pages={1--28},
  year={2022},
  publisher={ACM New York, NY, USA}
}

@article{czajkowski2015ideas,
  title={From ideas to efficacy: The ORBIT model for developing behavioral treatments for chronic diseases.},
  author={Czajkowski, Susan M and Powell, Lynda H and Adler, Nancy and Naar-King, Sylvie and Reynolds, Kim D and Hunter, Christine M and Laraia, Barbara and Olster, Deborah H and Perna, Frank M and Peterson, Janey C and others},
  journal={Health Psychology},
  volume={34},
  number={10},
  pages={971},
  year={2015},
  publisher={American Psychological Association}
}

@article{laranjo2021smartphone,
  title={Do smartphone applications and activity trackers increase physical activity in adults? Systematic review, meta-analysis and metaregression},
  author={Laranjo, Liliana and Ding, Ding and Heleno, Bruno and Kocaballi, Baki and Quiroz, Juan C and Tong, Huong Ly and Chahwan, Bahia and Neves, Ana Luisa and Gabarron, Elia and Dao, Kim Phuong and others},
  journal={British journal of sports medicine},
  volume={55},
  number={8},
  pages={422--432},
  year={2021},
  publisher={BMJ Publishing Group Ltd and British Association of Sport and Exercise Medicine}
}

%TC:ignore
\newpage

\appendix
\onecolumn

\begin{center}
{\Huge\bfseries Supplementary Materials\par}
\end{center}

\section{LLM Coaching Agent}
\label{appendix:agent}
In this section, we provide further technical and implementation details about our LLM agent's tools.

\subsection{Querying Health Data}
The \texttt{query\_health\_data} function allows the agent to fetch the user's wearable data using Apple's HealthKit API. This function is always available to the agent and has the following function signature:
\begin{flushleft}
\ttfamily
query\_health\_data(\\
\quad sample\_type: string, \\
\quad reference\_date: string = `today',\\
\quad aggregation\_level: enum[`day'|`week'|`month'] = `month',\\
\quad show\_user: boolean = false \\
)
\end{flushleft}
The HealthKit query is executed on device in Swift and returns the aggregated data as text along with a natural language description of the data source to the LLM. If \texttt{show\_user=true}, the data is displayed as a visualization in the chat (see Figure \ref{fig:system}D). We encountered issues with LLMs' ability to reliably perform date arithmetic, so we allow \texttt{reference\_date} to be parsed either as a natural language string (e.g., ``today'', ``last week'') or a date string. Unlike GPTCoach~\cite{jorke2025gptcoach}, we do not include a tool use chain (a prompt chain that forces the model to decide whether to augment its response with the user's health data), which led to many unnecessary data queries and greatly increased latency.

\subsection{Structured Plan Generation}
During onboarding and check-in chats, we expose the \texttt{generate\_plan} function to the agent. This function generates a weekly plan using an external prompt that takes the current conversation history and the user's plan history as input and produces a structured JSON object. After generating the plan as JSON, it is displayed to the user as a plan widget in the chat (see Figure \ref{fig:additional-screens}D; Figure \ref{fig:system}D) and saved to our database such that it can be rendered and modified in the app's UI. The plan generation JSON schema and prompt are provided in our GitHub repository.

The plan generation prompt includes guidelines for creating a well-rounded, stage-appropriate, and personalized exercise plan sourced from the Active Choices program and the Centers for Disease Control and Prevention's Physical Activity Guidelines~\cite{cdc2022}. 
The agent is instructed to call the \texttt{generate\_plan} function during the goal setting dialogue state. The preceding dialogue state prompts include explicit instructions to solicit relevant information and preferences from the user (e.g., goals, resources, injuries, past experience, preferred activities, schedule), ensuring that the goal setting process is collaborative, not prescriptive, and that the conversation history contains sufficient context for the \texttt{generate\_plan} function to generate a personalized plan. 
We do not allow the agent to advance out of the goal setting dialogue state until the \texttt{generate\_plan} function has been called. In the absence of this manual check, we encountered hallucination issues with the agent telling the user it would generate a plan without calling the underlying function.

Moreover, we expose \texttt{add\_workout} and \texttt{delete\_workout} functions to the at-will chat agent, allowing the at-will agent to make direct edits to the user's current plan via conversational interaction. We do not provide the at-will agent with access to the \texttt{generate\_plan} function---early tests indicated that hallucination and tool use errors could lead to highly destructive edits or deletions. Making small, granular edits to the current plan proved to be more reliable than generating a replacement plan directly.

\section{Ambient Display Logic}
\label{appendix:ambient-display}

Our ambient display encodings and logic were strongly inspired by UbiFit~\cite{consolvo2008activity, consolvo2008designing} and modified to match our four-week study progression.
The ambient display advances (i.e., a flower in the garden grows) as the user completes workouts in their weekly plan in 20\% increments. 
For example, a flower grows if completing a workout advances the plan progress from 30\% to 40\% completion, but not from 20\% to 30\%.
The flower fully blooms when the weekly plan is 100\% completed. If the user completes their plan in week $t$, the flower remains in the garden and a new flower starts growing in week $t+1$. If they do not complete their plan in week $t$, the flower starts growing again from scratch in week $t+1$. In addition to a fully bloomed flower, we add additional elements (week 2: bird on a branch; week 3: beehive on a branch; week 4: bird and birdhouse) to the display.

For each completed workout, a critter is drawn to the ambient display above the growing flowers. Users receive a bee for walking and a butterfly for other activities, with different colored butterflies for different activity types (red: cardio, orange: strength, green: team sports, yellow: flexibility \& dance, blue: outdoor recreation, purple: misc). A small (< 15min), medium (15-30 min), or large (30+ min) critter is drawn depending on the exercise duration. To avoid cluttering the display, critters are reset each week. After the final screen (week 4, 100\%), the ambient display remains fixed in subsequent weeks, but critters continue to appear. 

\section{Field Study}
\label{appendix:study}
% \subsection{Recruitment}
% 2,397 completed the screener, 342 were eligible, 188 were selected, and 56 enrolled. Two participants unenrolled after the first week due to non-compliance.

\subsection{Statistical Analysis}
\label{appendix:stats}

Let $Y_{i,w,t}$ represent the observed outcome (step count, energy burned, exercise time, or distance walking/running) for participant $i$, measured on day $t \in \{1, \dots, 7\}$ within study week $w \in \{0, \dots, 4\}$ (week 0 aggregates all baseline weeks). Define an indicator variable for the study period, $S_{w} = \mathbf{1}\{w > 0\}$, and let $T_i$ denote treatment assignment, where $T_i = 1$ for participants in the LLM condition and $T_i=0$ for control. Let $\boldsymbol{D}_{t}$ be an indicator for the weekday (Monday–Saturday, with Sunday as reference). We specify the following three-level linear mixed-effects model, with days ($t$) nested within weeks ($w$) nested within participants ($i$):

\begin{align*}
Y_{i,w,t} &= 
\underbrace{\beta_0}_{\text{baseline mean}} + 
\underbrace{\beta_1 \cdot S_{w}}_{\text{H1: baseline–study period difference}} + 
\underbrace{\beta_2 \cdot S_{w}\cdot T_i}_{\text{H2: treatment–control difference in change}} \\
&\quad + \underbrace{\beta_3 \cdot S_{w} \cdot w}_{\text{overall weekly trend}} +
\underbrace{\beta_4 \cdot S_{w}\cdot  T_i \cdot w}_{\text{H3: treatment–control difference in weekly trend}} \\
&\quad + \underbrace{\boldsymbol{\gamma}^{\!\top}\boldsymbol{D}_{t}}_{\text{day-of-week effects}}
+ \underbrace{b_{0,i} + b_{0,i,w}}_{\substack{\text{random intercepts for}\\\text{participant and participant–week}}}
+ \quad \varepsilon_{i,w,t}
\end{align*}
Random intercepts $b_{0,i} \sim \mathcal{N}(0, \tau_{\text{person}}^2)$ capture participant-level differences in mean activity.  
Random intercepts $b_{0,i,w} \sim \mathcal{N}(0, \tau_{\text{week}}^2)$ capture week-to-week deviations within each participant.  
We additionally allow for heteroskedasticity across participants by modeling the residuals as $\varepsilon_{i,w,t} \sim \mathcal{N}(0, \sigma_i^2)$, where $\sigma_i^2$ is a participant-specific variance.

In this model, hypothesis tests correspond directly to fixed effects:
coefficient $\beta_1$ tests whether average physical activity differs significantly between the baseline and study period (H1); coefficient $\beta_2$ assesses if the magnitude of this difference varies significantly between the treatment and control groups (H2); and coefficient $\beta_4$ evaluates whether the weekly rate of change (persistence) in activity over the study period differs significantly between groups (H3). 

To account for systematic data missingness (e.g., due to privacy permissions, data upload errors, or participants forgetting to wear their watch), we excluded participants with fewer than 30 baseline days or with three or more missing study days from our analysis. This yielded a sample of 54 participants for both step count and distance walking/running (6,348 observations), 41 participants for energy burned (4,615 observations), and 36 participants for exercise time analyses (3,811 observations). Notably, relaxing our exclusion criteria and including all participants in our model does not change the directionality or significance of our effects.

We explored a variety of models varying in complexity, from two-sample \emph{t}-tests and standard repeated-measures ANOVA, to more sophisticated mixed-effects models including multiple levels, random slopes, and nonlinear time trends. We also examined alternative coding schemes (e.g., treating weeks or days as factors) and aggregating data at the weekly rather than daily level.  
We selected this final model because it provided the best statistical fit to the data (lowest AIC and BIC), leveraged daily observations to maximize statistical power, and avoided the strong assumption that treatment effects vary linearly at the daily level. Instead, the model captures treatment effects as linear weekly trends, aligning well with our hypothesis concerning persistence (H3). Additionally, this approach explicitly accounts for heteroskedasticity by allowing participant-specific residual variances, acknowledging that observed changes in activity are more noteworthy for participants whose baseline behavior was less variable.
Importantly, the patterns of significance, directionality of effects, and substantive conclusions drawn from our results remained consistent across all modeling approaches tested.

\subsection{Power Considerations}
\label{appendix:power}

Given the exploratory nature of our study, we did not conduct a formal a priori power analysis. To aid in the interpretation of our quantitative results, we performed several post hoc analyses.
First, we conduct a \textit{sensitivity analysis} to estimate the sample sizes required to achieve 80\% power at a significance level of $\alpha = .05$ for a range of plausible effect sizes. We consider a small effect ($d=0.3$) commonly observed in meta-analyses of mHealth interventions~\cite{yang2019comparative, laranjo2021smartphone}, as well as medium ($d=0.5)$ and large ($d=0.8$) effects that might be expected in early-stage work. Second, given our actual sample size ($N=54$, assuming an equal split across conditions), we compute the \textit{minimally detectable effect size}. 

We use standard $t$-test approximations for both analyses because a full simulation of a three-level mixed-effects model would require specifying a complete generative model for all variance components, covariance structures, and missingness patterns. Instead, we opted for more transparent calculations that rely on fewer assumptions, treating H1 as a paired $t$-test and H2 as a two-sample $t$-test. This approximation provides an upper bound on the required sample sizes, as our actual mixed-effects model leverages repeated daily observations and partial pooling across weeks, which typically increases statistical efficiency. We exclude H3 because power for slope-difference interactions would similarly require full simulations of the multilevel model, and, given that higher-order interactions are generally harder to detect than mean differences (H2), would be expected to require an even larger sample. We use the R \texttt{pwr} package for our calculations.

Results indicate that detecting a small effect ($d = 0.3$) would require at least 90 participants for H1 and at least 352 participants for H2. A moderate effect ($d = 0.5$) would require at least 34 participants for H1 and at least 128 participants for H2. A large effect ($d=0.8$) would require at least 15 participants for H1 and at least 52 participants for H2.
Given our actual sample ($N = 54$), the minimally detectable effect sizes were $d \geq 0.39$ for H1 and $d \geq 0.78$ for H2. Thus, the study had adequate power to detect moderate or large within-person improvements (H1) but was underpowered to detect small or moderate between-condition effects (H2/H3). We observed consistently large within-person improvements for H1 ($d = 0.59$–$0.75$) and small between-condition and persistence effects for H2 and H3 (including small negative effects, $d = -0.22$–$0.23$), indicating that H1 had sufficient power but H2 and H3 were underpowered.

\subsection{Supplemental Analysis: Decomposing H1 into Within-Condition Effects}
In the main text, H1 is evaluated as an aggregate effect across both conditions, testing whether the study period showed meaningful improvements over the pre-study baseline regardless of condition. In this section, we decompose H1 into separate within-condition tests (H1a/H1b) to assess whether each group shows meaningful pre-post improvement.

\begin{itemize}
    \item[\textbf{H1a:}] Mean PA levels in the \emph{control group} will exceed their baseline levels during the study period.
    \item[\textbf{H1b:}] Mean PA levels in the \emph{treatment group} will exceed their baseline levels during the study period.
    \item[\textbf{H2:}] The treatment group's increase in mean PA levels from baseline to the study period will exceed that of the control group.
    \item[\textbf{H3:}] The treatment group will exhibit a smaller rate of decline (or greater persistence) in PA levels across the four-week study period compared to the control group.
\end{itemize}

\noindent
This decomposition allows H1b to provide early-stage evidence of a meaningful improvement in the treatment group, analogous to proof-of-concept evidence in the ORBIT model~\cite{czajkowski2015ideas}. We emphasize, however, that our study was not designed as a proof-of-concept trial and its primary aim was to surface design insights. We did not revise the hypothesis structure of the main text, as this framing reflects a post hoc interpretive consideration rather than the basis for our original study design.

These tests were conducted using the same three-level linear mixed-effects model described in \ref{appendix:stats}, except that H1a and H1b were fit separately within each condition with treatment interaction terms ($\beta_2$ and $\beta_4$) removed. In Table~\ref{tab:hk-poc-results}, we report coefficient means, standard errors, and significance levels for each hypothesis, analogous to Table~\ref{tab:hk-results} in the main text. Across all four PA outcomes (step count, energy burned, distance walked/run, and exercise time), both H1a and H1b were significant ($p < 0.01$), with meaningful improvements in the treatment condition (e.g., +994 steps/day or 81.9 minutes of exercise per week). 

\begin{table}[h!]
\centering
\small
\begin{tabularx}{\textwidth}{@{}p{0.14\textwidth} c X l@{}}
\toprule
\textbf{PA Outcome} & \textbf{Hyp.} & \textbf{Description} & \textbf{Estimate (SE)} \\
\midrule

Step Count 
    & \textbf{H1a} 
    & Difference in daily step count from baseline to study period (control group) 
    & \textbf{+1\,648\,(281)}***\\
    & \textbf{H1b} 
    & Difference in daily step count from baseline to study period (treatment group) 
    & \textbf{+994\,(337)}**\\
    & H2 
    & Treatment–control difference in daily step count change from baseline to study period 
    & –692\,(432) \\
    & H3 
    & Treatment–control difference in the weekly rate of change ($\Delta$ steps/day per study week) 
    & +229\,(137) \\
\midrule

Active Energy
    & \textbf{H1a} 
    & Difference in daily kcal burned from baseline to study period (control group) 
    & \textbf{+87.8\,(23.9)}*** \\
Burned (kcal) & \textbf{H1b} 
    & Difference in daily kcal burned from baseline to study period (treatment group) 
    & \textbf{+79.9\,(20.9)}*** \\
    & H2 
    & Treatment–control difference in daily kcal burned change from baseline to study period 
    & –7.09\,(31.3) \\
    & H3 
    & Treatment–control difference in the weekly rate of change ($\Delta$ kcal/day per study week) 
    & +7.80\,(9.82) \\
\midrule

Exercise Time (min) 
    & \textbf{H1a} 
    & Difference in daily exercise min from baseline to study period (control group) 
    & \textbf{+12.6\,(3.74)}**\\
    & \textbf{H1b} 
    & Difference in daily exercise min from baseline to study period (treatment group) 
    & \textbf{+11.7\,(3.49)}**\\
    & H2 
    & Treatment–control difference in daily exercise min change from baseline to study period 
    & –2.08\,(4.95) \\
    & H3 
    & Treatment–control difference in the weekly rate of change ($\Delta$ min/day per study week) 
    & +1.51\,(1.57) \\
\midrule

Distance Walking/  
    & \textbf{H1a} 
    & Difference in daily distance walked/run (km) from baseline to study period (control group) 
    & \textbf{+0.753\,(0.139)}*** \\
Running (km) & \textbf{H1b} 
    & Difference in daily distance walked/run (km) from baseline to study period (treatment group) 
    & \textbf{+0.519\,(0.144)}*** \\
    & H2 
    & Treatment–control difference in daily distance walked/run (km) change from baseline to study period 
    & –0.239\,(0.198) \\
    & H3 
    & Treatment–control difference in the weekly rate of change ($\Delta$ km/day per study week) 
    & +0.0712\,(0.0625) \\
\bottomrule
\end{tabularx}
\caption{\textbf{Within-Condition Wearable Data (Quantitative) Results.}  
We report mean (SE) parameter estimates from models fit separately within each condition (H1a/b) and from the combined model (H2/H3).  
*** denotes $p < 0.001$ and ** denotes $p < 0.01$, where $p$-values are Holm-adjusted for multiple comparisons.  
Significant results are \textbf{bolded}.}
\label{tab:hk-poc-results}
\end{table}

\clearpage
\newpage
\subsection{Survey Measures}

\begin{table}[htbp]
\begin{tabular}{@{\hspace{1.5em}}l@{\hspace{1.5em}}}
\toprule
\multicolumn{1}{@{\hspace{0.5em}}l}{\textit{Physical Activity  \& Health: General Assessment}}\\
In general, would you say your health is …\\
How important is health to you personally?\\
In general, how would you rate your physical fitness?\\
During the last 7 days, how much exercise did you get?\\[4pt]

\multicolumn{1}{@{\hspace{0.5em}}l}{\textit{Physical Activity  \& Health: Satisfaction}}\\
How satisfied are you with your current level of physical activities?\\
How satisfied are you with your current physical health?\\
How satisfied are you with your current mental health?\\
How satisfied are you with your current life situation in general?\\[4pt]

\multicolumn{1}{@{\hspace{0.5em}}l}{\textit{Physical Activity  \& Health: Motivation}}\\
How much do you intend to increase your current level of physical activity?\\
How motivated are you to adopt a healthier lifestyle?\\[2pt]
\midrule

\multicolumn{1}{@{\hspace{0.5em}}l}{\textit{User Experience \& Advice Quality (modified from~\cite{jorke2025gptcoach}})}\\
I received personalized physical activity advice.\\
I received actionable physical activity advice.\\
I received generic physical activity advice.\\
I felt comfortable sharing my concerns.\\
I felt supported in my physical activity.\\
I feel capable of overcoming challenges.\\
I feel more motivated to change.\\
I was asked for my opinion about what activities I would like to do.\\
I felt like my unique situation and concerns were understood.\\
I received unsolicited advice.\\
The system was empathetic.\\
The system used my data in a way that was relevant.\\
The system helped me identify obstacles to engaging in physical activity.\\
The system helped me reflect on what motivates me to be physically active.\\
The system helped make my own ideas about how to increase my physical activity more specific.\\
Interacting with the system provided me with new insights about my physical activity.\\[4pt]

\multicolumn{1}{@{\hspace{0.5em}}l}{\textit{Anthropomorphism (Post-only)}}\\
Interacting with the app felt like interacting with a human. \\
Interacting with the app made me feel like my wellbeing was cared for.\\
Beebo cared about my wellbeing. \\
\bottomrule
\end{tabular}
\caption{Full wording of every custom question included in the onboarding/offboarding survey.}
\label{table:custom-questions}
\end{table}

\begin{table*}
\centering
\footnotesize
\setlength{\tabcolsep}{2pt} 
\begin{tabular*}{\textwidth}{@{\extracolsep{\fill}} p{0.36\textwidth} rrr rrr r}
\toprule
& \multicolumn{3}{c}{\textbf{Treatment (LLM)}}
& \multicolumn{3}{c}{\textbf{Control (no LLM)}}
& \\
\cmidrule(lr){2-4}\cmidrule(lr){5-7}
Measure & Pre \phantom{aaa} 
    & Post \phantom{aaa} 
    & $\Delta_\text{T}$ \phantom{aaa}
    & Pre \phantom{aaa} 
    & Post \phantom{aaa} 
    & $\Delta_\text{C}$ \phantom{aaa}
    & $\Delta_\text{T} - \Delta_\text{C}$ \\
\midrule
Stage of Change~\cite{marcus1992self}
\textit{(1: pre-con., 5: maint.)}
    & 1.88 (0.59) & 2.85 (0.73) & 0.96 (0.96) & 1.89 (0.74) & 2.61 (1.07) & 0.71 (1.01) & 0.25 \\
IPAQ~\cite{craig2003international} \textit{(MET-min)}
    & 415.31 (329.00) & 749.73 (467.88) & 334.42 (505.20) & 548.18 (637.10) & 893.96 (741.87) & 345.79 (965.50) & -11.36 \\
IPAQ~\cite{craig2003international} \textit{(1: Low, 3: High)}
    & 1.27 (0.45) & 1.77 (0.51) & 0.50 (0.65) & 1.25 (0.44) & 1.68 (0.67) & 0.43 (0.79) & 0.07 \\
Self-Efficacy~\cite{sallis1988development}
    & 2.83 (0.67) & 2.82 (0.88) & -0.01 (0.93) & 2.61 (0.76) & 2.63 (0.86) & 0.02 (0.99) & -0.03 \\
Adequacy Mindset~\cite{zahrt2020effects} 
    & 2.39 (0.94) & 3.59 (0.74) & 1.21 (0.78) & 2.59 (0.85) & 3.35 (0.93) & 0.76 (0.85) & 0.44 \\
Process Mindset~\cite{boles2021can} 
    & 2.71 (0.61) & 3.04 (0.77) & 0.33 (0.48) & 2.62 (0.82) & 2.84 (0.74) & 0.22 (0.48) & 0.11 \\
Barriers to Being Active~\cite{cdc2022road} 
    & 2.59 (0.56) & 2.35 (0.63) & -0.24 (0.36) & 2.64 (0.62) & 2.38 (0.71) & -0.27 (0.64) & 0.02 \\
SUS~\cite{brooke1996sus} \textit{(0–100)}
    & 93.85 (8.89) & 82.21 (20.85) & -11.63 (18.64) & 93.75 (8.29) & 89.02 (8.91) & -4.73 (8.80) & -6.90 \\
SASSI~\cite{hone2000towards} 
    & 4.40 (0.43) & 4.07 (0.75) & -0.33 (0.63) & 4.38 (0.36) & 4.25 (0.44) & -0.13 (0.46) & -0.20 \\
TSRI~\cite{bentvelzen2021development} 
    & 3.80 (0.58) & 3.85 (0.86) & 0.06 (0.77) & 3.93 (0.53) & 3.61 (0.92) & -0.33 (0.80) & 0.38 \\
eHealth Literacy~\cite{norman2006eheals} 
    & 4.46 (0.50) & 4.35 (0.82) & -0.11 (0.80) & 4.44 (0.53) & 4.59 (0.52) & 0.15 (0.39) & -0.26 \\
(Adapted) SDM-Q-9~\cite{kriston20109} 
    & 4.06 (0.64) & 4.18 (0.79) & 0.12 (0.46) & 3.75 (0.85) & 3.48 (0.92) & -0.26 (0.69) & 0.38 \\[2pt]
\midrule
\multicolumn{8}{@{}l}{\textit{Physical Activity \& Health – General Assessment (Custom)}} \\[0.2em]
Overall Average 
    & 2.93 (0.47) & 3.31 (0.53) & 0.38 (0.39) & 2.69 (0.53) & 3.01 (0.60) & 0.32 (0.46) & 0.06 \\
\quad General Health Assessment
    & 2.81 (0.69) & 3.15 (0.61) & 0.35 (0.49) & 2.75 (0.80) & 2.93 (0.81) & 0.18 (0.55) & 0.17 \\
\quad Health Importance
    & 4.04 (0.60) & 4.15 (0.61) & 0.12 (0.52) & 3.71 (0.76) & 3.86 (0.85) & 0.14 (0.71) & -0.03 \\
\quad Physical Fitness Rating
    & 2.12 (0.77) & 2.38 (0.80) & 0.27 (0.67) & 1.96 (0.84) & 2.25 (0.84) & 0.29 (0.71) & -0.02 \\
\quad Exercise Past 7 Days
    & 2.69 (0.71) & 3.62 (0.70) & 0.92 (0.91) & 2.19 (0.66) & 3.00 (0.85) & 0.81 (0.91) & 0.11 \\
\midrule
\multicolumn{8}{@{}l}{\textit{Physical Activity \& Health – Satisfaction (Custom)}}\\[0.2em]
Overall Average 
    & 2.15 (0.42) & 2.64 (0.49) & 0.49 (0.39) 
    & 2.19 (0.50) & 2.53 (0.57) & 0.33 (0.40) & 0.15 \\
\quad Satisfaction: PA Level
    & 1.82 (0.81) & 3.23 (1.01) & 1.41 (0.93) 
    & 1.95 (0.71) & 2.98 (1.00) & 1.02 (0.97) & 0.39 \\
\quad Satisfaction: Physical Health
    & 2.28 (0.96) & 3.15 (1.04) & 0.87 (1.06) 
    & 2.52 (1.10) & 3.17 (1.05) & 0.64 (0.93) & 0.23 \\
\quad Satisfaction: Mental Health
    & 3.38 (0.93) & 3.82 (0.99) & 0.44 (0.80) 
    & 3.36 (1.11) & 3.50 (1.21) & 0.14 (0.91) & 0.29 \\
\quad Satisfaction: Life Situation
    & 3.44 (0.94) & 3.64 (0.95) & 0.21 (0.77) 
    & 3.33 (1.10) & 3.52 (1.03) & 0.19 (0.44) & 0.01 \\
\midrule
\multicolumn{8}{@{}l}{\textit{Physical Activity \& Health – Motivation (Custom)}}\\[0.2em]
Overall Average 
    & 3.50 (0.57) & 3.48 (0.70) & -0.02 (0.61) & 3.18 (0.78) & 3.38 (0.82) & 0.20 (0.72) & -0.22 \\
\quad Intention to Increase PA
    & 3.31 (0.62) & 3.23 (0.71) & -0.08 (0.69) & 3.04 (0.79) & 3.07 (0.86) & 0.04 (1.04) & -0.11 \\
\quad Motivation Healthier Lifestyle
    & 3.69 (0.68) & 3.73 (0.83) & 0.04 (0.72) & 3.32 (1.02) & 3.68 (1.09) & 0.36 (0.73) & -0.32 \\
\midrule
\multicolumn{8}{@{}l}{\textit{User Experience \& Advice Quality (drawn from~\cite{jorke2025gptcoach})}}\\
Overall Average 
    & 4.40 (0.47) & 4.04 (0.79) & -0.37 (0.52) & 4.12 (0.54) & 3.63 (0.74) & -0.50 (0.74) & 0.13 \\
\quad Actionable Advice 
    & 4.81 (0.40) & 4.42 (1.03) & -0.38 (0.80) & 4.71 (0.53) & 4.07 (0.98) & -0.64 (1.03) & 0.26 \\
\quad Personalized Advice 
    & 4.58 (0.58) & 4.23 (0.99) & -0.35 (1.02) & 3.82 (0.98) & 4.00 (1.25) & 0.18 (1.33) & -0.52 \\
\quad Generic Advice
    & 2.19 (1.17) & 2.58 (1.17) & 0.38 (1.50) & 2.79 (0.99) & 2.79 (1.17) & 0.00 (1.22) & 0.38 \\
\quad Comfortable Sharing Concerns
    & 4.73 (0.53) & 4.23 (0.99) & -0.50 (0.76) & 4.75 (0.52) & 3.96 (0.92) & -0.79 (0.99) & 0.29 \\
\quad Feel Supported
    & 4.46 (0.76) & 4.23 (0.99) & -0.23 (0.91) & 4.39 (0.69) & 4.14 (0.97) & -0.25 (0.93) & 0.02 \\
\quad Feel Capable of Overcoming Challenges
    & 4.35 (0.69) & 3.85 (1.19) & -0.50 (0.95) & 4.11 (0.88) & 3.54 (1.17) & -0.57 (1.20) & 0.07 \\
\quad Feel Motivated to Change
    & 4.54 (0.58) & 3.96 (1.37) & -0.58 (1.14) & 4.46 (0.64) & 3.89 (1.10) & -0.57 (1.32) & -0.01 \\
\quad Asked for my Opinion
    & 5.00 (0.00) & 4.81 (0.40) & -0.19 (0.40) & 4.46 (0.84) & 3.46 (1.57) & -1.00 (1.78) & 0.81 \\
\quad Understood Unique Situation \& Concerns
    & 4.38 (0.75) & 3.88 (1.24) & -0.50 (1.30) & 3.43 (1.14) & 2.71 (0.98) & -0.71 (1.54) & 0.21 \\
\quad Unsolicited Advice
    & 1.42 (0.81) & 1.96 (1.11) & 0.54 (1.33) & 2.18 (1.33) & 1.86 (1.08) & -0.32 (1.61) & 0.86 \\
\quad Empathetic
    & 4.23 (0.99) & 4.12 (0.99) & -0.12 (0.82) & 3.54 (1.04) & 2.79 (1.03) & -0.75 (1.24) & 0.63 \\
\quad Used Data in a Relevant Way
    & 4.65 (0.63) & 4.27 (0.96) & -0.38 (0.98) & 4.50 (0.88) & 4.36 (0.78) & -0.14 (1.11) & -0.24 \\
\quad Helped Identify Obstacles
    & 4.23 (0.91) & 3.73 (1.12) & -0.50 (1.07) & 4.36 (0.95) & 3.11 (1.31) & -1.25 (1.43) & 0.75 \\
\quad Helped Reflect on Motivation
    & 4.08 (1.09) & 3.65 (1.20) & -0.42 (1.39) & 4.57 (0.74) & 3.43 (1.32) & -1.14 (1.18) & 0.72 \\
\quad Helped Make Own Ideas More Specific
    & 4.35 (1.06) & 3.92 (1.13) & -0.42 (1.03) & 4.07 (0.94) & 3.68 (1.19) & -0.39 (1.17) & -0.03 \\
\quad Provided New Insights
    & 3.65 (1.16) & 3.81 (1.13) & 0.15 (1.19) & 3.79 (1.17) & 3.54 (1.35) & -0.25 (1.29) & 0.40 \\
\midrule
\multicolumn{8}{@{}l}{\textit{Anthropomorphism (Post-only, Custom)}}\\[0.2em]
Overall Average 
    & — & 3.51 (1.17) & — & — & 2.83 (0.98) & — & —\\
\quad Human-Like Interaction
    & — & 2.88 (1.37) & — & — & 1.89 (1.20) & — & —\\
\quad Wellbeing Cared For
    & — & 3.85 (1.22) & — & — & 3.25 (1.21) & — & —\\
\quad Beebo Cared
    & — & 3.81 (1.41) & — & — & 3.36 (1.06) & — & —\\

\bottomrule
\vspace{0pt}
\end{tabular*}
\caption{\textbf{Pre/Post Survey Results.} We report mean (SD) survey measures for participants in the treatment (LLM) and control (no LLM) conditions for pre- and post-study surveys. $\Delta$ refers to the mean (SD) of participants' post-pre difference in scores within each condition while $\Delta_\text{T} - \Delta_\text{C}$ refers to the treatment-control difference in post-pre differences. For all measures except stage of change, IPAQ, and SUS, responses are standardized to fall on a 5pt-Likert scale (1: Strongly disagree; 5: Strongly agree). For all surveys except Barriers to Being Active (where lower scores indicate lower perceived barriers to activity), responses are coded such that higher scores are more desirable. \\
}
\label{table:pre-post-survey-results}
\end{table*} 
\begin{table*}
\centering
\footnotesize
\setlength{\tabcolsep}{2pt}
\begin{tabular*}{\textwidth}{@{\extracolsep{\fill}} p{0.42\textwidth} rrrrr rrrrr}
\toprule
& \multicolumn{5}{c}{\textbf{Treatment (LLM)}} 
& \multicolumn{5}{c}{\textbf{Control (no LLM)}} \\
\cmidrule(lr){2-6}\cmidrule(lr){7-11}
\textbf{Question} 
& Mean & Day 1 & Day 28 & $\beta_\text{T}$ & $\Delta_\text{T}$
& Mean & Day 1 & Day 28 & $\beta_\text{C}$ & $\Delta_\text{C}$ \\
\midrule

I am satisfied with my current level of physical activity.
& 3.17 (1.09) & 2.59 (1.18) & 3.45 (1.01) & 0.022 & 0.62 
& 3.06 (1.07) & 2.71 (1.10) & 2.91 (1.16) & 0.006 & 0.17 \\[0.2em]

I am a physically active person. 
& 3.15 (1.03) & 2.77 (1.02) & 3.32 (0.95) & 0.030 & 0.84
& 3.02 (1.05) & 2.53 (0.87) & 3.00 (1.00) & 0.009 & 0.25 \\[0.2em]

I am committed to my physical activity goal. 
& 4.27 (0.66) & 4.23 (0.53) & 4.09 (0.81) & -0.007 & -0.21
& 3.96 (0.84) & 3.82 (0.73) & 3.91 (0.90) & -0.000 & -0.01 \\[0.2em]

How hopeful are you about your physical health today? 
& 4.07 (0.70) & 3.91 (0.68) & 4.14 (0.71) & 0.005 & 0.14
& 3.84 (0.85) & 3.88 (0.60) & 3.87 (0.87) & -0.003 & -0.09 \\[0.2em]

How was your mood today? (unpleasant/pleasant)
& 3.93 (0.80) & 4.14 (0.64) & 4.32 (0.84) & 0.004 & 0.12
& 3.91 (0.83) & 4.00 (0.71) & 3.96 (0.77) & 0.002 & 0.06 \\

\bottomrule
\end{tabular*}
\caption{\textbf{Daily Survey Results.} We report mean (SD) values by condition across all days, for day 1, and for day 28.
$\beta$ is the per-day slope from a linear regression of score on day. 
$\Delta$ is the predicted 28-day change computed as $28 \times \beta$.
All responses map to a 5-point Likert scale.}
\label{table:daily-survey-results}
\end{table*}
\begin{table*}
\centering
\footnotesize
\setlength{\tabcolsep}{2pt}
\begin{tabular*}{\textwidth}{@{\extracolsep{\fill}} p{0.42\textwidth} rrrrr rrrrr}
\toprule
& \multicolumn{5}{c}{\textbf{Treatment (LLM)}} 
& \multicolumn{5}{c}{\textbf{Control (no LLM)}} \\
\cmidrule(lr){2-6}\cmidrule(lr){7-11}
\textbf{Question} 
& Mean & Week 1 & Week 4 & $\beta_\text{T}$ & $\Delta_\text{T}$
& Mean & Week 1 & Week 4 & $\beta_\text{C}$ & $\Delta_\text{C}$ \\
\midrule

Overall App Rating
& 3.84 (0.74) & 3.90 (0.67) & 3.98 (0.81) & -0.005 & -0.02
& 4.02 (0.69) & 3.94 (0.68) & 4.02 (0.68) & 0.040 & 0.16 \\[0.2em]

Visual Wallpaper Rating
& 3.78 (1.12) & 3.70 (1.18) & 3.93 (1.22) & 0.077 & 0.31
& 3.94 (0.94) & 3.73 (0.98) & 4.14 (0.81) & 0.153 & 0.61 \\[0.2em]

Exercising is ... (difficult/easy)
& 3.11 (0.82) & 2.92 (0.78) & 3.35 (1.11) & 0.139 & 0.56
& 2.93 (1.00) & 2.90 (1.17) & 2.87 (0.98) & 0.024 & 0.10 \\[0.2em]

Exercising is ... (boring/fun)
& 3.17 (0.91) & 3.13 (0.86) & 3.27 (1.03) & 0.045 & 0.18
& 2.96 (1.06) & 2.81 (1.10) & 2.87 (1.10) & 0.056 & 0.22 \\[0.2em]

Barriers To Being Active (4 item subset)
& 2.54 (0.70) & 2.51 (0.74) & 2.53 (0.71) & -0.009 & -0.04
& 2.79 (0.85) & 2.92 (0.86) & 2.60 (0.96) & -0.059 & -0.24 \\

\bottomrule
\end{tabular*}
\caption{\textbf{Weekly Survey Results.} We report mean (SD) values by condition across all weeks, for week 1, and for week 4. 
$\beta$ is the per-week slope from a linear regression of score on week. 
$\Delta$ is the predicted four-week change, computed as $4 \times \beta$. 
All responses were rescaled to a 5-point Likert scale.}
\label{table:weekly-survey-results}
\end{table*}

%TC:endignore

\end{document}